\documentclass{aa}  
\usepackage{multirow}
\usepackage{xcolor}
\usepackage{lscape}
\usepackage{graphicx}
\usepackage{ulem}
\usepackage{txfonts}
\begin{document}

\title{Catalogue of exoplanets accessible in reflected starlight to the Nancy Grace Roman Space Telescope}
\subtitle{A population study and prospects for phase-curve measurements}

   \author{\'O. Carri\'on-Gonz\'alez\inst{1}\thanks{ \email{o.carriongonzalez@astro.physik.tu-berlin.de, oscar.carrion.gonzalez@gmail.com}}
          \and
          A. Garc\'ia Mu\~noz\inst{1,2}
          \and
          N. C. Santos\inst{3,4}
          \and
          J. Cabrera\inst{5}
          \and
          Sz. Csizmadia\inst{5}
          \and
          H. Rauer\inst{1,5,6}
          }

   \institute{Zentrum f\"ur Astronomie und Astrophysik, Technische Universit\"at Berlin, Hardenbergstraße 36, D-10623 Berlin, Germany
         \and AIM, CEA, CNRS, Université Paris-Saclay, Université de Paris, F-91191 Gif-sur-Yvette, France
         \and
             Instituto de Astrof\'isica e Ci\^encias do Espa\c{c}o, Universidade do Porto, CAUP, Rua das Estrelas, 4150-762 Porto, Portugal
         \and
            Departamento de F\'isica e Astronomia, Faculdade de Ci\^encias, Universidade do Porto, Rua do Campo Alegre, 4169-007 Porto, Portugal
         \and
             Deutsches Zentrum f\"ur Luft- und Raumfahrt, Rutherfordstraße 2, D-12489 Berlin, Germany
         \and
             Institute of Geological Sciences, Freie Universit\"at Berlin, Malteserstraße 74-100, D-12249 Berlin, Germany
                 }

   %\date{Received September 15, 1996; accepted March 16, 1997}

% \abstract{}{}{}{}{} 
% 5 {} token are mandatory
 
  \abstract
   {Reflected starlight measurements will open a new path in the characterization of directly imaged exoplanets. However, we still lack a population study of known targets amenable to this technique.   
   }
   {
   Here, we investigate which of the about 4300 exoplanets confirmed to date are accessible to the Roman Space Telescope’s coronagraph (CGI) in reflected starlight at reference wavelengths $\lambda$=575, 730 and 825 nm. We carry out a population study and also address the prospects for phase-curve measurements.
   }
   {We used the NASA Exoplanet Archive as a reference for planet and star properties, and explored the impact of their uncertainties on the exoplanet's detectability by applying statistical arguments.
   We define a planet as Roman-accessible on the basis of the instrument's inner and outer working angles and its minimum planet-to-star constrast (IWA, OWA, $C_{min}$). We adopt for these technical specifications three plausible configurations, labeled as pessimistic, intermediate and optimistic.
   Our key outputs for each exoplanet are its probability of being Roman-accessible ($P_{access}$), the range of observable phase angles, the evolution of its equilibrium temperature, the number of days per orbit that it is accessible and its transit probability.
   }
   {In the optimistic scenario, we find 26 Roman-accessible exoplanets with $P_{access}$>25\% and host stars brighter than $V$=7 mag. This population is biased towards planets more massive than Jupiter but also includes the super-Earths tau Cet e and f which orbit near their star’s habitable zone. A total of 13 planets are part of multiplanetary systems, 3 of them with known transiting companions, offering opportunities for contemporaneous atmospheric characterization.
   The intermediate and pessimistic scenarios yield 10 and 3 Roman-accessible exoplanets, respectively.
   We find that inclination estimates (e.g. with astrometry) are key for refining the detectability prospects.
   }
   {A science phase of the CGI has a remarkable potential to characterize the atmospheres of exoplanets that cannot be studied with other techniques.}

   %\keywords{catalogs --
               %planets and satellites: detection --
               %planets and satellites: fundamental parameters --
               %planets and satellites: gaseous planets --
               %planets and satellites: terrestrial planets
               %}

   \maketitle
%
%-------------------------------------------------------------------

\section{Introduction}\label{sec:introduction}

The population of more than 4000 exoplanets confirmed to date
shows a vast diversity of worlds, many of which have no analogue in the Solar System.
Yet, a large number of them, in particular those that orbit far from their host stars, are still not amenable to atmospheric characterization with the available techniques.
Upcoming direct imaging space telescopes
observing at optical wavelengths will enable the investigation of cold and temperate exoplanets on long-period orbits by measuring the starlight that they reflect.
The atmospheres of these planets remain largely unexplored but they may represent a key piece in the exoplanet diversity puzzle, helping trace the planets' history and evolution.

The Nancy Grace Roman Space Telescope\footnote{Formerly the Wide Field Infrared Survey Telescope, WFIRST.} \citep{spergeletal2013} (hereon, the Roman Telescope) will be the first space-borne facility designed to directly image exoplanets in reflected starlight.
Planned for launch in the mid 2020s, it will be equipped with an optical coronagraph and a set of filters for imaging and spectroscopy for technology demonstration \citep{akesonetal2019, mennessonetal2020}.
This instrument will be able to characterize far-out non-transiting exoplanets, most of them presumably discovered in radial velocity (RV) searches.
For long-period planets, reflected starlight measurements will provide insight into lower atmospheric layers than the layers probed during transit, which are masked by refraction
\citep{garciamunozetal2012, misraetal2014}.
Probing deep down in the atmosphere will be particularly relevant in the search for biosignatures \citep{raueretal2011}, which is a main goal of future direct-imaging missions targeting Earth-like exoplanets, such as LUVOIR \citep{bolcaretal2016} or HabEx \citep{mennessonetal2016}.

The question of which exoplanets will be observable by the Roman Telescope and next-generation direct imaging space telescopes is timely. Answering it will provide technical context for future designs, will motivate new and follow-up RV ans astrometric measurements, and will encourage modelers to build tools with which to interpret the prospective spectra. 
Understanding this population of exoplanets will help plan the observations and select the most interesting targets.
Several works have addressed the possible science outcome of direct-imaging missions and discussed potential criteria to define observation strategies
\citep[e.g.][]{traubetal2014, brown2015, greco-burrows2015, kaneetal2018, lacy-burrows2020, starketal2020}.

For instance, \citet{traubetal2014} studied the detection yield of different coronagraph architectures proposed for the WFIRST-AFTA mission based on a population of over 400 confirmed RV exoplanets, assuming for them circular orbits and sky-projected orbital inclinations $i$=60$^\circ$.
Depending on the specific coronagraph architecture, their predictions resulted in detection yields between 0 and 31 exoplanets.
\citet{brown2015} analysed also over 400 RV exoplanets lacking an inclination determination and tried to infer this value from simulated direct-imaging measurements to constrain the planets' true masses.
That study concluded that the uncertainties in the orbital parameters may prevent an accurate estimate of $i$. 
\citet{kaneetal2018} computed the maximum angular separation between planet and star ($\Delta \theta_{max}$) for a subset of 300 RV exoplanets.
That work identified those planets with the largest $\Delta \theta_{max}$ and estimated their orbital position and uncertainty as of 2025-01-01.

For exoplanets with incomplete orbital information, \citet{kaneetal2018} assumed inclination $i=90^\circ$,  eccentricity $e=0$ or argument of periastron $\omega=90^\circ$ when the corresponding parameter was missing.
However, they did not consider other factors affecting the detectability such as the planet-to-star contrast ratio ($F_p/F_\star$).
\citet{greco-burrows2015} studied how $F_p/F_\star$ changes with the orbital configuration of an exoplanet and its position on the orbit, and found that the contrast is indeed a major limitation for the detectability of direct-imaging exoplanets in reflected starlight.

Focusing on thermal emission rather than reflected starlight, and with the aim of specifying possible targets for the Roman Telescope, \citet{lacy-burrows2020} provided a list of 14 known self-luminous planets and brown-dwarf companions that might be observable in the optical wavelength range.
These objects will have larger contrasts than mature planets at the same orbital distance.
Although their study discusses the prospects to observe a reflected-light component in the spectra of such objects, their masses, temperatures and orbital distances in practice limit the eventual observations of these targets to primarily thermal emission.

Our first goal in this work is to determine which of the currently confirmed exoplanets could be observable in reflected starlight by the Roman Telescope.
For those planets whose orbital solution is not completely known, we compute
the likelihood of the exoplanet to be accessible based on a statistical analysis rather than assuming fixed values for the unconstrained parameters.
Our second goal is to understand the main properties of the population of known exoplanets that will be potentially detectable with the Roman Telescope.
We compare this subset to the whole population of confirmed exoplanets as well as to those that have been observed in transit.
This way we outline how direct-imaging space missions will contribute to completing the big picture of exoplanet diversity.

In addition, we explore the possibility of measuring the phase curve of these exoplanets.
To that end, we compute the planet-star-observer phase angles ($\alpha$) that would be observable and the corresponding uncertainties for each planet.
Optical phase-curve observations have proven valuable to constrain the atmospheric properties of Solar System planets \citep[e.g.][]{arking-potter1968, mallamaetal2006, garciamunozetal2014, dyudinaetal2016, mayorgaetal2016} and their energy budget \citep[e.g.][]{pollacketal1986, lietal2018}.
Optical phase curves have also been used to investigate the atmospheres of transiting exoplanets and infer their thermal properties and the presence of clouds \citep[e.g.][]{demoryetal2013, angerhausenetal2015, estevesetal2015, garciamunoz-isaak2015, huetal2015}.
According to recent theoretical investigations \citep{nayaketal2017, damianoetal2020}, observing at multiple phases will help better characterize directly-imaged exoplanets in reflected starlight.
Remarkably, no previous work has addressed the feasibility and limitations of such optical phase-curve measurements for the confirmed exoplanets, which is essential to prioritise the best targets for atmospheric characterization.

Finally, we discuss the benefits of constraining the orbital inclination by means of astrometric measurements or dynamical stability studies.
We do so by comparing, for a selection of exoplanets that have estimates of $i$ available, the detectability prospects if $i$ is assumed constrained or unconstrained.
Future data releases from the Gaia mission \citep{perrymanetal2001, gaia2016} and ensuing enhanced astrometry will strengthen these synergies.

The paper is structured as follows.
In Sect. \ref{sec:directimagingtechnical} we describe the general conditions under which an exoplanet would be accessible.
Section \ref{sec:orbit} contains the definition of the orbital geometry and the parameters determining the position and brightness of an exoplanet.
In Sect. \ref{sec:dataprocessing} we outline the dataset of planet and star properties
used in our study and the assumptions that we adopted.
We present our results in Sect. \ref{sec:results} and discuss more thoroughly in Sect. \ref{sec:discussion_selectedtargets} the observational prospects for a selection of particularly interesting targets, as well as the implications for their atmospheric characterization.
Section \ref{sec:conclusions} contains the summary and conclusions.

\section{Direct imaging of exoplanets. Technical requirements}
\label{sec:directimagingtechnical}

The technique of direct imaging applied to exoplanets relies on suppressing the light from their host stars with optical devices such as coronagraphs or starshades. 
In this way, the faint planetary point source can be distinguished from the stellar glare.
As the star is masked, a certain region around it is also masked. 
This region is defined by the inner working angle (IWA), and prevents the detection of planets at smaller star-planet angular separations.
Coronagraphs also have an outer working angle (OWA) that sets an outer limit to the observable region.

Another factor that affects the detectability of exoplanets is the minimum contrast ($C_{min}$) of the instrument.
The planet needs to be bright enough to be distinguished from background noise.
The usual way to quantify the planet brightness is through the 
contrast ratio between the flux from the planet and that from the star at a certain 
wavelength $\lambda$ and observing condition, given by: 
\begin{equation}
\label{eq:contrast}
\frac{F_p}{F_{\star}}= \left(\frac{R_p}{r(t)} \right)^2  A_g(\lambda) \, \Phi(\alpha, \lambda)
\end{equation}
where $R_p$ is the planet radius, $r$ is the star-planet distance at the orbital position being considered and $\alpha$ is the corresponding phase angle.
$A_g$ is the exoplanet's geometrical albedo and $\Phi$ is its normalized scattering phase law. 
Both $A_g$ and $\Phi$ depend on the properties of the planetary atmosphere.
These properties are discussed in more detail in Sect. \ref{sec:orbit}.

From this perspective, the limitations set by the IWA, OWA and $C_{min}$ shape the population of exoplanets that can be directly imaged.
For instance, hot and ultra-hot short-period planets orbit too close to their host star and thus inside the IWA of any realistic coronagraph, which means that they are undetectable. 
In turn, exoplanets on long-period orbits and inclinations close to face-on may fall outside the OWA during their whole orbit, which prevents them from being observed.
In addition, the planet-to-star contrast decreases as the planet-star distance increases and hence observing planets in reflected starlight will become progressively difficult for the longer-period ones.
This is particularly important for small exoplanets, as the amount of photons reflected by them scales with the object's cross section.

In this work, we consider as a basis the mission design of the Roman Telescope as envisioned in \citet{spergeletal2015}, with a telescope diameter of $D$=2.4 m.
It will be equipped with a Coronagraph Instrument (CGI) including an optical hybrid Lyot coronagraph and a shaped pupil coronagraph \citep{traugeretal2016}, as a technology demonstrator for future direct-imaging missions targeting Earth-like planets.
The original design aimed at a minimum planet-to-star contrast ratio $C_{min}$ on the order of 10$^{-9}$ after post-processing \citep{spergeletal2015, douglasetal2018}.
More up-to-date expectations according to the Nancy Grace Roman Space Telescope on the IPAC (Roman-IPAC) website\footnote{\url{https://roman.ipac.caltech.edu/sims/Param_db.html}} aim for $C_{min}$ of about 2-3$\times$10$^{-9}$ at a moderate signal-to-noise ratio S/N=5.

At the time of writing, only one spectroscopy filter, centred at 730 nm, and two imaging filters, centred at 575 and 825 nm, are planned for full commissioning.
However, other filters which are not officially supported will fly with the coronagraph and might be commissioned for science operations if the 3-month technology demonstration phase is successful and a potential science phase is funded \citep{akesonetal2019}.
The three currently official observing modes according to the Roman-IPAC website are: Imaging Mode N (IWA=$3\, \lambda / D$, OWA=$9.7\, \lambda / D$, $C_{min}$=$2.94\times10^{-9}$), Spectroscopy Mode (IWA=$3\, \lambda / D$, OWA=$9.1\, \lambda / D$, $C_{min}$=$2.2\times10^{-9}$) and Imaging Mode W (IWA=$5.9\, \lambda / D$, OWA=$20.1\, \lambda / D$, $C_{min}$=$1.95\times10^{-9}$).
The latter mode will be mainly devoted to debris discs observations \citep{akesonetal2019}.

As these figures and the $C_{min}$ requirement will likely evolve as the mission design progresses, in this work we will adopt three possible configurations of IWA, OWA and $C_{min}$ for the exoplanet observing modes (Table \ref{table:instrument_scenarios}). 
We define a pessimistic scenario with: IWA=$4\, \lambda / D$, OWA=$8\, \lambda / D$, $C_{min}$=$5\times10^{-9}$; an intermediate scenario with IWA=$3.5\, \lambda / D$, OWA=$8.5\, \lambda / D$, $C_{min}$=$3\times10^{-9}$; and an optimistic scenario with: IWA=$3\, \lambda / D$, OWA=$9\, \lambda / D$, $C_{min}$=$1\times10^{-9}$.
These are not officially-bounded scenarios and different performances of the instrument (e.g. worse than our pessimistic scenario) cannot be ruled out.
However, the cases proposed herein are representative of a plausible range of performances within the CGI capabilities considered realistic at this point.
Table \ref{table:instrument} summarizes the available CGI filters and corresponding IWA and OWA for the optimistic scenario.
Unless noted otherwise, we assume as a reference in this work the imaging filter centred at 575 nm.

We acknowledge that additional factors will limit the detectability of exoplanets by the Roman Telescope.
For instance, the most recent update on the Roman-IPAC website (14.01.2021) states a CGI host star requirement of 
$V\le$5 mag but also notes that stars with $V=6-7$ could potentially be targeted.
The performance of the instrument on such fainter stars is still to be determined after the technology demonstration phase.
The solar or anti-solar telescope pointing at the time of the observation may also affect any proposed target list \citep[e.g.][]{brown2015}, although zodiacal light will not be as determinant as in future instruments with 10-100 times more contrast sensitivity.
This effect, however, will depend on the final launch date and mission schedule.
Exo-zodiacal dust may also prevent the detection of certain targets but this noise source will have to be analysed on a one-by-one basis through follow-up observations of each planetary system and will not be considered here.

For the sake of generality, we adopt the IWA, OWA and $C_{min}$ at $\lambda$=575 nm as our main detectability criteria.
For those exoplanets meeting these criteria, we coin the term Roman-accessible.
Given that our current focus is on the geometrical constraints for exoplanet detectability, we leave for future work the computation of the S/N that could be achieved for each Roman-accessible planet or the required integration times.

\begin{table}
\caption{Plausible configurations of CGI exoplanet observing modes that will be considered in this work. These scenarios are not officially bounded but are within the range of realistic CGI performances according to current predictions.}
\label{table:instrument_scenarios}
\centering 
\begin{tabular}{c c c c}
\hline \hline
    Scenario        &   $C_{min}$   &       IWA              &  OWA  \\
\hline
    Pessimistic     &   $5\times10^{-9}$ &   $4\, \lambda / D$  &   $8\, \lambda / D$ \\
    Intermediate    &   $3\times10^{-9}$ &   $3.5\, \lambda / D$ &   $8.5\, \lambda / D$ \\
    Optimistic      &   $1\times10^{-9}$ &   $3\, \lambda / D$ &   $9\, \lambda / D$ \\
\hline
\end{tabular}
\end{table}

\begin{table}
\caption{Filters of the CGI of the Roman Telescope and corresponding IWA and OWA for the optimistic configuration scenario (Table \ref{table:instrument_scenarios}). Engineering filters could be used as regular broadband imaging filters if commissioned for exoplanet observations after the technology demonstration phase. }
\label{table:instrument}
\centering 
\begin{tabular}{c c c c c}
\hline \hline
    $\lambda_{center}$ & Mode       & Commissioned  & IWA    & OWA \\
    $[$nm]             &            &             & [mas]    & [mas] \\
\hline
550 & Engineering  & No  & 142    & 425 \\
575 & Imaging  & Yes  & 148    & 445 \\
575 & Engineering  & No  & 148    & 445 \\
599 & Engineering  & No  & 154    & 463 \\
615 & Engineering  & No  & 159    & 476 \\
638 & Engineering  & No  & 164    & 493 \\
656.3 & Engineering  & No  & 169    & 508 \\
660 & Spectroscopy  & No  & 170    & 511 \\
681 & Engineering  & No  & 176    & 527 \\
704 & Engineering  & No  & 182    & 545 \\
727 & Engineering  & No  & 187    & 562 \\
730 & Spectroscopy  & Yes  & 188    & 565 \\
752 & Engineering  & No  & 194    & 582 \\
754 & Engineering  & No  & 194    & 583 \\
777.5 & Engineering  & No  & 200    & 601 \\
792 & Engineering  & No  & 204    & 613 \\
825 & Imaging  & Yes  & 213    & 638 \\
825 & Engineering  & No  & 213    & 638 \\
857 & Engineering  & No  & 221    & 663 \\
\hline
\end{tabular}
\end{table}

\section{Theoretical setting: planet detectability along the orbit} \label{sec:orbit}

In this section, we lay out the equations for the trajectory of a planet in the three-dimensional space and the evolution of $\Delta \theta$, $\alpha$ and $F_p/F_\star$ with time.
We base the description of the planet orbit on the book chapter by \citet{hatzes2016}. 
Figure \ref{fig:sketch_orbit_XYZ} sketches the geometry and main elements of the orbit and is based on Fig. 1.36 in that chapter, with additional information specific to the reference axes.

For a general elliptic orbit, the distance between planet and star at each orbital position is given by:
\begin{equation} \label{eq:kepler_planet-star_distance}
r = \frac{a\, (1-e^2)}{1+e\, cosf}
\end{equation}
Here, $e$ is the eccentricity, $a$ is the semi-major axis and $f$ is the true anomaly.
A more thorough description of the orbital equations and parameters can be found in Appendix \ref{sec:appendix_equations_orbit}.

The orbit can be given in a three-dimensional space with the host star at the origin, the $X$ and $Y$ axes defining the plane of the sky and $Z$ oriented away from the observer.
The three coordinates of the planet's position vector $\boldsymbol{r_p}$ are:
\begin{equation} \label{eq:kepler_XYZ}
\begin{aligned}
X = r\, cos\,(\omega_p+f) \\
Y = r\, cos\, i\,\,  sin\,(\omega_p+f) \\
Z = r\, sin\, i\,\, sin\,(\omega_p+f)
\end{aligned}
\end{equation}
where $i$ is the orbital inclination and $\omega_p$ is the planet's argument of periastron. 
In this work, the longitude of ascending node is assumed $\Omega$ = 0 without loss of generality.

\subsection{Angular separation} \label{subsec:orbit_maxangsep}
The sky-projected distance between planet and star is given by:, 
\begin{equation} \label{eq:kepler_rsky}
\sqrt{X^2+Y^2} = r\, \sqrt{1-sin^2(\omega_p + f)\, sin^2i}
\end{equation}

If the stellar system is located at a distance $d$ from the observer, the apparent angular separation is:
\begin{equation} \label{eq:kepler_angular_separation}
\Delta \theta = \frac{\sqrt{X^2+Y^2}}{d}
\end{equation}

\subsection{Observed phase angles} \label{subsec:orbit_alphas}

\begin{figure}
   \centering
   \includegraphics[width=9.cm]{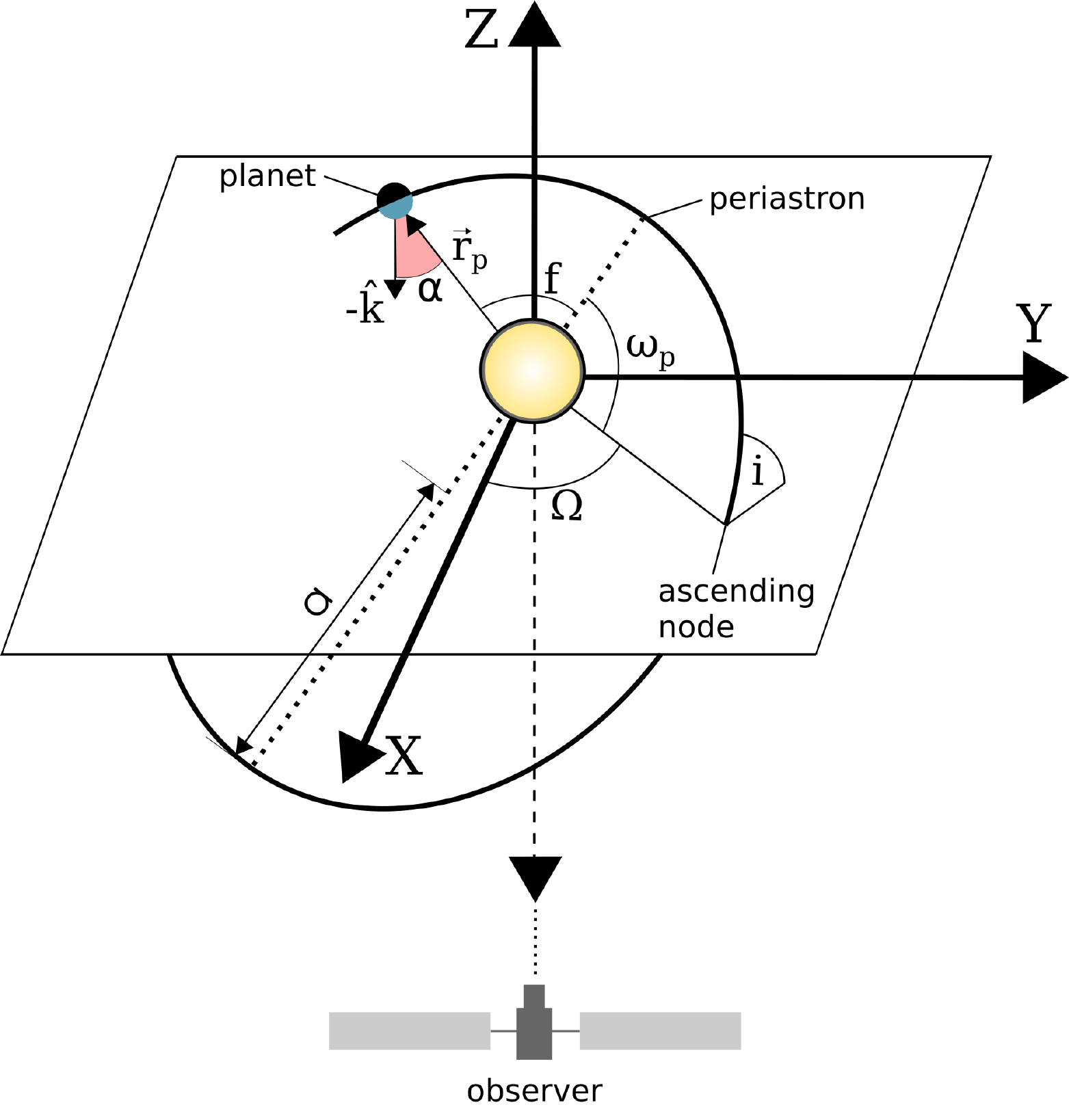}
      \caption{\label{fig:sketch_orbit_XYZ} 
      Sketch of the orbital geometry and graphical definition of the planet-star-observer phase angle $\alpha$, for a planet at a certain position on its orbit.
      }
   \end{figure}

The phase angle $\alpha$ is the planetocentric angle between the directions to the star and to the observer (see Fig. \ref{fig:sketch_orbit_XYZ}).
It can be computed at each orbital position from the dot product of the reversed planet's position vector ($-\boldsymbol{r_p}$) and a unit vector in the direction of the observer ($-\boldsymbol{\hat{k}}$, as $d \gg r$).
With the components of $\boldsymbol{r_p}$ defined in Eq. (\ref{eq:kepler_XYZ}):

\begin{equation} \label{eq:kepler_alpha}
\alpha = cos^{-1}(sin\, i\,\, sin \,(\omega_p+f) )
\end{equation}

\subsection{Scattering and planet-to-star contrast} \label{subsec:orbit_contrast}
To compute the brightness of the planet at each orbital position, we substitute the expressions given 
above for $r$ and $\alpha$ into Eq. (\ref{eq:contrast}).
We assume for the planet a Lambertian scattering phase law:

\begin{equation} \label{eq:lambertian}
    \Phi(\alpha) = \frac{sin\,\alpha + (\pi - \alpha)\, cos\,\alpha}{\pi}
\end{equation}
and a geometrical albedo $A_g$=0.3. 
Both $A_g$ and $\Phi(\alpha)$ are assumed to be wavelength-independent and to
represent the planet's reflecting properties over the operational spectral range of the Roman Telescope.

Our assumed albedo provides a reasonable representation of the outer planets in the Solar System \citep{karkoschka1994, karkoschka1998}. 
Other works investigating the prospects for reflected-starlight measurements of exoplanets have also assumed or predicted values of $A_g$ between 0.3 and 0.5 for Neptune and Jupiter analogues \citep[e.g][]{cahoyetal2009,cahoyetal2010,traubetal2014,greco-burrows2015}.
Larger values of $A_g$ will potentially increase the number of exoplanets exceeding 
the $C_{min}$ of the instrument, and vice versa.

The Lambertian scattering phase law is a simple yet pragmatic approximation to the scattering of planetary atmospheres.
It has been frequently applied in studies planning the science outcome of reflected-starlight observations of exoplanets \citep[e.g.][]{starketal2014, guimond-cowan2018}. 
At small phase angles, the Lambertian function yields brighter values than other models such as isotropic or Rayleigh-like scattering.
The results under a Lambertian assumption may differ slightly from those obtained with other phase laws.
Nevertheless, in reality the scattering properties of a planet will depend on the specifics of its atmosphere, which will be unknown a priori.

\subsection{Time dependence of the orbital position} \label{subsec:orbit_time}
The relation between the true anomaly and time ($t$) can be derived from Kepler's equation (Appendix \ref{sec:appendix_equations_orbit}). 

For an exoplanet with orbital period $P$ and time of periastron passage $t_p$:

\begin{equation}    \label{eq:kepler_time_dependence}
    \frac{t-t_p}{P} = \frac{1}{2\,\pi} \left[ 2\, arctan\left( \sqrt{\frac{1-e}{1+e}}\, tan\frac{f}{2} \right) - e\,\frac{sin\,f\, \sqrt{1-e^2}}{1+e\, cos\,f} \right]
\end{equation}

This equation combined with Eqs. (\ref{eq:kepler_planet-star_distance}), (\ref{eq:kepler_angular_separation}) and (\ref{eq:kepler_alpha}) yields, respectively, the planet-star distance, angular separation and phase angle at a given time.
Leaving aside $\Omega$ and $t_{\rm{p}}$, which are not important here (Appendix \ref{sec:appendix_equations_orbit}), the planet's orbit is specified by 5 parameters, namely: $a$, $i$, $\omega_p$, $e$ and $P$.

\section{Building a complete set of confirmed exoplanets}
\label{sec:dataprocessing}

We downloaded the complete set of confirmed exoplanets from the NASA Exoplanet Archive\footnote{\url{https://exoplanetarchive.ipac.caltech.edu/}} \citep{akesonetal2013}, that we used as our main source of known planets and corresponding planet-star properties.
As of 16$^{th}$ of September (2020), it contains 4276 confirmed planets.
For specific targets, complementary information was obtained from the original references, from correspondence with the paper authors or from other resources such as the Extrasolar Planets Encyclopaedia \citep{schneideretal2011}.
Hereon, we refer to this compilation mainly based on the NASA Exoplanet Archive as the input catalogue, shown in Table \ref{table:NASA_database}.

\longtab{
\tiny
\begin{landscape}
\begin{longtable}{ l  c  c  c  c  c  c  c  c  c  c  c  c  c } 
\caption{\label{table:NASA_database} 
Main planetary and stellar properties from the input catalogue for the confirmed exoplanets that are Roman-accessible.
For reference, we list all planets accessible in the optimistic CGI scenario, including all possible values of $P_{access}$ and host-star magnitudes.
The quoted values correspond mainly to the information in the NASA Exoplanet Archive as of 16$^{th}$ of September, 2020 (see Sect. \ref{sec:dataprocessing}).
The parameters for bet Pic c were obtained from \citet{nowaketal2020} because the quoted values in the NASA Archive show no upper or lower uncertainties.
For DE CVn b, the inclination value was obtained from \citet{hanetal2018} and the corresponding uncertainties, from \citet{besselaaretal2007}.
} \\
 \hline 
 \hline
  Planet & {$d$} & {$P$} & {$a$} & {$M_p$} & {$R_p$} & {$i$} & $e$ & {$\omega_p$} & $T_{\star}$ & St. type & {$M_\star$} & {$V$} & Age \\
  & [pc] & [days] & [AU] & [$M_{J}$] & [$R_{J}$] & [deg] & & [deg] & K & & [$M_\odot$] & [mag] & [Gyr]  \\
\hline
  \endfirsthead
\caption{continued.}\\
\hline
\hline                 
  Planet & {$d$} & {$P$} & {$a$} & {$M_p$} & {$R_p$} & {$i$} & $e$ & {$\omega_p$} & $T_{\star}$ & St. type & {$M_\star$} & {$V$} & Age \\
  & [pc] & [days] & [AU] & [$M_{J}$] & [$R_{J}$] & [deg] & & [deg] & K & & [$M_\odot$] & [mag] & [Gyr]  \\
\hline
\endhead
\hline
\endfoot
GJ 687 c 		&  4.55$^{ +0.00 }_{ 0.00 }$ 		&  727.6$^{ +12.2 }_{ -12.2 }$ 		&  1.17$^{ +0.02 }_{ -0.02 }$ 		&  0.050$^{ +0.013 }_{ -0.013 }$ $^\dagger$ 		&  $-$ 		&  $-$ 		&  0.40$^{ +0.22 }_{ -0.22 }$ 		&  356.0$^{ +63.0 }_{ -63.0 }$ 		& 3826 		&  M3 V 		&  0.40$^{ +0.02 }_{ -0.02 }$ 		&  9.150 		&  $-$ \\ 
55 Cnc d 		&  12.59$^{ +0.01 }_{ -0.01 }$ 		&  4825.0$^{ +39.0 }_{ -39.0 }$ 		&  5.50$^{ +0.03 }_{ -0.03 }$ 		&  3.878$^{ +0.068 }_{ -0.068 }$ $^\dagger$ 		&  $-$ 		&  90.0$^{ +60.0 }_{ -60.0 }$ 		&  0.02$^{ +0.01 }_{ -0.01 }$ 		&  224.0$^{ +31.0 }_{ -31.0 }$ 		& 5196 		&  G8 V 		&  0.91$^{ +0.01 }_{ -0.01 }$ 		&  5.960 		&  10.20$^{ +2.50 }_{ -2.50 }$ \\ 
HD 114613 b 		&  20.29$^{ +0.09 }_{ -0.09 }$ 		&  4000.0$^{ +120.0 }_{ -120.0 }$ 		&  5.34$^{ +0.26 }_{ -0.26 }$ 		&  0.357$^{ +0.032 }_{ -0.032 }$ $^\dagger$ 		&  $-$ 		&  $-$ 		&  0.46$^{ +0.09 }_{ -0.09 }$ 		&  16.0$^{ +11.0 }_{ -11.0 }$ 		& 5641 		&  G4 IV 		&  1.27$^{  }_{  }$ 		&  4.850 		&  $-$ \\ 
ups And d 		&  13.41$^{ +0.06 }_{ -0.06 }$ 		&  1276.5$^{ +0.6 }_{ -0.6 }$ 		&  2.51$^{ +0.00 }_{ -0.00 }$ 		&  4.132$^{ +0.029 }_{ -0.029 }$ $^\dagger$ 		&  $-$ 		&  $-$ 		&  0.30$^{ +0.01 }_{ -0.01 }$ 		&  78.8$^{ +0.4 }_{ -0.4 }$ 		& 6105 		&  F8 V 		&  1.30$^{  }_{  }$ 		&  4.100 		&  5.00$^{  }_{  }$ \\ 
HD 217107 c 		&  20.07$^{ +0.02 }_{ -0.02 }$ 		&  4270.0$^{ +220.0 }_{ -220.0 }$ 		&  5.32$^{ +0.38 }_{ -0.38 }$ 		&  2.600$^{ +0.150 }_{ -0.150 }$ $^\dagger$ 		&  $-$ 		&  $-$ 		&  0.52$^{ +0.03 }_{ -0.03 }$ 		&  18.6$^{ +6.0 }_{ -6.0 }$ 		& 5622 		&  G8 IV 		&  1.00$^{ +0.05 }_{ -0.05 }$ 		&  6.160 		&  $-$ \\ 
GJ 179 b 		&  12.36$^{ +0.01 }_{ -0.01 }$ 		&  2288.0$^{ +59.0 }_{ -59.0 }$ 		&  2.41$^{ +0.04 }_{ -0.04 }$ 		&  0.820$^{ +0.070 }_{ -0.070 }$ $^\dagger$ 		&  $-$ 		&  $-$ 		&  0.21$^{ +0.08 }_{ -0.08 }$ 		&  333.0$^{ +24.0 }_{ -24.0 }$ 		& 3370 		&  M3.5 V 		&  0.36$^{ +0.03 }_{ -0.03 }$ 		&  12.000 		&  $-$ \\
\hline
\end{longtable}
\tablefoot{\\
 $^\dagger$ indicates that the $M_p$ value corresponds to $M_p sin(i)$. \\
We note that, in those cases where the quoted uncertainties are 0.00, this is a result of insufficient significant figures in the rounding. We have not included upper or lower uncertainties for those cases in which they were not reported in the NASA Archive.\\
In those cases where the quoted uncertainties of $e$ reach nonphysical values (e.g. $e$<0 for HD 154345 b, HD 114783 c or HD 62509 b), we truncate our sampling to a physical range of values.\\
\tablefoottext{a}{2MASS J21402931+1625183 A b}\\
\tablefoottext{b}{WISEP J121756.91+162640.2 A b}\\
\tablefoottext{c}{CFBDSIR J145829+101343 b}\\
\tablefoottext{d}{VHS J125601.92-125723.9 b}\\
\tablefoottext{e}{2MASS J02192210-3925225 b}\\
\tablefoottext{f}{2MASS J01225093-2439505 b}\\
}
\end{landscape}
}

\subsection{Completing missing orbital information} \label{subsec:dataprocessing_orbits}

Not all of the Keplerian elements are known or listed in the input catalogue for each of the confirmed exoplanets.
If any orbital parameter is missing, we need to make additional assumptions in order to compute the orbital solution.
For 246 exoplanets, $a$ is missing but $P$ as well as the masses of the star ($M_{\star}$) and the planet ($M_p$) are available.
For 124 of them, $P$ is missing but $a$, $M_{\star}$ and $M_p$ are available.
In such cases, we compute the missing value by means of Kepler's third law.

Still, there is a significant number of exoplanets (2513) with no information on $M_p$ or $M_p\,sin\,i$.
For these, we approximate $M_{\star}+M_p$$\approx$$M_{\star}$, which results in a negligible underestimation of $a$ for planetary-mass objects \citep{stevens-gaudi2013}.
There are 119 exoplanets with no available information on at least two of the three critical parameters in Kepler's third law ($M_\star$, $P$, $a$), making it impossible to include them in our study.

When the values of the orbital inclination or the argument of periastron are not available in the NASA Exoplanet Archive, we assigned them random values assuming that the possible orbital orientations are isotropically distributed with respect to the observer.
We therefore assume $cos\, i$ and $\omega_p$ to be distributed uniformly over the intervals [$-1$,$1$] and $[0, 2\pi]$, respectively.

\subsubsection{A note of caution about $\omega_p$} \label{subsubsec:dataprocessing_orbits_longperiast}
There is no homogeneous convention in the literature to report the argument of periastron.
This has been noted previously \citep[e.g.][]{brown2015, xuan-wyatt2020} but stands out as a particularly relevant issue for our work and for the direct imaging of RV planets.
In some cases the reported $\omega$ refers to the argument of periastron of the planet ($\omega_p$) as it orbits around the system's barycenter, while in others it refers to the argument of periastron of the star ($\omega_\star$).
There is a shift of 180$^\circ$ between $\omega_p$ and $\omega_\star$ ($\omega_p = \omega_\star + 180^\circ$) \citep{perryman2011}.
In addition, the assumed location of the observer with respect to the $+Z$ axis and the definition of the origin for the argument of periastron may also introduce additional  180$^\circ$$-$shifts in $\omega$.

The lack of a homogeneous convention and the fact that it is not always stated how the reported $\omega$ is defined potentially complicate a systematic analysis as proposed in this work.
We verified that both the NASA Exoplanet Archive and the Extrasolar Planets Encyclopaedia quote, for each exoplanet, the $\omega$ given in the original reference without assessing the actual definitions used in them.\footnote{Exoplanet Archive Service Desk and J. Schneider respectively, private comm. }

The value of $\omega$ has no impact on the range of angular separations over the orbit (see Eq. \ref{eq:kepler_angular_separation}) but it does affect the position of an exoplanet at a given time (Eq. \ref{eq:kepler_XYZ}), its phase angle (Eq. \ref{eq:kepler_alpha}) and therefore the value of $F_p/F_\star$. 
$\omega$ will also have an impact on the probability that a planet will transit its host star (see Sect. \ref{subsec:dataprocessing_transitprobability}).
As the design of direct-imaging missions and the corresponding target selection progress, it would be desirable to have clearly defined conventions for all the reported orbital parameters.
We therefore urge efforts towards a standardisation of the data available in the exoplanet catalogues and towards the compilation of self-consistent catalogues \citep[e.g.][]{hollisetal2012} which are updated with new discoveries.
We discuss in Appendix \ref{sec:appendix_effectomega} how mistakenly using the value of $\omega_\star$ instead of $\omega_p$ affects the detectability of exoplanets and the prospects for measuring their optical phase curves.

In this work, we will generally assume that the $\omega$ reported by  the NASA Exoplanet Archive corresponds to the argument of periastron of the star, which is the prevailing convention for RV \citep{perryman2011, hatzes2016}.
For all the exoplanets that we find to be Roman-accessible (see Sect. \ref{sec:results}), we checked the corresponding reference papers or contacted the authors to confirm the values of $\omega$ as quoted in Table \ref{table:NASA_database}. 
Extending this case-by-case inspection to the 4276 confirmed exoplanets is out of the scope of this paper.

\subsubsection{Eccentricity distribution} \label{subsubsec:dataprocessing_orbits_ecc}

For those exoplanets without a measurement of eccentricity, we draw it from a uniform distribution in $e \in [0, 1)$. 
This is a simplification to the reality, which suggests that short-period exoplanets tend to have small eccentricities while long-period ones show broader $e$ distributions \citep{winn-fabrycky2015}.
However, empirically-derived distributions of $e$ might be affected by observational biases, especially for long-period planets, whose orbits are more challenging to characterize and for which the discovery numbers are relatively low.
For reference, uniform distributions of $e$ have been used in previous works that analysed the detection yield of direct-imaging missions \citep[e.g.][]{starketal2014}.

We note however that this is not the only approach considered in the literature.
For instance, \citet{steffenetal2010} used both Rayleigh and exponential probability distributions to describe the eccentricity, and \citet{wang-ford2011} used a distribution with both uniform and exponential components.
\citet{kipping2013} described the observed dispersion of $e$ with two Beta probability distributions, for short- and long-period planets ($P$<382.3 and >382.3 days, respectively).
In Sect. \ref{subsec:results_populationstudy} we compare the $e$ distributions of exoplanets with short and long periods, as described by \citet{kipping2013}, with that of the Roman-accessible exoplanets.
Future studies with access to a larger sample of long-period planets will result in refined representations of the $e$ distribution.

\subsection{Planet radius} \label{subsec:dataprocessing_radius}
The value of $R_p$ can only be measured for transiting exoplanets.
It may be estimated from thermal emission measurements, as for instance with young, self-luminous exoplanets, but these estimates are by definition model dependent \citep[e.g.][]{mawetetal2019, lacy-burrows2020}.
Hence, the population of exoplanets suitable for direct imaging in reflected starlight will generally lack an estimate of $R_p$.

To assign a value of $R_p$ to the planets in our input catalogue, 
we use the mass-density relationship from \citet{hatzes-rauer2015} for giant planets, defined in term of Jupiter's mass ($M_{J}$) as those with $0.3M_{J}<M_p<65M_{J}$:
\begin{equation}    \label{eq:mass-density_giants}
    log_{10}(\rho)\, [g\,cm^{-3}] = (1.15 \pm 0.03)\, log_{10}(M_p/M_J) - (0.11 \pm 0.03)
\end{equation}
Eq. (\ref{eq:mass-density_giants}) is therefore valid for planets more massive than Saturn, approximately.
A priori, we cannot rule out that lower-mass exoplanets will be detectable \citep{robinsonetal2016} (see also Sect. \ref{subsec:results_populationstudy}).
Thus, for planets less massive than 120 Earth masses ($M_\oplus$), we use the mass-radius relationships in \citet{otegietal2020}.
They distinguish between rocky and volatile-rich exoplanets, and obtain two different mass-radius relationships depending on the planet density ($\rho$):
\begin{subequations}    \label{eq:mass-radius_lowmass}
\begin{align}
    R_p/R_\oplus = (1.03 \pm 0.02)\, (M_p/M_\oplus)^{0.29 \pm 0.01},\, \rm{if}\, \rho > 3.3\,g\,cm^{-3} & \label{eq:mass-radius_lowmass_rocky} \\ 
    R_p/R_\oplus = (0.70 \pm 0.11)\, (M_p/M_\oplus)^{0.63 \pm 0.04},\, \rm{if}\, \rho < 3.3\,g\,cm^{-3} & \label{eq:mass-radius_lowmass_volatiles} 
\end{align}
\end{subequations}

Although \citet{otegietal2020} note that the $M_{\rm{p}}$--$R_{\rm{p}}$ statistics suggest a lower limit of 5$M_\oplus$ for volatile-rich planets, we extend the mass-radius relationship to 3.1$M_\oplus$ in order to achieve a continuous coverage in $M_p$. 
This causes that some exoplanets with $\rho > 3.3\,\rm{g}\,\rm{cm}^{-3}$ (those with 3.1$M_\oplus$<$M_p$<5$M_\oplus$) are modeled in our case with Eq. (\ref{eq:mass-radius_lowmass_volatiles}).

\begin{figure}
   \centering
   \includegraphics[width=9.cm]{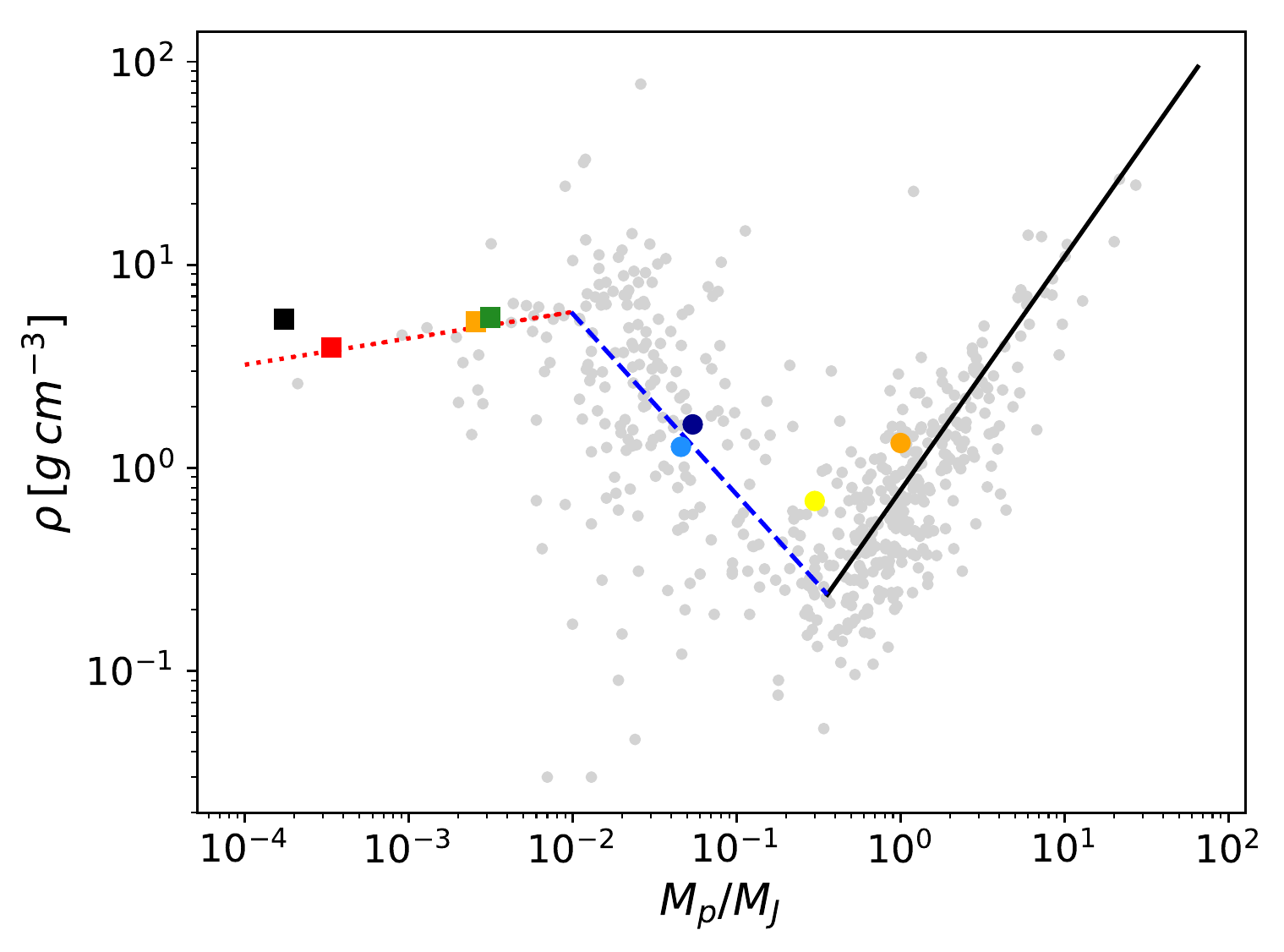}
      \caption{\label{fig:mass_density} 
      Grey dots: confirmed exoplanets with a known value of both $M_p$ and $\rho$.
      Planets for which only $M_p\,sin\,i$ is known are not included. 
      Solid black line: mass-density relationship for giant exoplanets, obtained from \citet{hatzes-rauer2015}.
      Dashed blue line: mass-density relationship for low-mass exoplanets with volatile-rich composition, obtained from the $M_p$-$R_p$ relations in \citet{otegietal2020}.
      Dotted red line: mass-density relationship for low-mass exoplanets with rocky composition, obtained from the $M_p$-$R_p$ relations in \citet{otegietal2020}.
      Coloured squares: rocky planets of the Solar System (Mercury, black; Venus, orange; Earth, green; Mars, red).
      Coloured dots: giant planets of the Solar System (Jupiter, orange; Saturn, yellow; Uranus, light blue; Neptune, dark blue).
      }
   \end{figure}

In summary, for planets with $M_p$<3.1$M_\oplus$ we use the rocky $M_p$-$R_p$ relationship in Eq. (\ref{eq:mass-radius_lowmass_rocky}), for 3.1$M_\oplus$<$M_p$<0.36$M_{J}$ we use the volatile-rich relationship in Eq. (\ref{eq:mass-radius_lowmass_volatiles}) and for $M_p$>0.36$M_{J}$ we use the giant-planet relationship in Eq. (\ref{eq:mass-density_giants}).
In all cases, we account for the quoted uncertainties to estimate $R_{\rm{p}}$ (see Sect. \ref{subsec:dataprocessing_bootstraping}).
Figure \ref{fig:mass_density} shows these relationships together with all of the confirmed exoplanets with measurements of both $M_p$ and $\rho$ in the NASA Exoplanet Archive.
For reference, we added the Solar System planets to the diagram.
We find an overall good fit to the observed population of both Solar System and extrasolar planets.

\begin{table*}
\caption{Summary of the parameters used to compute the exoplanet detectability.}
\label{table:params_description}
\centering
    \begin{tabular}{c c l l}
    \hline
    \hline
    Parameter   &   Units   &   Description             &   Comments\\ 
    \hline
    $d$         &   pc      &   Distance to the star  &   \\
    $a$         &   AU      &   Orbital semi-major axis &           \\
    $P$         &   days    &   Orbital period          &   \tiny{Used to compute $a$ through Kepler's third law, if not available in the NASA Archive}.
    \\
    $R_\star$   &   $R_\odot$   &   Stellar radius      &   \\
    $M_\star$   &   $M_\odot$   &   Stellar mass        &   \tiny{Used to compute $a$ through Kepler's third law.}\\
    $T_\star$   &   K   &   Stellar effective temperature      &   \tiny{Used to compute $T_{eq}$ through Eq. (\ref{eq:Teq}).}\\
    $T_{eq}$    &   K   &   Planet equilibrium temperature      &   \\
    $R_p$       &   $R_J$   &   Planet radius           &   \\
    $M_p$       &   $M_J$   &   Planet mass             &   \tiny{Used to compute $R_p$ through $M_p$-$R_p$ relationships, if not available in the input catalogue}\\
    $i$         &   deg     &   Orbital inclination     &   \tiny{If unknown, we draw the value from a uniform distribution $cos\,i \in [-1,1]$ }\\
    $\omega_p$    &   deg     &   Argument of periastron  &   \tiny{If unknown, we draw the value from a uniform distribution $\omega_p \in [0, 2\pi]$.}\\
    $e$         &   $-$     &   Orbital eccentricity    &   \tiny{If unknown, we draw the value from a uniform distribution $e \in [0,1)$.}\\
\hline
\end{tabular}
\tablefoot{In our computations we have considered the uncertainties of all these parameters as explained in Sect. \ref{subsec:dataprocessing_bootstraping}.
}
\end{table*}

\subsection{Transit requirements} \label{subsec:dataprocessing_transitprobability}

Exoplanets that are suitable for both direct imaging and transit spectroscopy will become prime targets for atmospheric characterization \citep{carriongonzalezetal2020, starketal2020}.
Given their special interest, we computed the transit probability ($P_{tr}$) of the Roman-accessible exoplanets (Section \ref{sec:results}).

Based on Eq. (\ref{eq:kepler_rsky}) the eventual eclipses (transits and occultations) will take place when the planet-star distance in the sky plane $\sqrt{X^2+Y^2}$ is a local minimum.
Following \citet{winn2010}, we consider that transits happen at inferior conjunctions (that is, when $X$=0 and the planet is in front of the star). 
With our viewing geometry (Fig. \ref{fig:sketch_orbit_XYZ}) this means:

\begin{equation} \label{eq:f_transits}
f_{tr}=+\frac{3 \pi}{2}-\omega_p
\end{equation}

The impact parameter is defined as the distance between the centres of the planet and the star, projected onto the plane of the sky and normalized to the stellar radius.
Substituting Eq. (\ref{eq:f_transits}) in Eq. (\ref{eq:kepler_rsky}), the impact parameter at transit is given by:
\begin{equation} \label{eq:b_transits}
b_{tr}=\frac{a}{R_\star}\, \left(\frac{1-e^2}{1-e\, sin\,\omega_p} \right)\, cos\,i
\end{equation}

The condition for a full transit to be observed is therefore:
\begin{equation} \label{eq:transit_condition}
|b_{tr}|<\frac{R_\star-R_p}{R_\star}.
\end{equation}
We use $R_\star-R_p$ to exclude grazing transits from the analysis because these only provide a lower limit for $R_p$.
For those systems without a $R_\star$ determination in the input catalogue, we extracted its value from the Planetary Systems database in the NASA Exoplanet Archive.
Preferentially, we used the value from the source referencing Gaia DR2 \citep{gaia2018} or, if unavailable, from the one referencing the Revised TESS Input Catalog \citep{stassunetal2019}. 
If $R_\star$ was not available in any of these sources either, the transit probability could not be computed for that system.

The mass of a planet discovered in RV cannot be unlimitedly large, and this sets a limit on the range of physically realistic inclinations for a measured $M_p\,sin\,i$.
In this respect, \citet{stevens-gaudi2013} note that the prior distribution of possible $M_p$ affects the prior distribution of $i$, thereby affecting the calculated transit probabilities.
For generality, we will not consider here any prior information on the $M_p$ distribution.

\subsection{Planetary equilibrium temperature} \label{subsec:dataprocessing_Teq}
The equilibrium temperature of a planet $T_{eq}$ provides an indication of its possible atmospheric structure and the potential conditions for habitability. 
For each orbital position $r$, we computed $T_{eq}$ by assuming a Bond albedo ($A_B$) of 0.45 and applying:
\begin{equation} \label{eq:Teq}
T_{eq} = \left( \frac{1-A_B}{4\,f} \right)^{1/4} \left( \frac{R_\star}{r} \right)^{1/2} T_\star
\end{equation}
where the factor $f$ accounts for the heat redistribution of the planet.
We assume $f\,=\,1$, consistent with rapid rotators \citep{traub-oppenheimer2010}.

$T_{eq}$ bears no impact on the detectability criteria in our methodology. 
Given its importance for atmospheric modeling, however, we compute $T_{eq}$ throughout the planet's orbit. 
In future work, it could be used to investigate the temporal variability of the atmosphere and to estimate the emitted radiation from the planet.

\subsection{Statistical analysis of detectability}  \label{subsec:dataprocessing_bootstraping}

For a given orbit specified by its Keplerian parameters, we assess if the detectability criteria for IWA, OWA and $C_{min}$ described in Sect. \ref{sec:directimagingtechnical} are met at any orbital position.
We repeat this for each of the pessimistic, intermediate and optimistic scenarios described in Table \ref{table:instrument_scenarios}.
To describe the orbit, we divide it into 360 points with a step in the true anomaly $\Delta f$=1$^\circ$, which is related to time through Eq. (\ref{eq:kepler_time_dependence}). 
We checked a posteriori for a few selected cases that the adopted $\Delta f$ step affects negligibly our findings.
The planetary and orbital parameters used in this work are summarized in Table \ref{table:params_description}.

For each parameter from Table \ref{table:params_description}, we considered the upper and lower uncertainties quoted in the NASA Exoplanet Archive.
We also considered the uncertainties in the coefficients of the mass-radius relationships in Eqs. (\ref{eq:mass-density_giants}) and (\ref{eq:mass-radius_lowmass}). 
All these uncertainties are taken into account when producing random realizations of the planet orbits and corresponding planet-to-star contrasts.
For each planet, we accounted for all the uncertainties simultaneously and computed 10000 independent realizations of both the orbital and non-orbital parameters.
When the value of a specific parameter is not available in the NASA Archive but instead must be estimated through e.g. Kepler's third law or the $M_p$-$R_p$ relationships of Eqs. (\ref{eq:mass-density_giants}) and (\ref{eq:mass-radius_lowmass}), our treatment ensures that the uncertainties are properly propagated.
We use this bootstrap-like method to derive statistical conclusions \citep{pressetal2007} on properties of interest such as $\Delta \theta$, $\alpha$ and $F_p/F_\star$.

Some of the parameters in Table \ref{table:params_description} are indeed correlated through the specific techniques with which they were originally estimated and hence their uncertainties are not independent.
We also note that the uncertainties in the NASA Archive are extracted from references with no homogeneous criteria in the statistical treatment of the data.
A re-evaluation of the orbital parameters to obtain their joint confidence intervals is beyond the scope of this paper, and for simplicity we sample each of them independently from uniform probability distributions
between the quoted uncertainty limits.

We consider an exoplanet to be Roman-accessible if the detectability criteria defined by the IWA, OWA and $C_{min}$
are met over at least one point in the numerically discretised orbit of at least one of the 10000 independent orbital realizations.
The probability of a planet to be Roman-accessible ($P_{access}$) is given by the number of orbital realizations in which the exoplanet is accessible, compared to the total of 10000 realizations. 
The transit probability ($P_{tr}$) is computed as the fraction of orbital realizations in which the condition in Eq. (\ref{eq:transit_condition}) is met.
For a particular orbit, the amount of days that the planet remains observable ($t_{obs}$) can be computed with Eq. (\ref{eq:kepler_time_dependence}) by time-integration along the orbit.
We compute this for each accessible orbital realization to derive a statistical distribution of $t_{obs}$. 
We infer the median value of this distribution and upper and lower uncertainties corresponding to the percentiles 16\% and 84\%, equivalent to $\pm 1 \sigma$ for Gaussian errors.
In addition, for each accessible orbit we compute the interval of observable phase angles ($\alpha_{obs}$) with Eq. (\ref{eq:kepler_alpha}).
We will refer to the minimum and maximum phase angles ($\alpha_{obs(min)}$, $\alpha_{obs(max)}$), together with the corresponding $\pm 1 \sigma$ uncertainties.
We emphasize that the distributions of $t_{obs}$ and $\alpha_{obs}$ are based only on the accessible orbital realizations.
This results in intrinsically biased statistics, since the null detections are not accounted for.
However, we opted for these definitions to have metrics that describe specifically the accessible orbits given that, for instance, $\alpha_{obs}$ is not defined in a non-accessible orbit.
The corresponding $P_{access}$ quantifies to some extent the bias introduced in these metrics.

For each planet in the input catalogue, this statistical method produces posterior distributions for each of the sampled parameters in Table \ref{table:params_description}.
With this, we create an output catalogue (Table \ref{table:output_catalogue}) with the resulting median values of each parameter and their corresponding uncertainties.

The above definition of $P_{access}$ is however flawed because there are planets with very small associated values of this metric for which it is difficult to justify a future observational effort.
In order to keep our findings useful for target prioritisation, in what follows we will only consider planets that in the optimistic CGI configuration have $P_{access}>25\%$ (Table \ref{table:instrument_scenarios}). 
In addition, we restrict our analysis to targets orbiting stars brighter than $V$=7 mag, according to the updated CGI possible performances.
These additional vetting criteria determine the population of planets studied in Sects. \ref{sec:results} and \ref{sec:discussion_selectedtargets}.
For reference, the complete list of Roman-accessible exoplanets including those with $P_{access}<25\%$ or $V>7$ mag is kept in the input and output catalogues (Tables \ref{table:NASA_database} and \ref{table:output_catalogue}).

\section{Results: Roman-accessible exoplanets} \label{sec:results}
We next identify the Roman-accessible exoplanets that meet the additional vetting criteria ($P_{access}>25\%$, $V$<7 mag) in the optimistic CGI configuration.
We compare their properties to the complete set of confirmed exoplanets, as well as to those that have been observed in transit (Sect. \ref{subsec:results_populationstudy}).
Afterwards, we describe their overall detectability conditions ($P_{access}$, $\alpha_{obs}$, $t_{obs}$, $P_{tr}$) as well as the main limiting factors (Sect. \ref{subsec:results_general_detectability}) in the the different CGI scenarios from Table \ref{table:instrument_scenarios}.
Finally, we report the equilibrium temperatures computed for these planets and the variation of $T_{eq}$ along their orbit (Sect. \ref{subsec:results_Teq}).

\subsection{Population analysis: the subset of direct-imaging exoplanets} \label{subsec:results_populationstudy}

\longtab{
\tiny
\begin{landscape}
\begin{longtable}{ l  c  c  c  c  c  c  c  c  c  c  c  c  c  c } 
\caption{\label{table:output_catalogue} 
Output catalogue of the Roman-accessible exoplanets listed in Table \ref{table:NASA_database}.
The quoted values are the median values of the distributions obtained with our statistical methodology (Sect. \ref{sec:dataprocessing}) and the corresponding uncertainties given by the 16\% and 84\% percentiles.} \\
 \hline \hline
  Planet & {$d$} & {$P$} & {$a$} & {$M_p$} & {$R_p$} & {$i$} & $e$ & {$\omega_p$} & $T_{eq}$ & $V$ & $P_{tr}$ & $P_{access}$ & $P_{access}$ & $P_{access}$ \\
  & [pc] & [days] & [AU] & [$M_{J}$] & [$R_{J}$] & [deg] & & [deg] & K & [mag] & \% & \%$_{Optim.}$ & \%$_{Interm.}$ & \%$_{Pessim.}$ \\
  \hline
  \endfirsthead
\caption{continued.}\\
\hline \hline       
  Planet & {$d$} & {$P$} & {$a$} & {$M_p$} & {$R_p$} & {$i$} & $e$ & {$\omega_p$} & $T_{eq}$ & $V$ & $P_{tr}$ & $P_{access}$ & $P_{access}$ & $P_{access}$ \\
  & [pc] & [days] & [AU] & [$M_{J}$] & [$R_{J}$] & [deg] & & [deg] & K & [mag] & \% & \%$_{Optim.}$ & \%$_{Interm.}$ & \%$_{Pessim.}$ \\
\hline
\endhead
\hline
\endfoot
GJ 687 c		& 4.55$^{ +0.00 }_{ -0.00 }$		& 727.3$^{ +8.5 }_{ -8.1 }$		& 1.17$^{ +0.02 }_{ -0.02 }$		& 0.061$^{ +0.034 }_{ -0.013 }$		& 0.41$^{ +0.14 }_{ -0.07 }$		& 90.7$^{ +41.9 }_{ -43.6 }$		& 0.40$^{ +0.15 }_{ -0.15 }$		& 356.2$^{ +42.6 }_{ -42.8 }$		& [79$^{ +5 }_{ -4 }$, 120$^{ +19 }_{ -13 }$]		& 9.150		& 0.20		& 100.0		& 95.0		& 56.17		 \\  
55 Cnc d		& 12.59$^{ +0.01 }_{ -0.01 }$		& 4824.4$^{ +27.2 }_{ -26.1 }$		& 5.50$^{ +0.02 }_{ -0.02 }$		& 4.487$^{ +1.540 }_{ -0.545 }$		& 1.11$^{ +0.02 }_{ -0.02 }$		& 90.1$^{ +40.8 }_{ -40.4 }$		& 0.02$^{ +0.01 }_{ -0.01 }$		& 224.2$^{ +20.7 }_{ -21.4 }$		& [88$^{ +1 }_{ -1 }$, 90$^{ +1 }_{ -1 }$]		& 5.960		& 0.06		& 100.0		& 0.0		& 0.0		 \\  
HD 114613 b		& 20.29$^{ +0.06 }_{ -0.06 }$		& 4000.9$^{ +81.3 }_{ -82.8 }$		& 5.34$^{ +0.18 }_{ -0.18 }$		& 0.416$^{ +0.259 }_{ -0.056 }$		& 1.24$^{ +0.03 }_{ -0.03 }$		& 89.9$^{ +43.4 }_{ -43.1 }$		& 0.46$^{ +0.06 }_{ -0.06 }$		& 16.0$^{ +7.4 }_{ -7.4 }$		& [123$^{ +3 }_{ -3 }$, 202$^{ +12 }_{ -10 }$]		& 4.850		& 0.17		& 100.0		& 65.57		& 0.0		 \\  
ups And d		& 13.41$^{ +0.04 }_{ -0.04 }$		& 1276.5$^{ +0.4 }_{ -0.4 }$		& 2.51$^{ +0.00 }_{ -0.00 }$		& 4.761$^{ +2.820 }_{ -0.578 }$		& 1.10$^{ +0.02 }_{ -0.03 }$		& 89.9$^{ +43.4 }_{ -42.6 }$		& 0.30$^{ +0.00 }_{ -0.00 }$		& 78.8$^{ +0.3 }_{ -0.3 }$		& [175$^{ +3 }_{ -3 }$, 237$^{ +4 }_{ -4 }$]		& 4.100		& 0.20		& 100.0		& 100.0		& 0.0		 \\  
HD 217107 c		& 20.07$^{ +0.01 }_{ -0.01 }$		& 4270.5$^{ +151.3 }_{ -147.6 }$		& 5.32$^{ +0.26 }_{ -0.26 }$		& 3.011$^{ +1.725 }_{ -0.364 }$		& 1.13$^{ +0.02 }_{ -0.03 }$		& 89.8$^{ +42.4 }_{ -42.7 }$		& 0.52$^{ +0.02 }_{ -0.02 }$		& 18.6$^{ +4.1 }_{ -4.1 }$		& [87$^{ +2 }_{ -2 }$, 154$^{ +5 }_{ -5 }$]		& 6.160		& 0.10		& 100.0		& 9.62		& 0.0		 \\
\hline
\end{longtable}
\tablefoot{
We note that, in those cases where the quoted uncertainties are 0.00, this is a result of insufficient significant figures in the rounding.
}
\tablefoottext{a}{Our results for HR 8799 $b$, $c$ and $d$ correspond to those with an unconstrained orbital inclination. Different works suggest values of $i$ between $20^\circ$ and $30^\circ$ \citep{soummeretal2011, konopackyetal2016, wangetal2018, gravity2019} but these are not currently included as default parameters in the NASA Exoplanet Archive for these planets.}
\end{landscape}
}

We analysed all confirmed exoplanets as described in Sect. \ref{subsec:dataprocessing_bootstraping} and found that 26 of the total 4276 meet the criteria of angular separation and planet-to-star contrast for the optimistic CGI configuration, with the additional vetting criteria $P_{access}>25\%$ and $V$<7 mag.
The number of planets meeting these criteria in the intermediate and pessimistic scenarios drops to 10 and 3, respectively.
Focusing on the optimistic scenario, we study below the main properties, as listed in our input catalogue (Table \ref{table:NASA_database}) of this subset of Roman-accessible objects.

\begin{figure}
   \centering
   \includegraphics[width=9.cm]{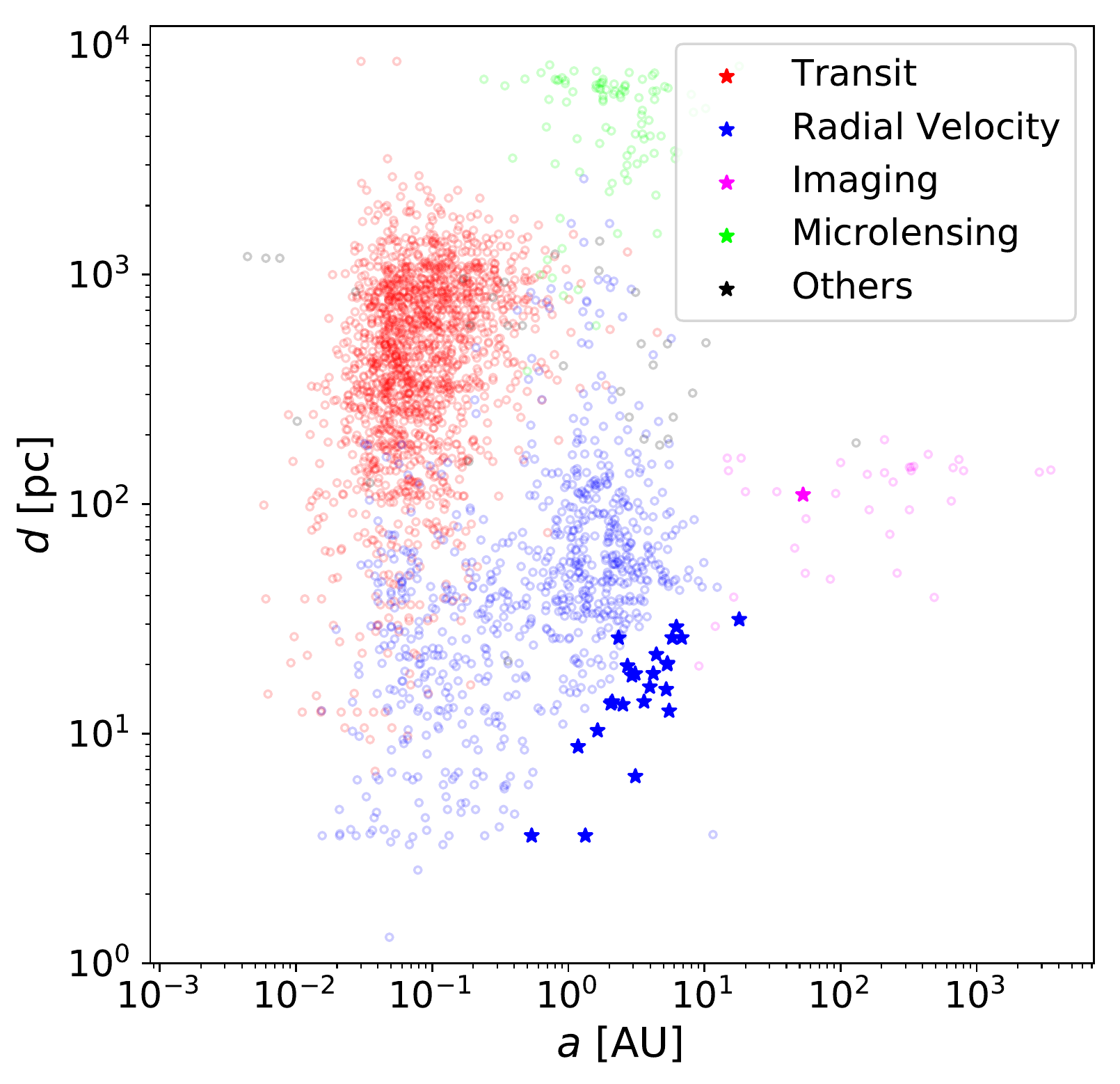}
      \caption{\label{fig:aVSd_confirmed} 
      Semi-transparent dots: confirmed exoplanets for which we know $d$ and can derive $a$ as explained in Sect. \ref{subsec:dataprocessing_orbits}.
      Solid stars: confirmed exoplanets that we find Roman-accessible in the optimistic CGI configuration, with $P_{access}>25\%$ and orbiting stars brighter than $V$=7 mag.
      Colour code indicates the corresponding discovery technique (that by which the planet was first identified), as detailed in the legend.
      "Others" refers to all other possible discovery techniques considered in the NASA Exoplanet Archive.
      HD 100546 b appears as the only Roman-accessible discovered in Imaging, although its existence is marked as controversial in the NASA Archive.
      }
   \end{figure}

Figure \ref{fig:aVSd_confirmed} displays the semi-major axis and distance to the Earth of all confirmed exoplanets, showing how different discovery techniques are sensitive to different ranges of these parameters.
The population of Roman-accessible exoplanets is composed of objects discovered in radial-velocity, with the exception of HD 100546 b which was discovered in imaging \citep{quanzetal2015}.
The existence of this protoplanet with $R_p=6.9_{-2.9}^{+2.7}\, R_J$ is however controversial, as indicated in the NASA Archive.
Despite the transit method is the most fruitful technique so far in terms of number of planets discovered (76\% of the total), none of them is Roman-accessible.
New transit missions with long baselines and focusing on nearby stars such as TESS or PLATO \citep{rickeretal2014, raueretal2014} are expected to yield additional transiting planets amenable to direct-imaging \citep{starketal2020}.
Other planets may be accessible in thermal emission to the Roman Telescope \citep{lacy-burrows2020}.
Computing the contribution of thermal emission for each confirmed exoplanet, which depends on the age of the system and the evolutionary models assumed, is out of the scope of this work.

\begin{figure}
   \centering
   \includegraphics[width=9.cm]{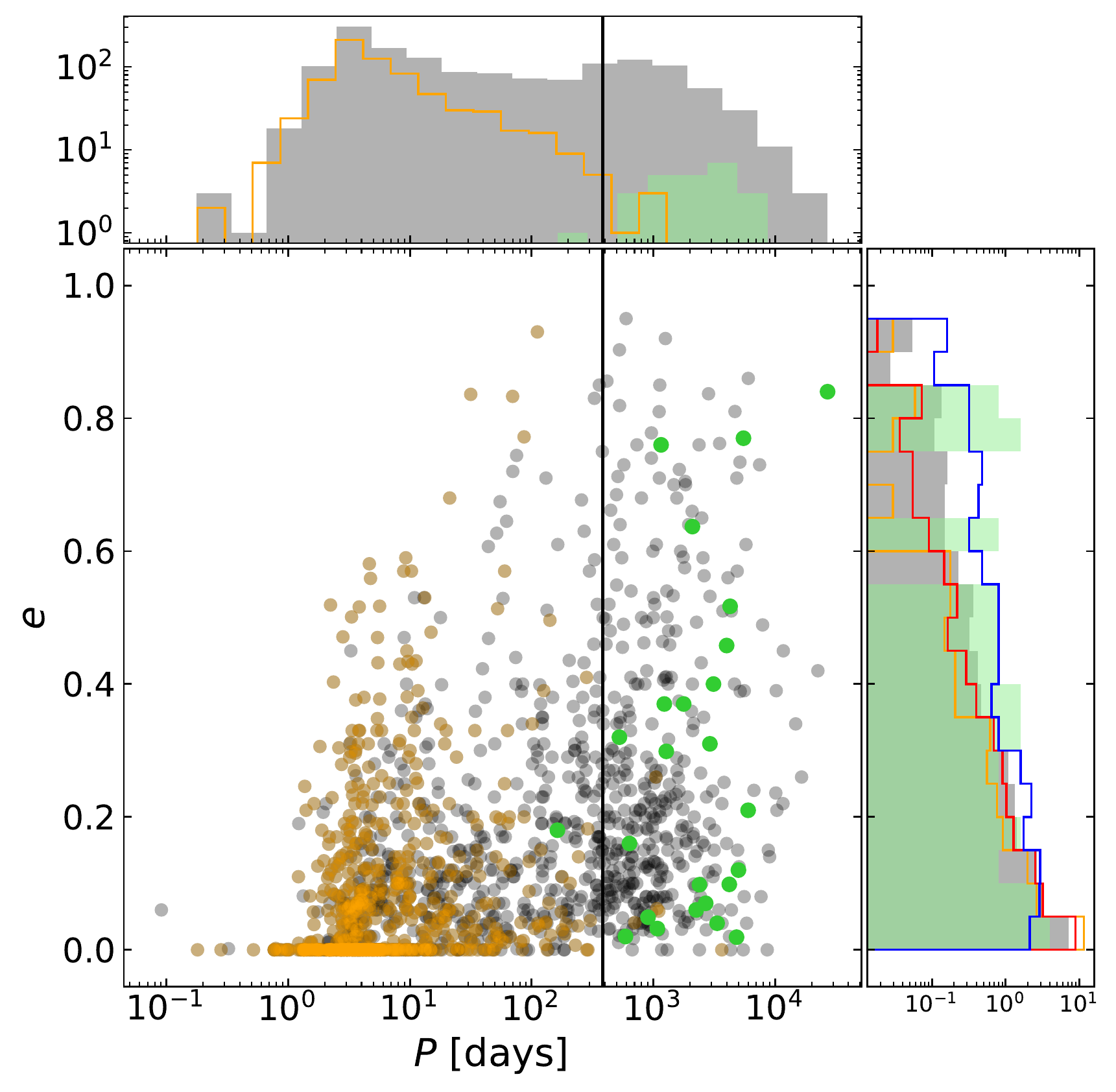}
      \caption{\label{fig:eccen_distrib}
      Main panel.
      Eccentricity and orbital period for all confirmed exoplanets (grey dots), those that have been observed in transit (whether discovered by that method or not)
      (orange dots) and those that are Roman-accessible in the optimistic CGI configuration, with $P_{access}>25\%$ and $V$<7 mag (green dots).
      We only consider those planets for which $e$ is known and $P$ can be derived as explained in Sect. \ref{subsec:dataprocessing_orbits}.
      The black line shows the limit between short- and long-period exoplanets ($P$=382.3 days) as defined in \citet{kipping2013} (see Sect. \ref{subsubsec:dataprocessing_orbits_ecc}).
      Top panel. It shows the $P$ distribution of all confirmed exoplanets (grey), those observed in transit (orange line) and those that are Roman-accessible (semi-transparent green).
      Right panel. Normalized distribution of $e$ such that it shows the relative frequency instead of the total count of planets.
      The same colour code as for the top panel applies.
      Given that the green bars are semi-transparent (so that the grey distribution underneath can also be seen), the overall graph becomes either darker or lighter green depending on whether both histograms overlap.
      For reference, we include the eccentricity for the subsets of short- and long-period exoplanets in \citet{kipping2013} (red and blue lines, respectively). 
      }
   \end{figure}

Long-period planets typically have larger eccentricities than short-period ones, and this has an impact on the median eccentricity of the ensemble of Roman-accessible planets.
Figure \ref{fig:eccen_distrib} displays the statistics of orbital period and eccentricity (when it is reported in the NASA Archive).
The top panel shows the total number of planets in different ranges of orbital periods.
Correspondingly, the right panel shows the normalized distributions of $e$, such that the integral under the histogram is equal to one for the selected bin size.\footnote{Multiplying the value of the bin size by the value of the normalized distribution at that bin yields the fraction of planets if the total number of planets is normalized to one. This implies that, for histogram bin sizes smaller than one such as in the $e$ histogram of  Fig. \ref{fig:eccen_distrib} (with a bin size of 0.05), the value of the normalized distribution may be greater than one (as seen in the figure).}
The key informative of the normalized distributions is their shape, enabling a more evident comparison of populations with different total counts.
We find that the Roman Telescope will be able to detect a relatively large proportion of highly eccentric planets, with the median value of this distribution being $e$= $0.21^{+ 0.33 }_{- 0.16 }$.
In comparison, the total population of confirmed exoplanets with a measurement of $e$ has a median eccentricity of $e$=$0.10^{+ 0.21 }_{- 0.10 }$ and the subset of those that have been observed in transit (even if discovered by other methods), $e$=$0.02^{+ 0.17 }_{- 0.02 }$.
The observed $e$ distribution for the Roman-accessible exoplanets behaves similarly to the long-period planets defined by \citet{kipping2013}.
However, this remains a modest sample and therefore more long-period exoplanets need to be followed up to understand the biases existing in the observed $e$ distributions.

\begin{figure}
   \centering
   \includegraphics[width=9.cm]{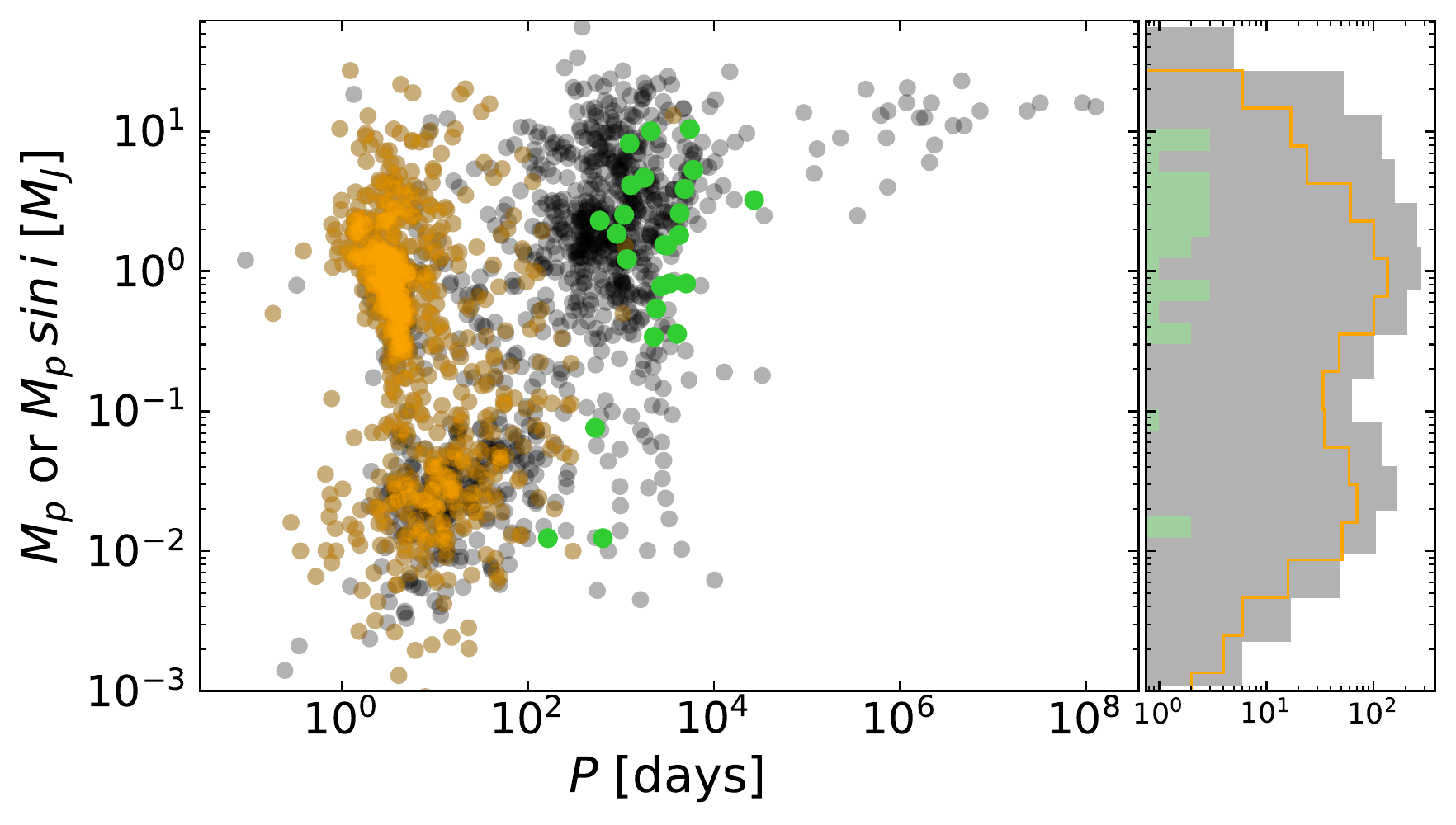}
      \caption{\label{fig:Mp_VS_P} 
      Distribution of mass and orbital period for all confirmed exoplanets (semi-transparent grey), those observed in transit (semi-transparent orange) and those that are Roman-accessible in the optimistic CGI configuration, with $P_{access}>25\%$ and $V$<7 mag (green).
      The plot considers without distinction planets for which either $M_p$ or $M_p\,sin\,i$ are known.
      }
   \end{figure}

\begin{figure}
   \centering
   \includegraphics[width=9.cm]{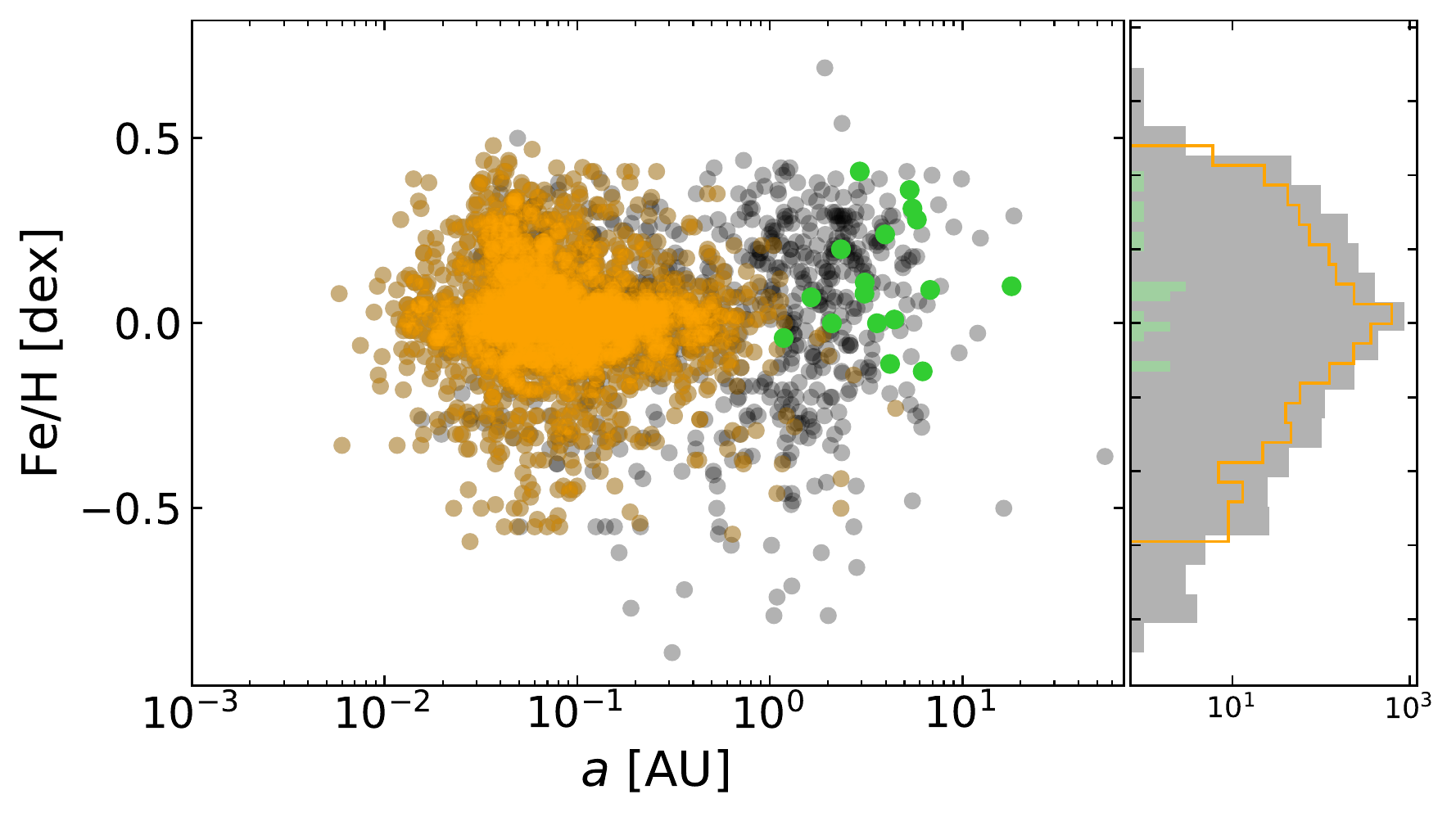}
      \caption{\label{fig:metallicity_VS_a} 
      Distribution of stellar metallicity and semi-major axis of the planet for all confirmed exoplanets (semi-transparent grey), those observed in transit (semi-transparent orange) and those Roman-accessible in the optimistic CGI configuration, with $P_{access}>25\%$ and $V$<7 mag (green).
      }
   \end{figure}

\begin{figure}
   \centering
   \includegraphics[width=9.cm]{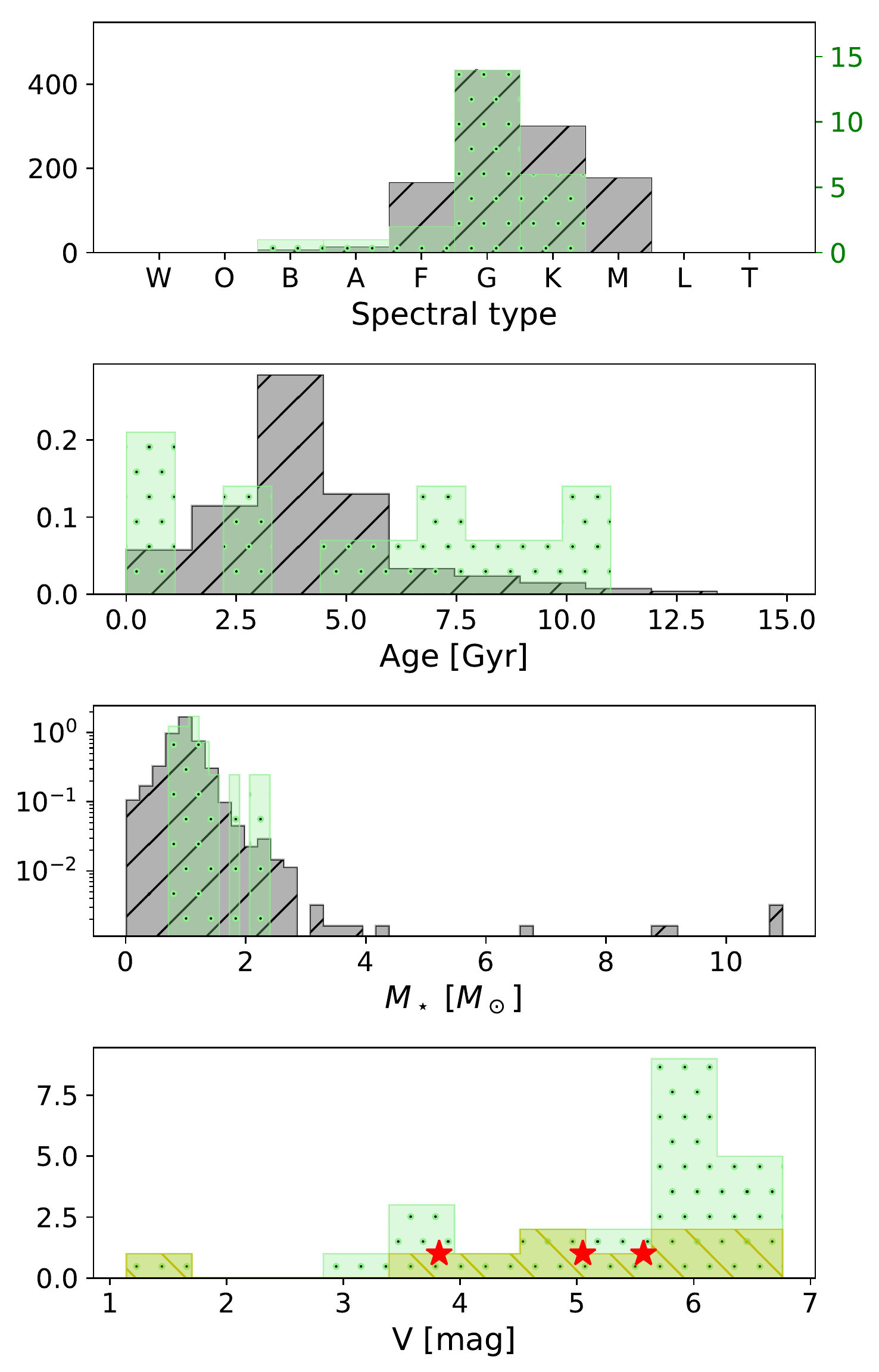}
      \caption{\label{fig:population_stellar_properties} 
      From top to bottom. \textit{First}: total count of planet-hosting stars of each spectral type.
      \textit{Second}: normalized distributions of the stellar age.
      \textit{Third}: normalized distributions of the stellar mass.
      Grey bars with '/' hatch correspond to the total population of confirmed exoplanets.
      Over-plotted semi-transparent green bars with dotted hatch correspond to those exoplanets that we find Roman-accessible in the optimistic CGI configuration, with $P_{access}>25\%$ and $V$<7 mag.
      We note that these parameters are not available for all of the confirmed exoplanets in the NASA Exoplanet Archive.
      The spectral type is available for all of the 24 Roman-accessible-planet host stars; the stellar age, for 13 of them and the metallicity, for 16.
      \textit{Fourth}: count of Roman-accessible-planet host stars of different optical magnitudes in each CGI configuration.
      Green bars with dotted hatch correspond to the optimistic scenario.
      Semi-transparent yellow bars with '$\setminus$' hatch correspond to the intermediate scenario.
      Red stars mark the three stars hosting Roman-accessible planets in the pessimistic scenario.
      }
   \end{figure}

Figure \ref{fig:Mp_VS_P} shows that the statistics of known exoplanets is dominated by giant ones because they are generally easier to detect.
This bias is however particularly noticeable in the Roman-accessible population. 
Given that most of these exoplanets lack an estimate of $i$ and we only know their minimum mass (Table \ref{table:NASA_database}), some of these objects may actually be at the boundary between giant exoplanets and brown dwarfs.
Interestingly, we also find that the Roman Telescope may be able to detect tau Cet $e$ and $f$, both with minimum masses of $3.9\,M_\oplus$ and thus in the super-Earth to mini-Neptune mass regime (see Sect. \ref{subsec:results_general_detectability}). 
The ongoing efforts to discover low-mass exoplanets around nearby stars \citep{pepeetal2021, quirrenbachetal2016} as well as the future development of direct imaging missions with lower $C_{min}$ and smaller IWA will expectedly reduce this observational bias.

Host-star properties such as the spectral type or the mass may be of interest to test hypotheses on the formation and evolution of an exoplanet \citep{laughlinetal2004, boss2006}. 
The spectral type also determines the chemistry of the star, which has an impact on the plausible structure and composition of its exoplanets \citep{santosetal2017}.
Furthermore, both the age of the star and its spectral type set constraints on the stellar activity, which affects the eventual exoplanetary atmospheres.

Regarding the host stars of the Roman-accessible exoplanets, we find that the median value of their metallicity is $Fe/H=0.09^{+ 0.20 }_{- 0.11 }$.
This shows a mild but not significant bias towards super-solar metallicities (Fig. \ref{fig:metallicity_VS_a}) compared to the total population of confirmed exoplanets, with $Fe/H=0.02^{+ 0.16 }_{- 0.14 }$.
The bias is consistent with the observed trend of giant planet hosts to be more metal-rich than low-mass-planet hosts \citep{santos-buchhave2018}.
The stars hosting Roman-accessible planets are currently dominated by G-type stars, similar to the total population of confirmed planet hosts (Fig. \ref{fig:population_stellar_properties}).
In turn, this figure shows an under-representation of F, K and M stars for the Roman-accessible exoplanets in comparison to the complete population. 
We find that this lack of K and M stars in the Roman-accessible targets is mainly caused by the $V$<7 mag threshold.
Indeed, if the condition on the stellar magnitude was omitted, we would obtain an overabundance of M-type stars hosting Roman-accessible targets (see Table \ref{table:output_catalogue}).

We also find that the stars hosting Roman-accessible planets show no clear bias to a particular stellar age, whereas in the total set of planet-hosting stars there is a clear bias favouring ages of 3 to 4 Gyr.
The Roman-accessible planets in the youngest systems are HD 100546 b (0.005 Gyr), discovered in imaging, eps Eri b (0.5 Gyr) and HD 62509 b (0.980 Gyr), the latter two discovered by radial velocity.
Figure \ref{fig:population_stellar_properties} (bottom) shows similar $M_\star$ distributions in the direct-imaging subset and in the total population of host stars.
The lack of low-mass stars is again due to the $V$<7 mag threshold that rules out M stars from the target list.
However, we do not find any Roman-accessible exoplanet orbiting a star more massive than $2\, M_\odot$.
This might be caused partly by the difficulties of searching for RV planets around early-type stars.
In future work, we will compare these trends in stellar properties with those from self-consistently computed stellar catalogues such as SWEET-Cat \citep{santosetal2013}.

The above findings show that the population of Roman-accessible exoplanets does indeed differ from the general population of confirmed exoplanets or from those observed in transit.
These differences are partly influenced by the sensitivity of different discovery techniques to reveal amenable targets.
Hence, reflected-starlight measurements will enable the atmospheric characterization of exoplanets that are not accessible with other techniques.

\subsection{General detectability conditions} \label{subsec:results_general_detectability}

\begin{figure*}
	\centering
	\includegraphics[width=6cm]{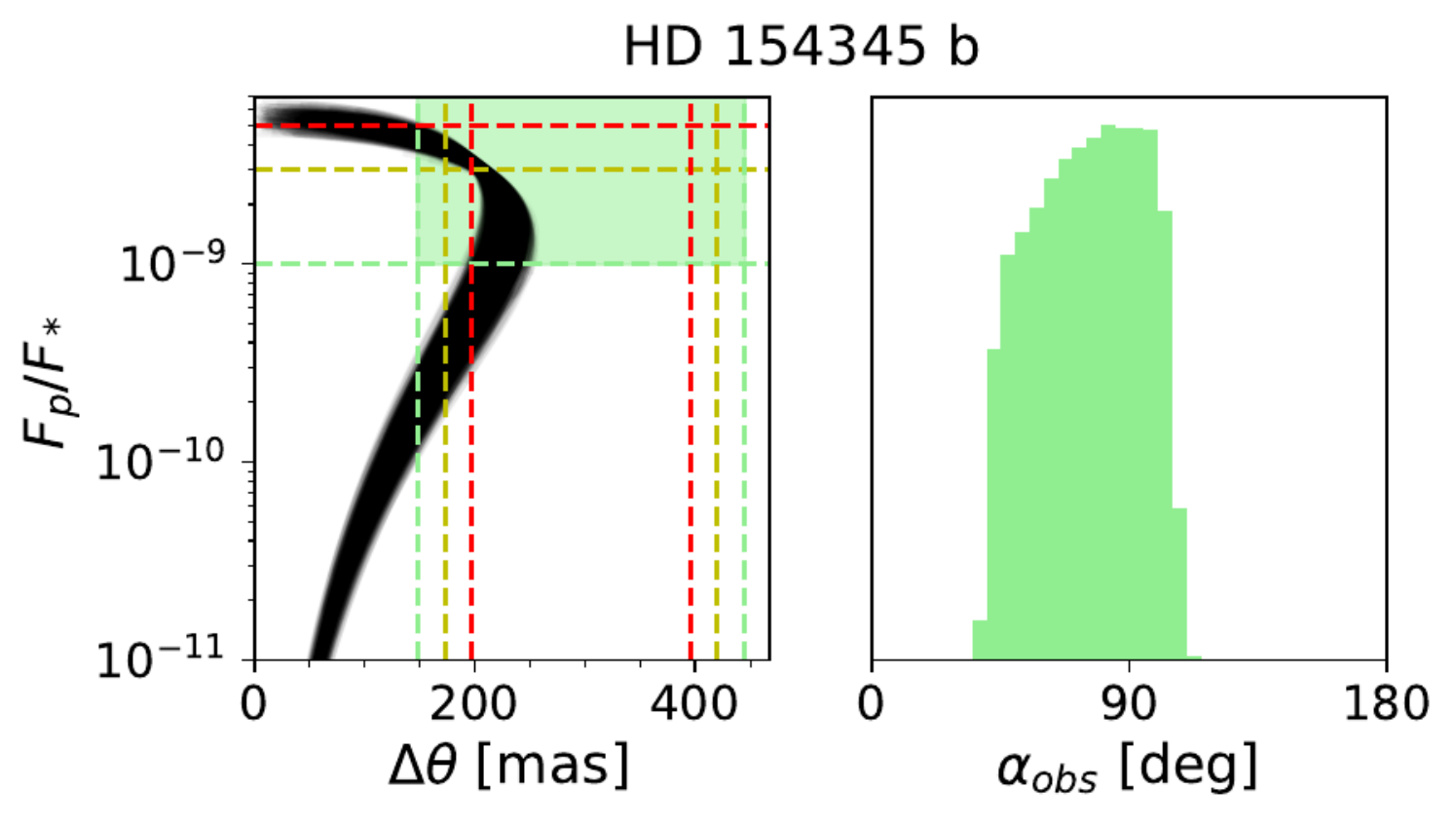} 
	\hfill
	\includegraphics[width=6cm]{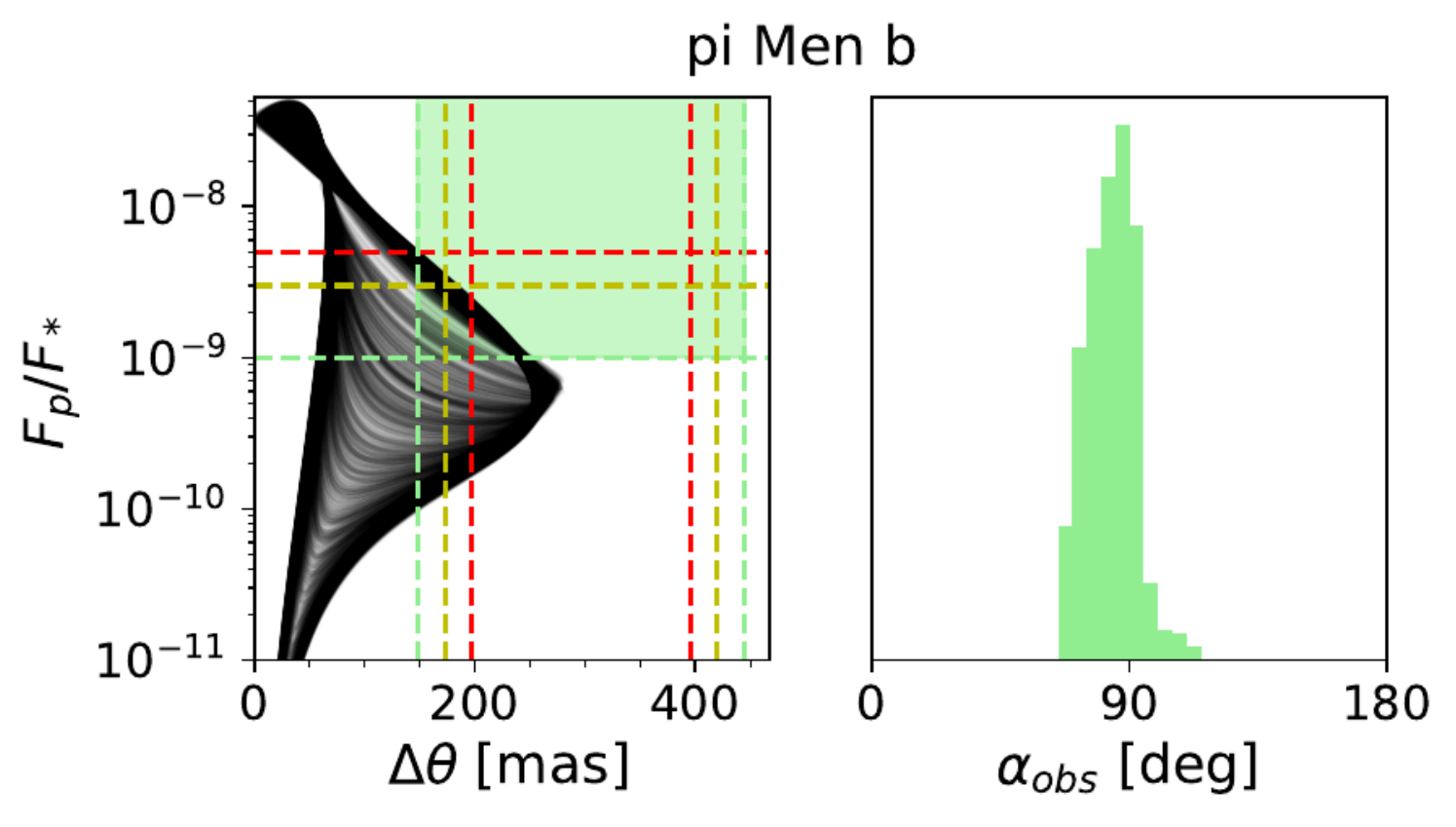} 
	\hfill
	\includegraphics[width=6cm]{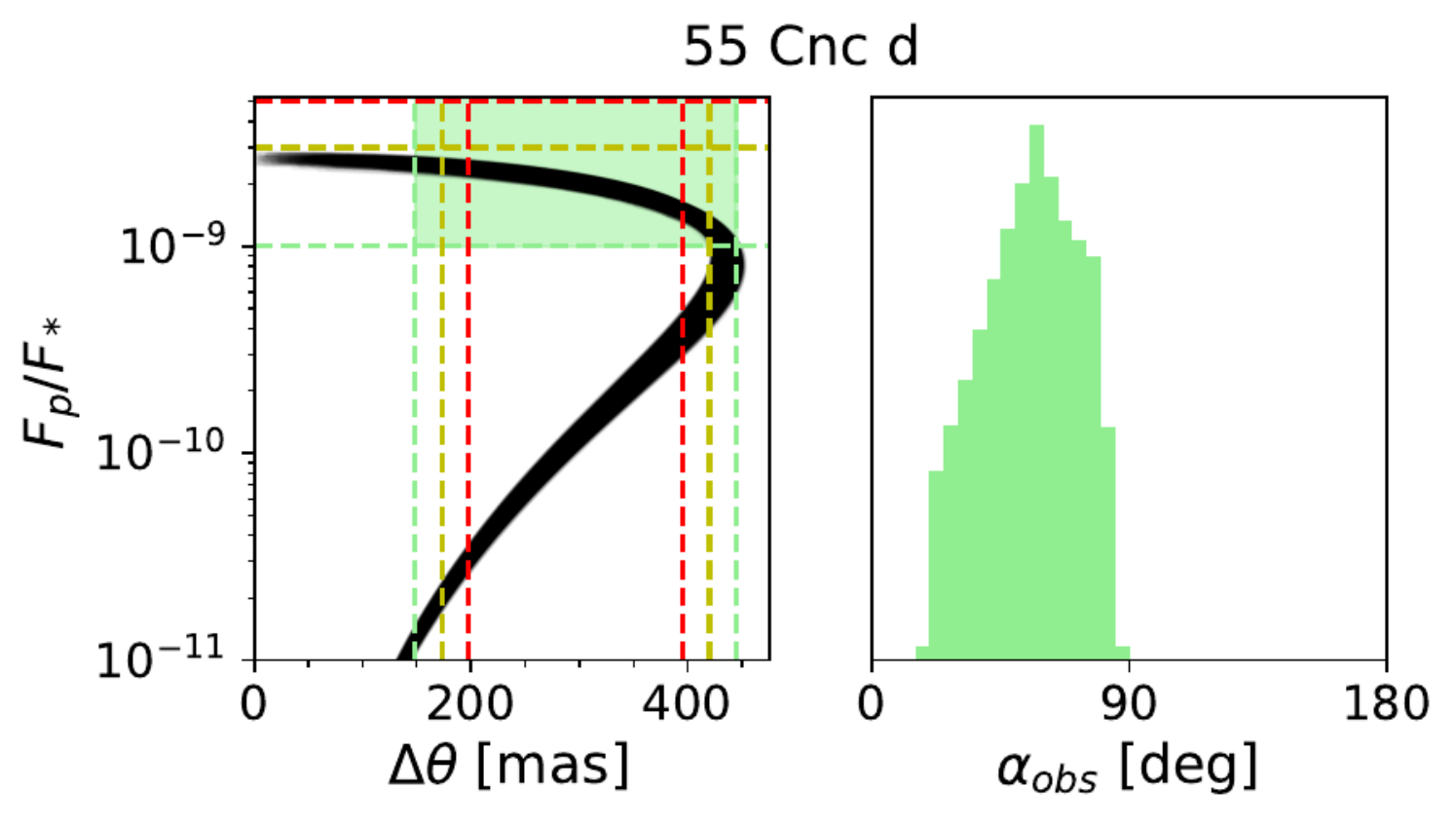} 
   \vspace{-0.1cm}
	   \\
	\includegraphics[width=6cm]{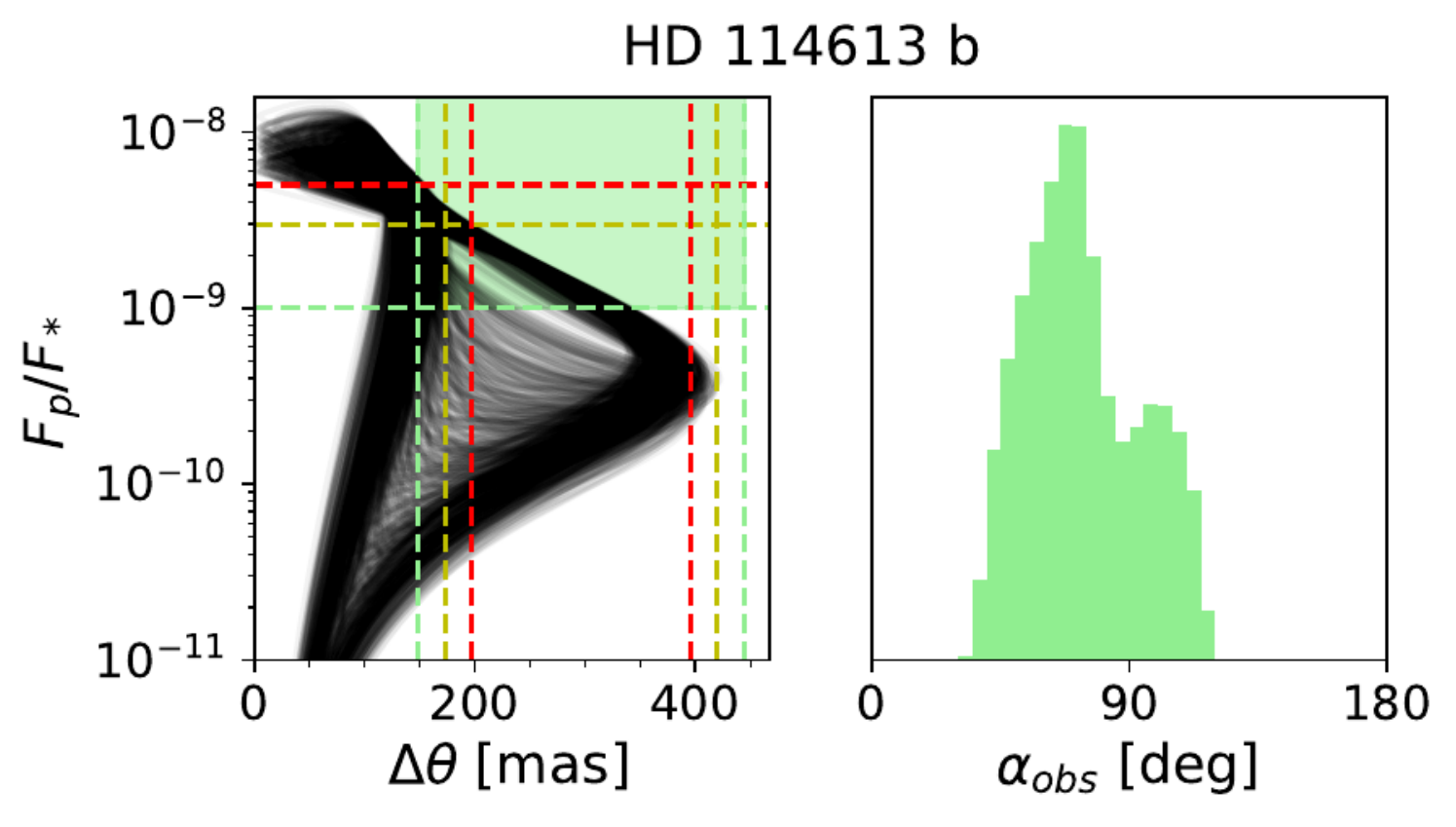} 
	\hfill
	\includegraphics[width=6cm]{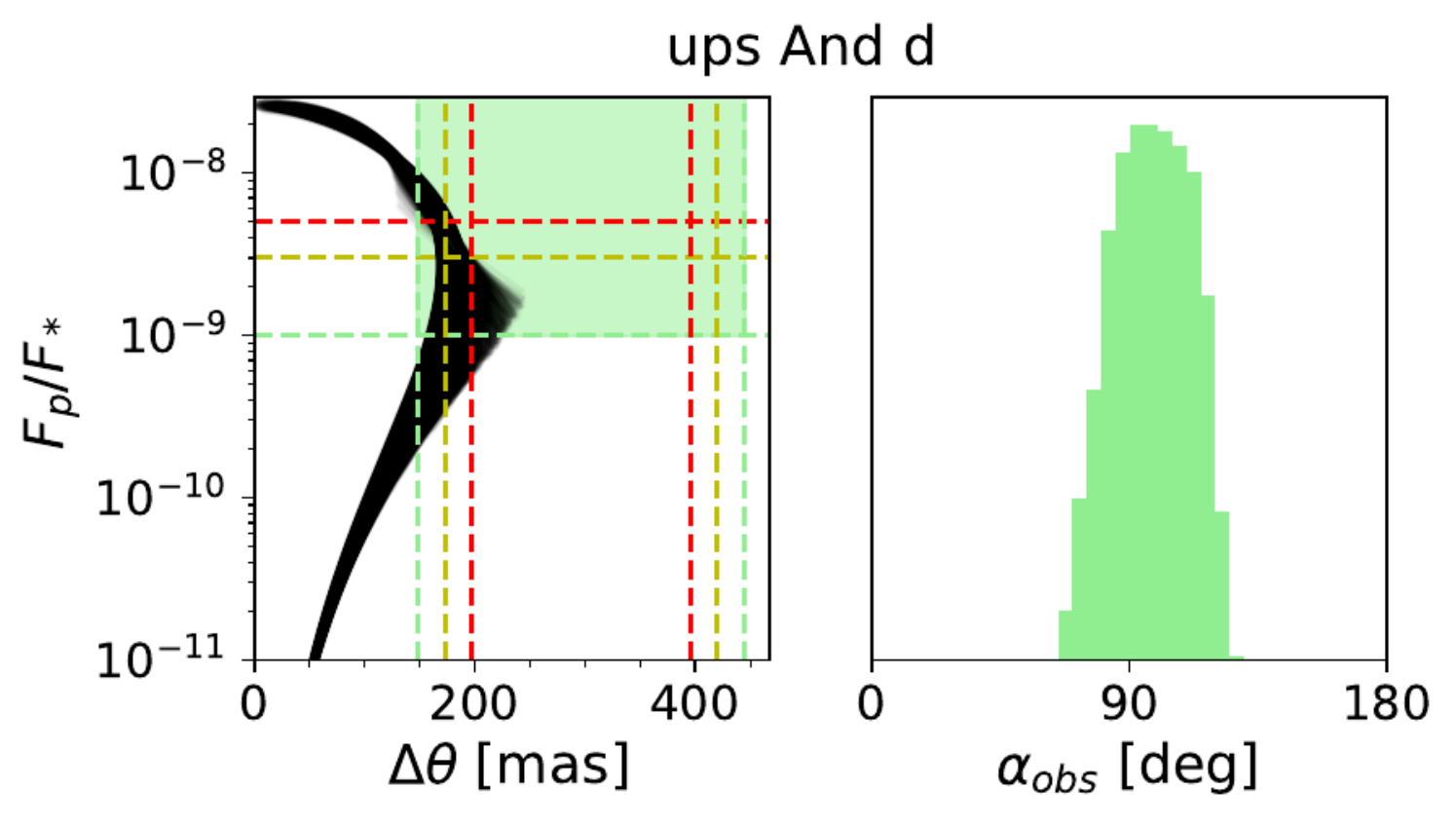} 
	\hfill
	\includegraphics[width=6cm]{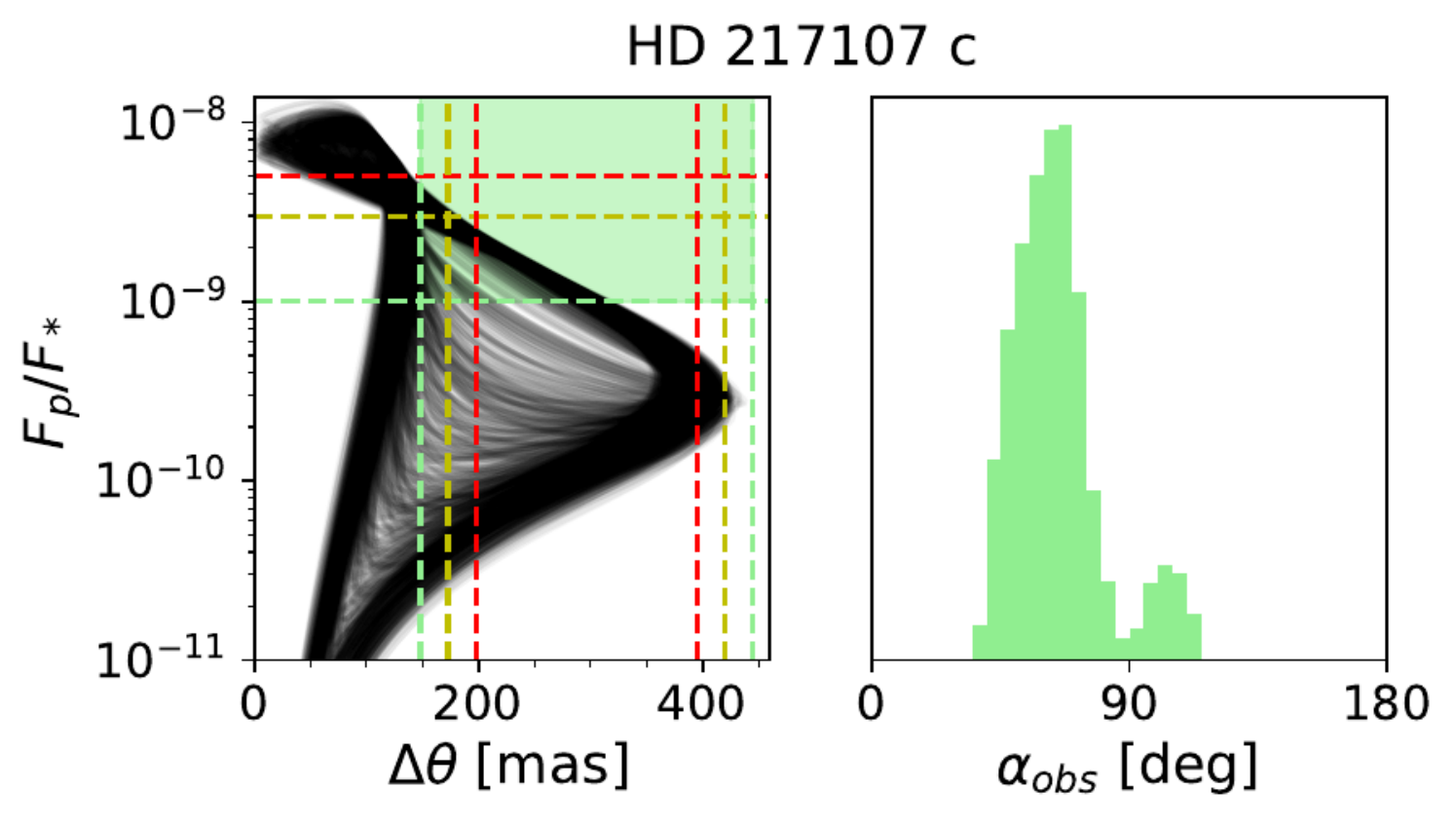} 
   \vspace{-0.1cm}
	   \\
	\includegraphics[width=6cm]{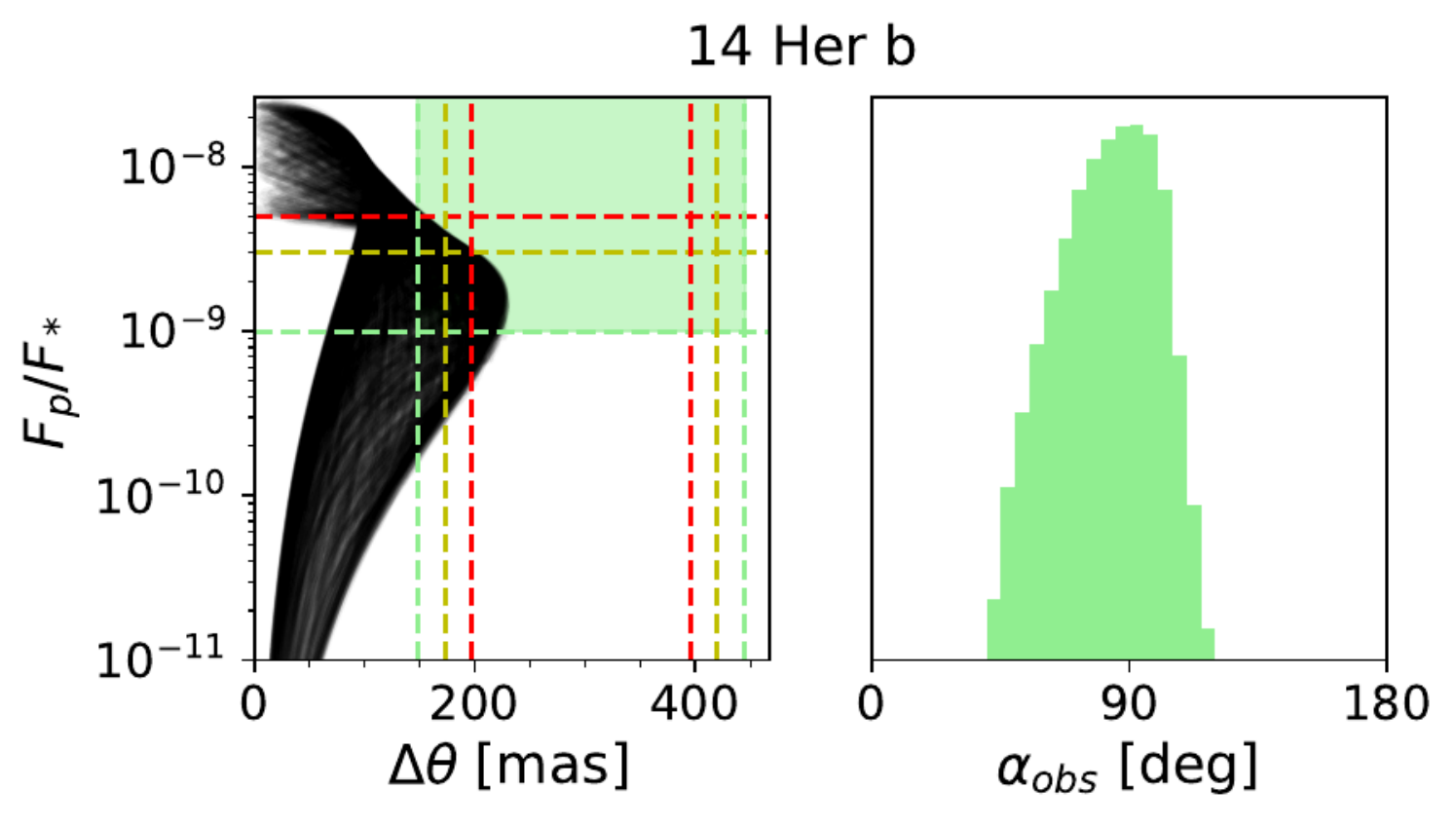} 
	\hfill
	\includegraphics[width=6cm]{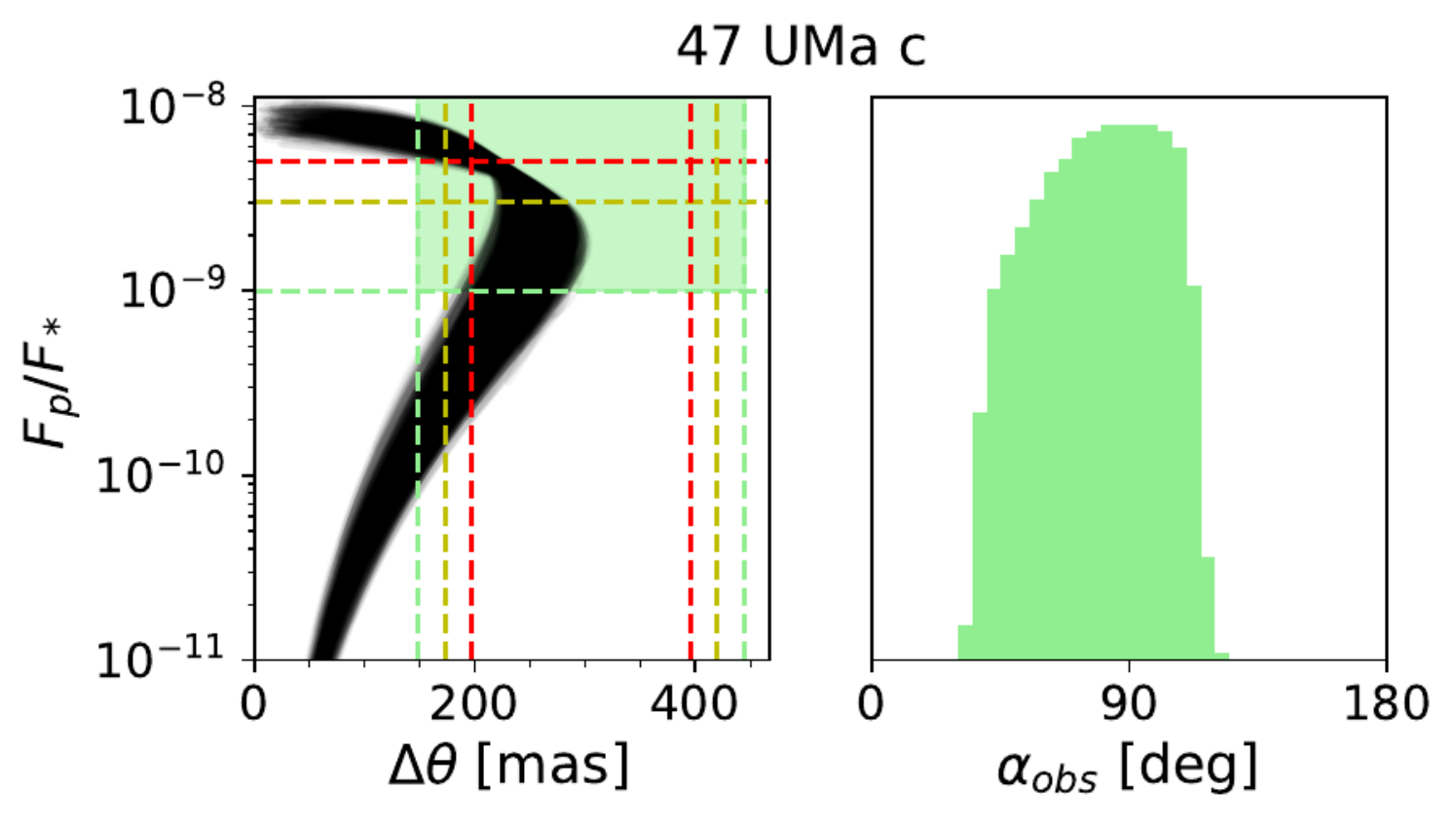} 
	\hfill
	\includegraphics[width=6cm]{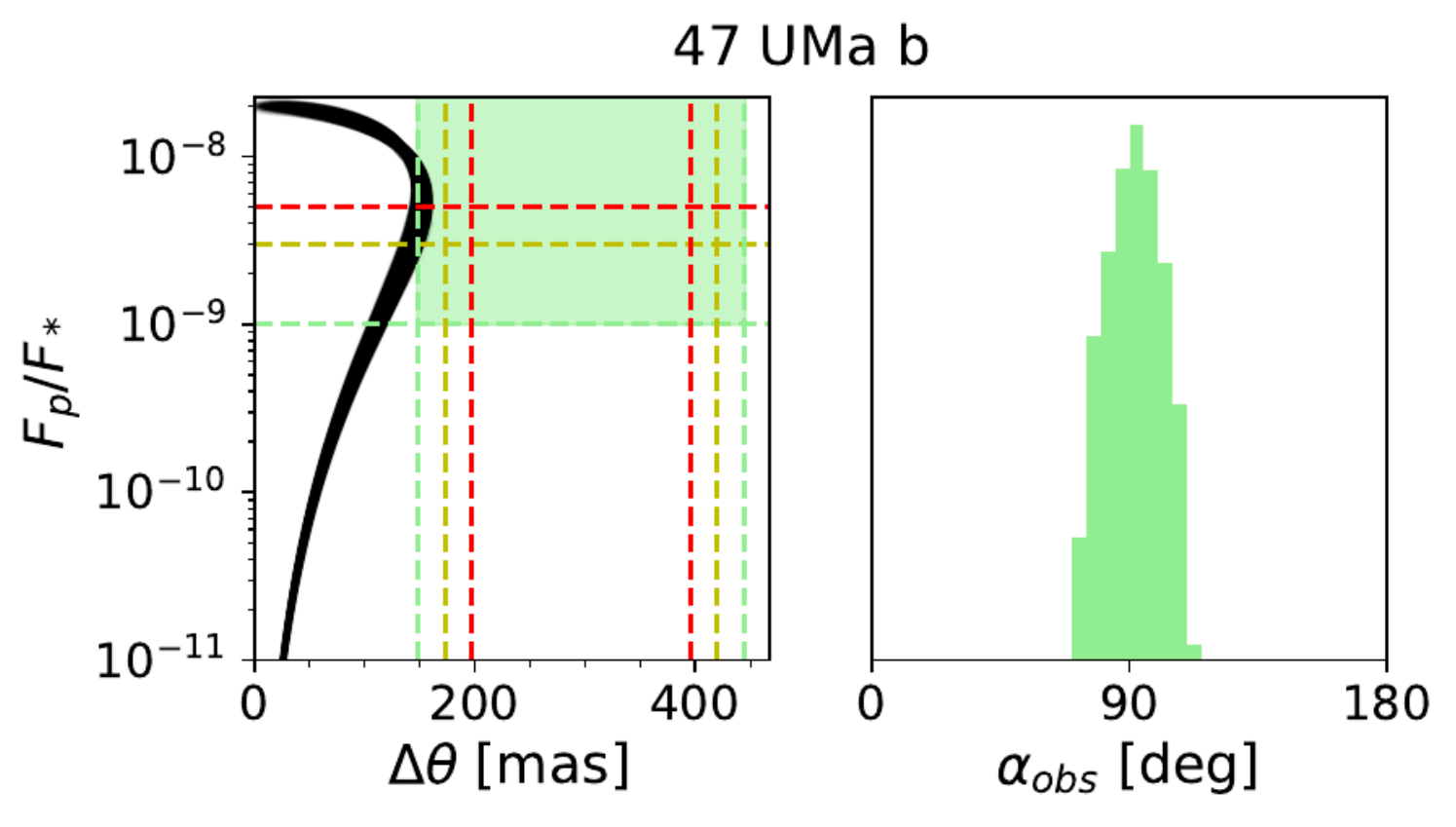} 
   \vspace{-0.1cm}
	   \\
	\includegraphics[width=6cm]{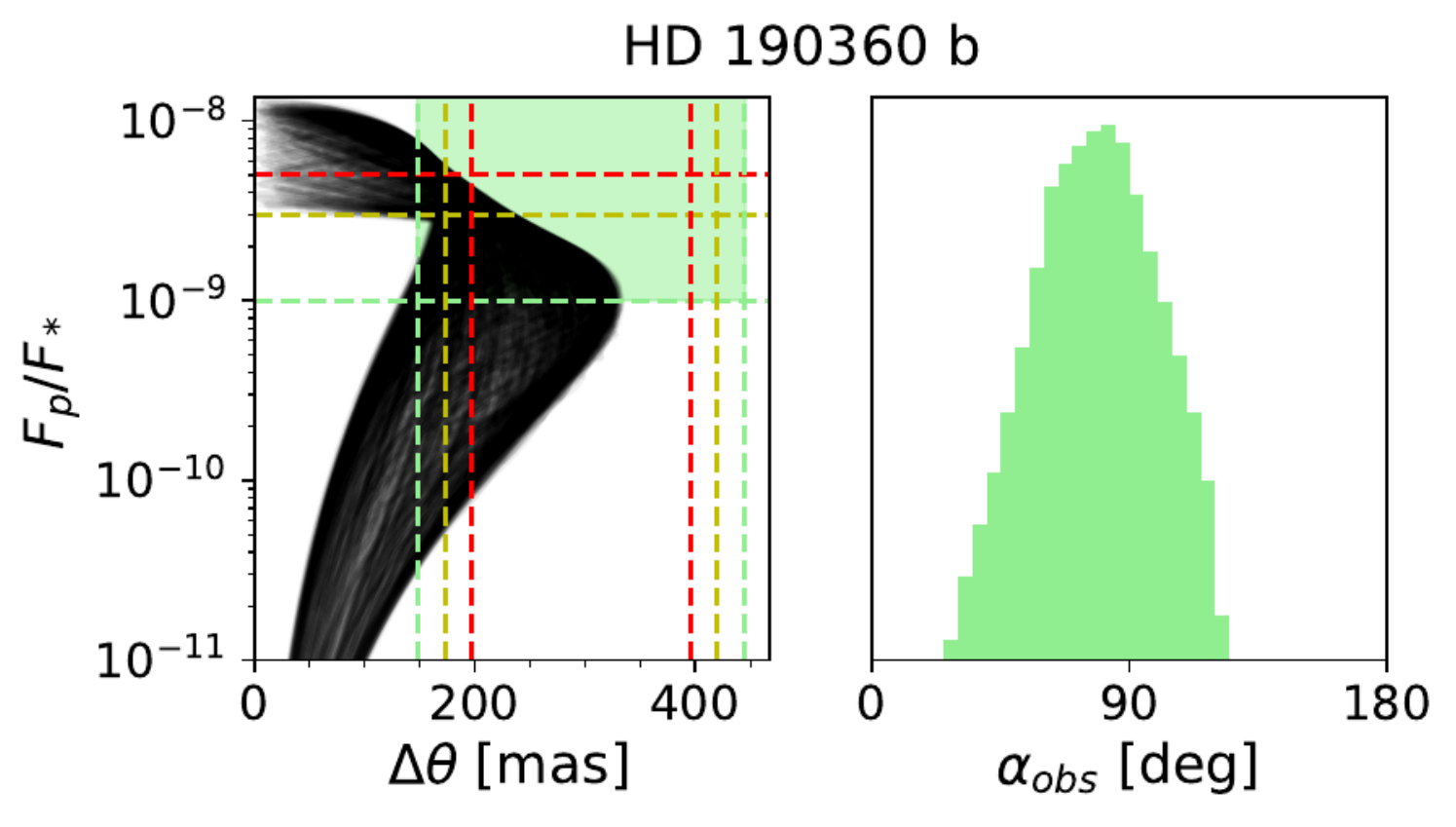} 
	\hfill
	\includegraphics[width=6cm]{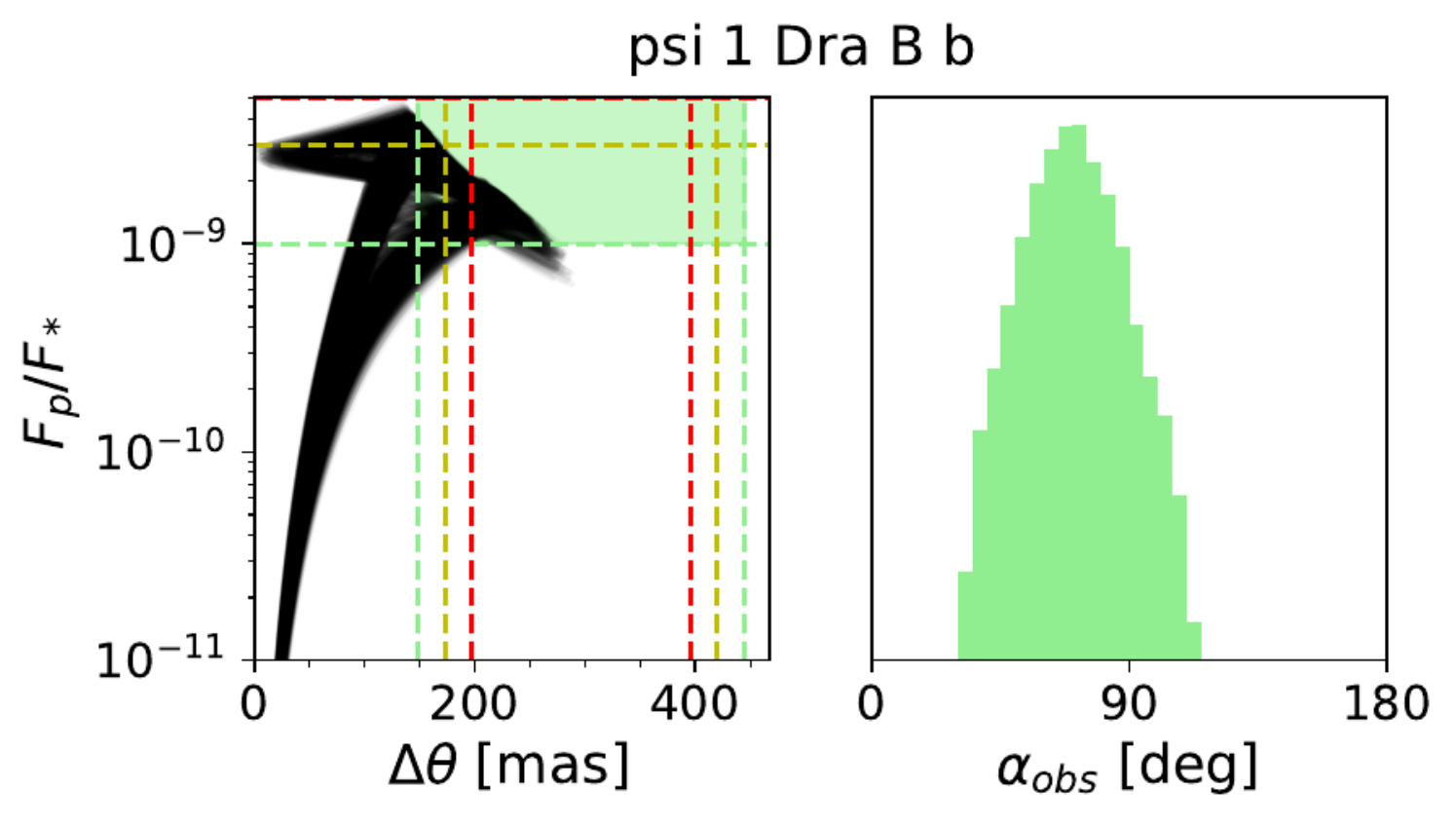} 
	\hfill
	\includegraphics[width=6cm]{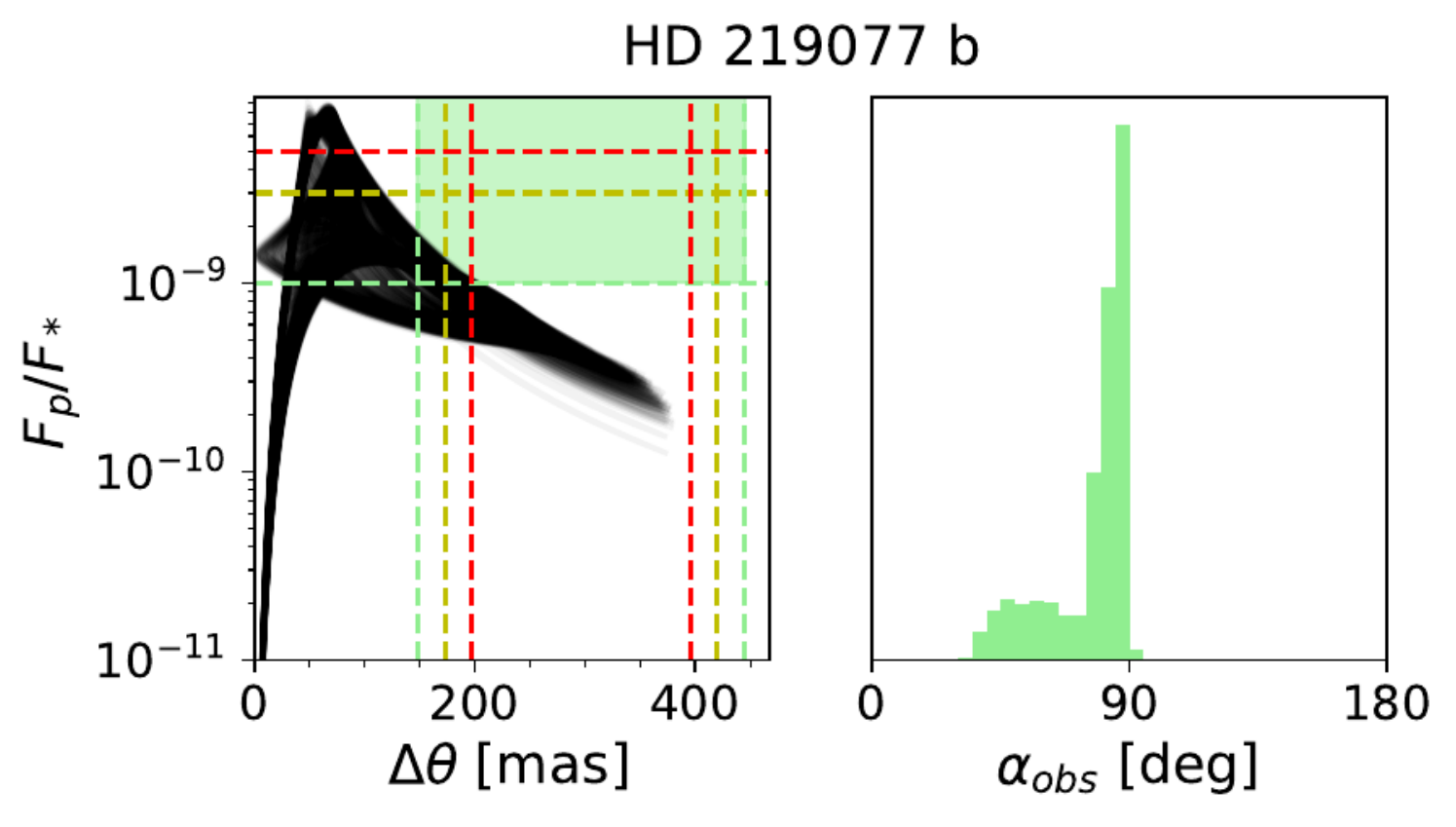} 
   \vspace{-0.1cm}
	   \\
	\includegraphics[width=6cm]{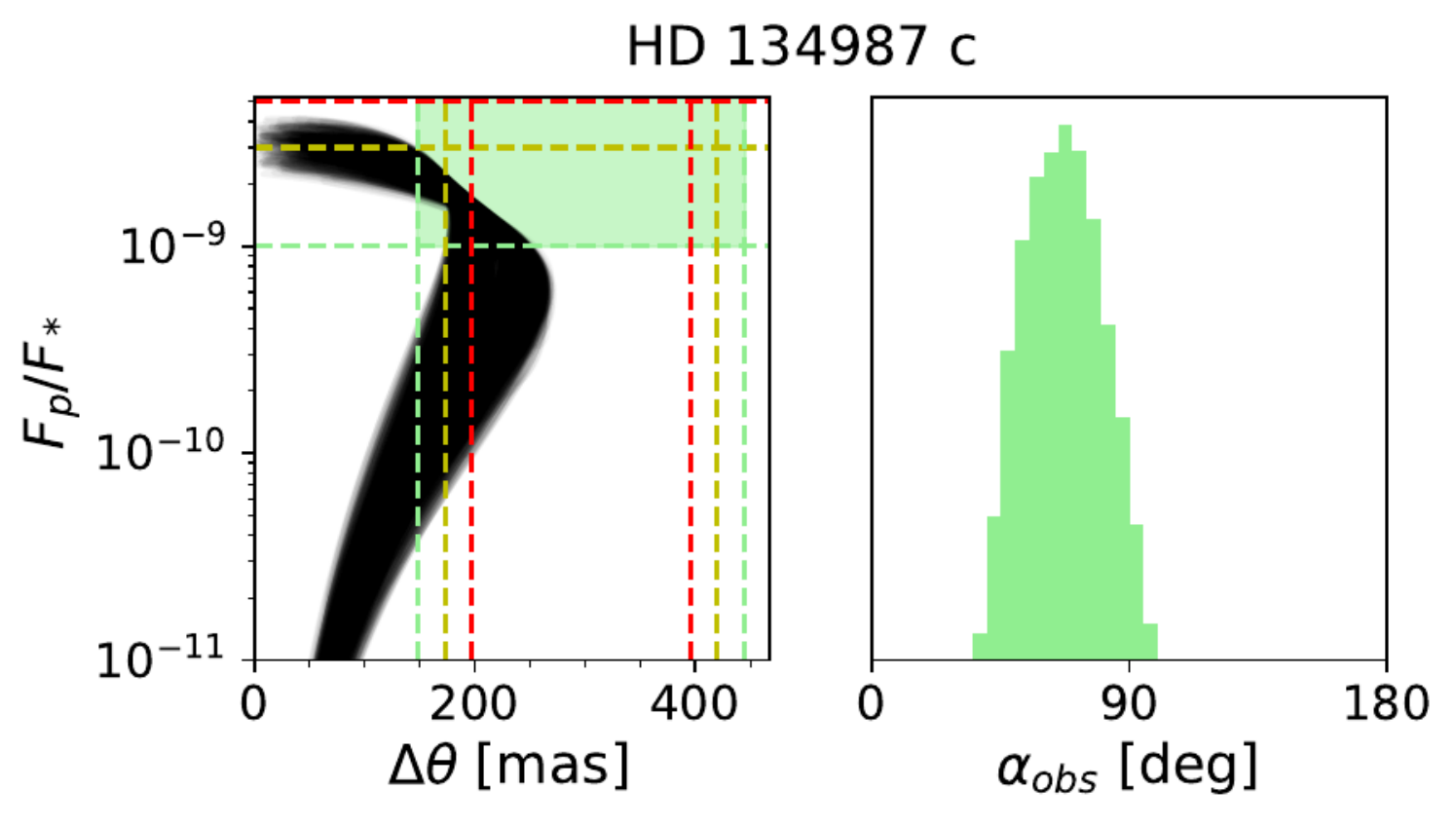} 
	\hfill
	\includegraphics[width=6cm]{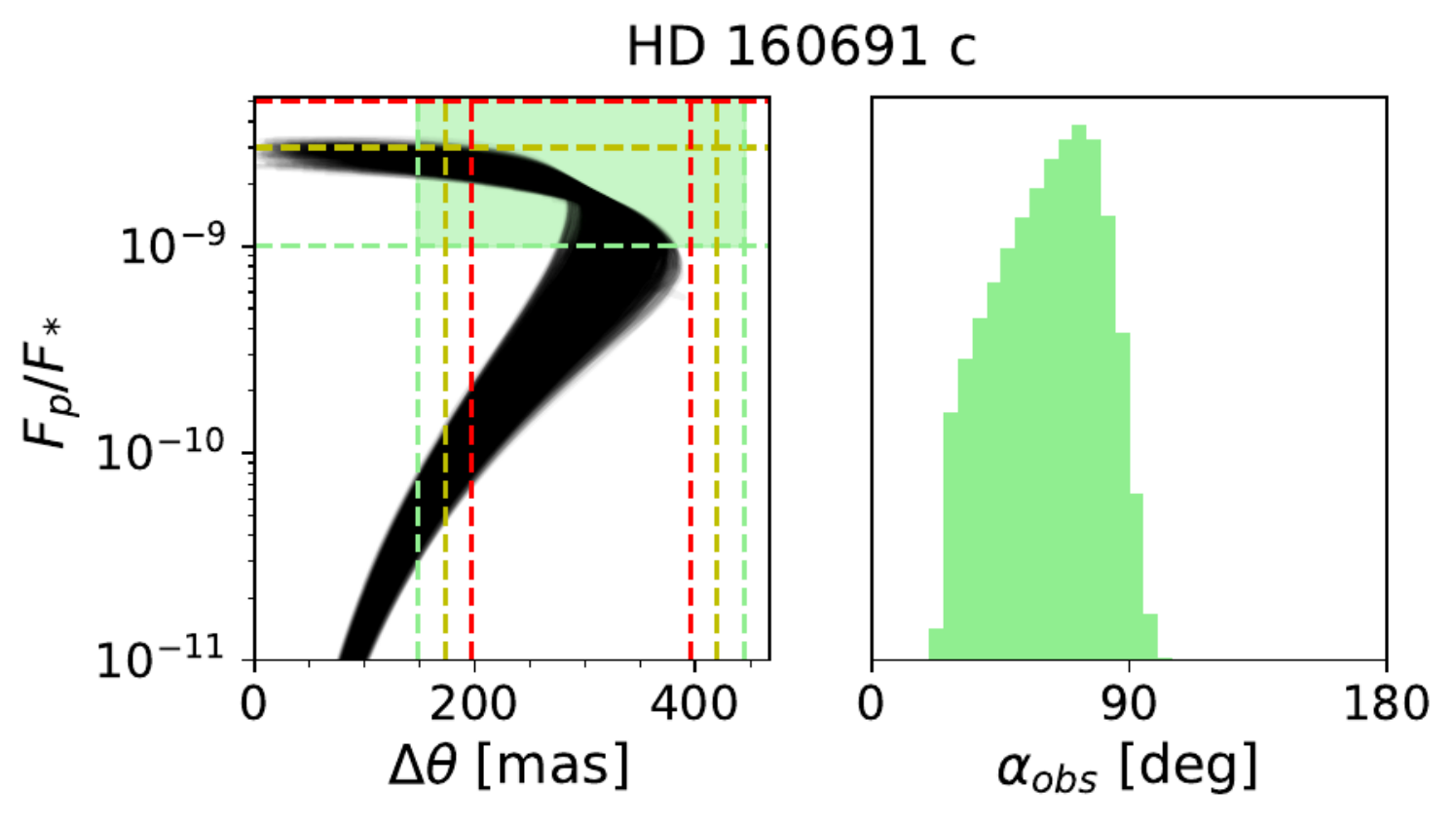} 
	\hfill
	\includegraphics[width=6cm]{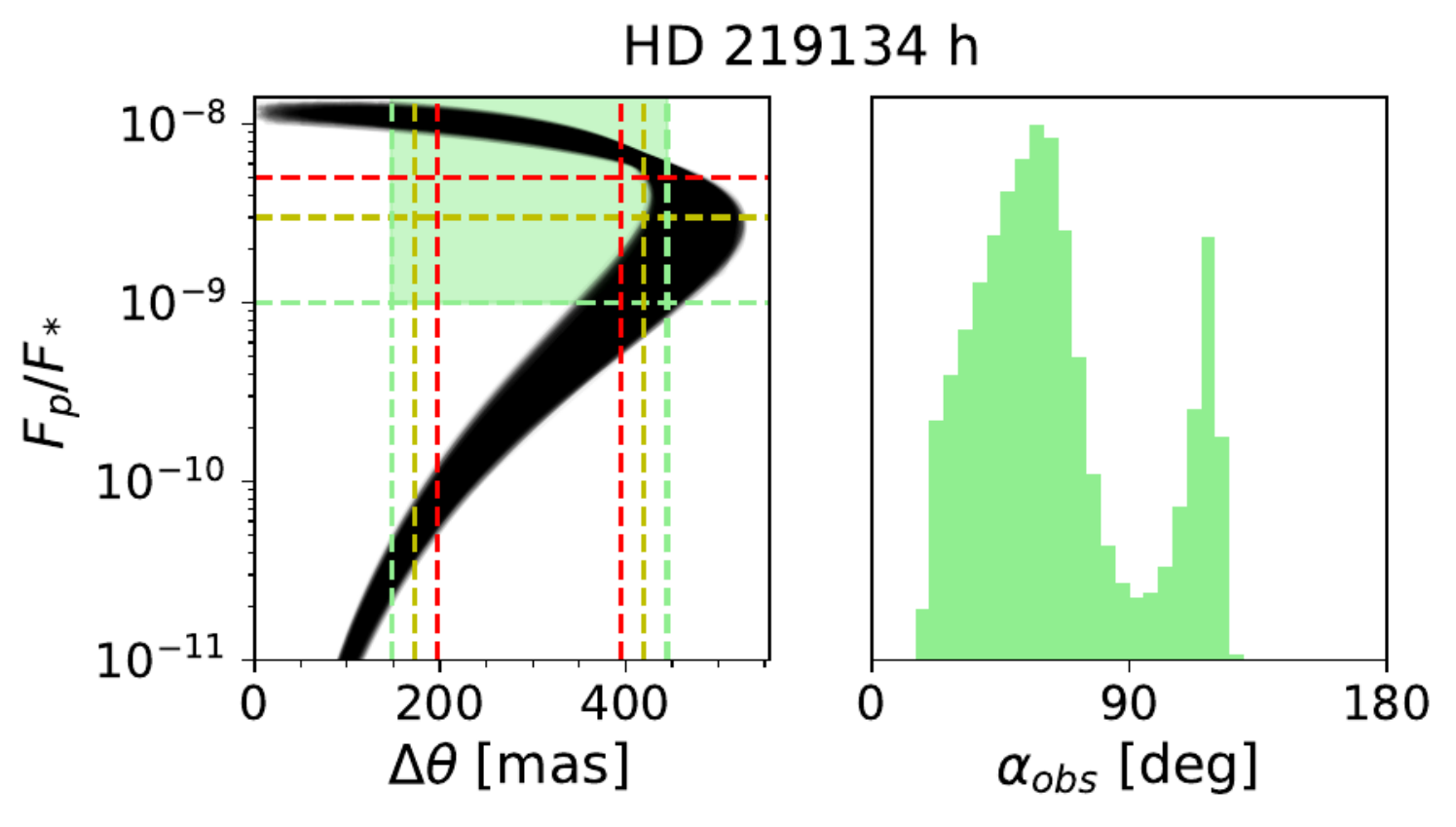} 
   \vspace{-0.1cm}
	   \\
	\includegraphics[width=6cm]{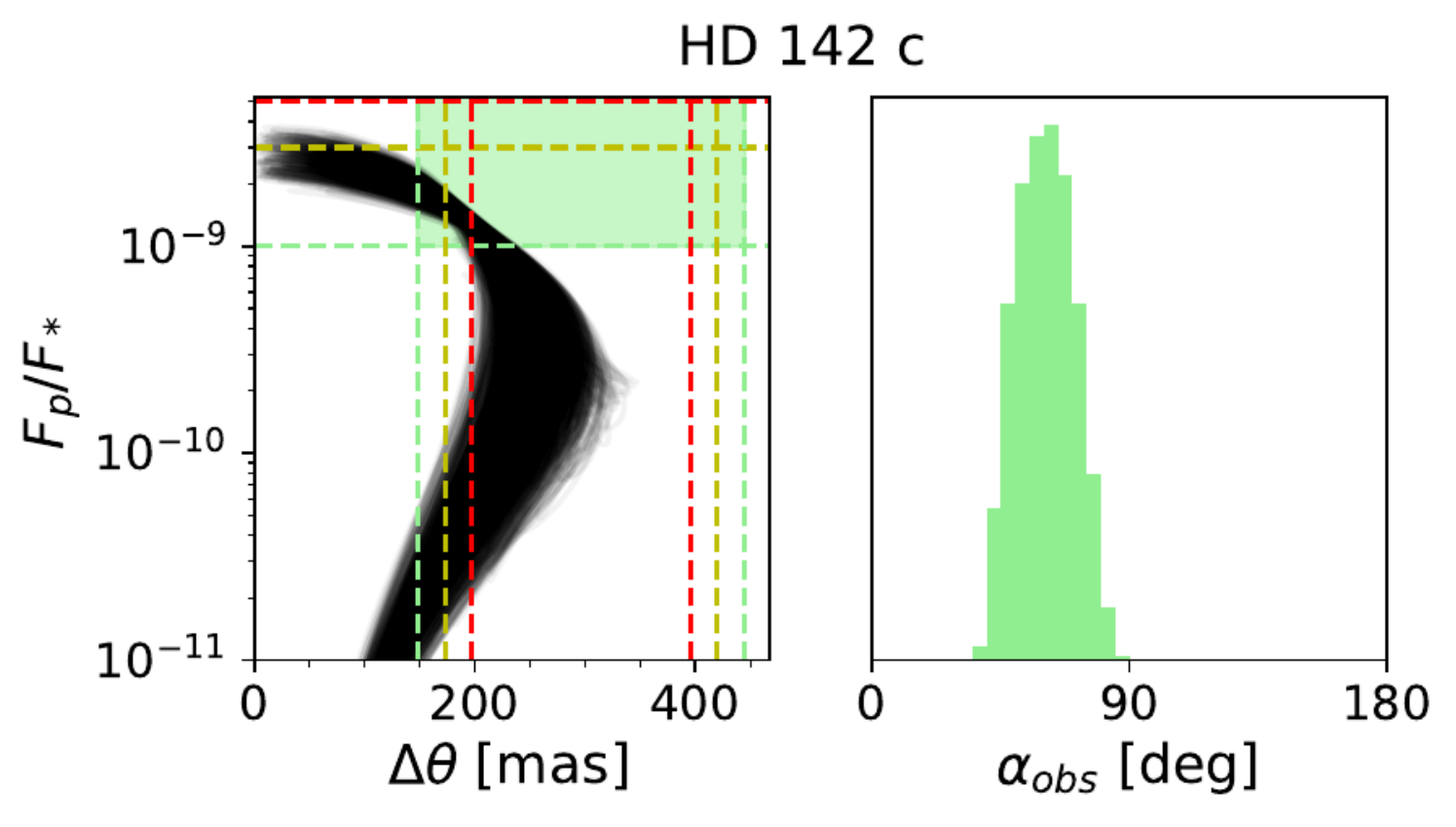} 
	\hfill
	\includegraphics[width=6cm]{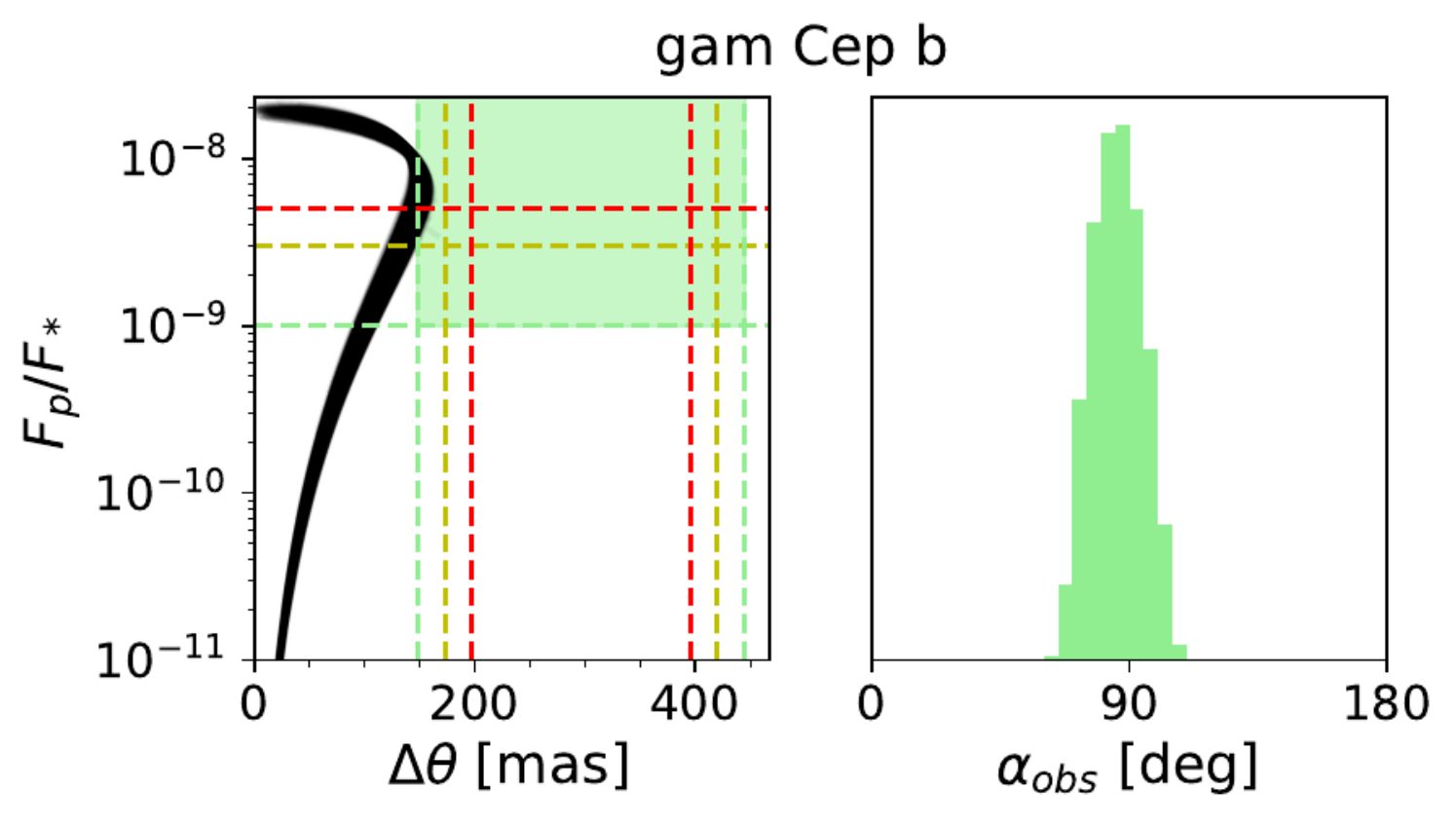} 
	\hfill
	\includegraphics[width=6cm]{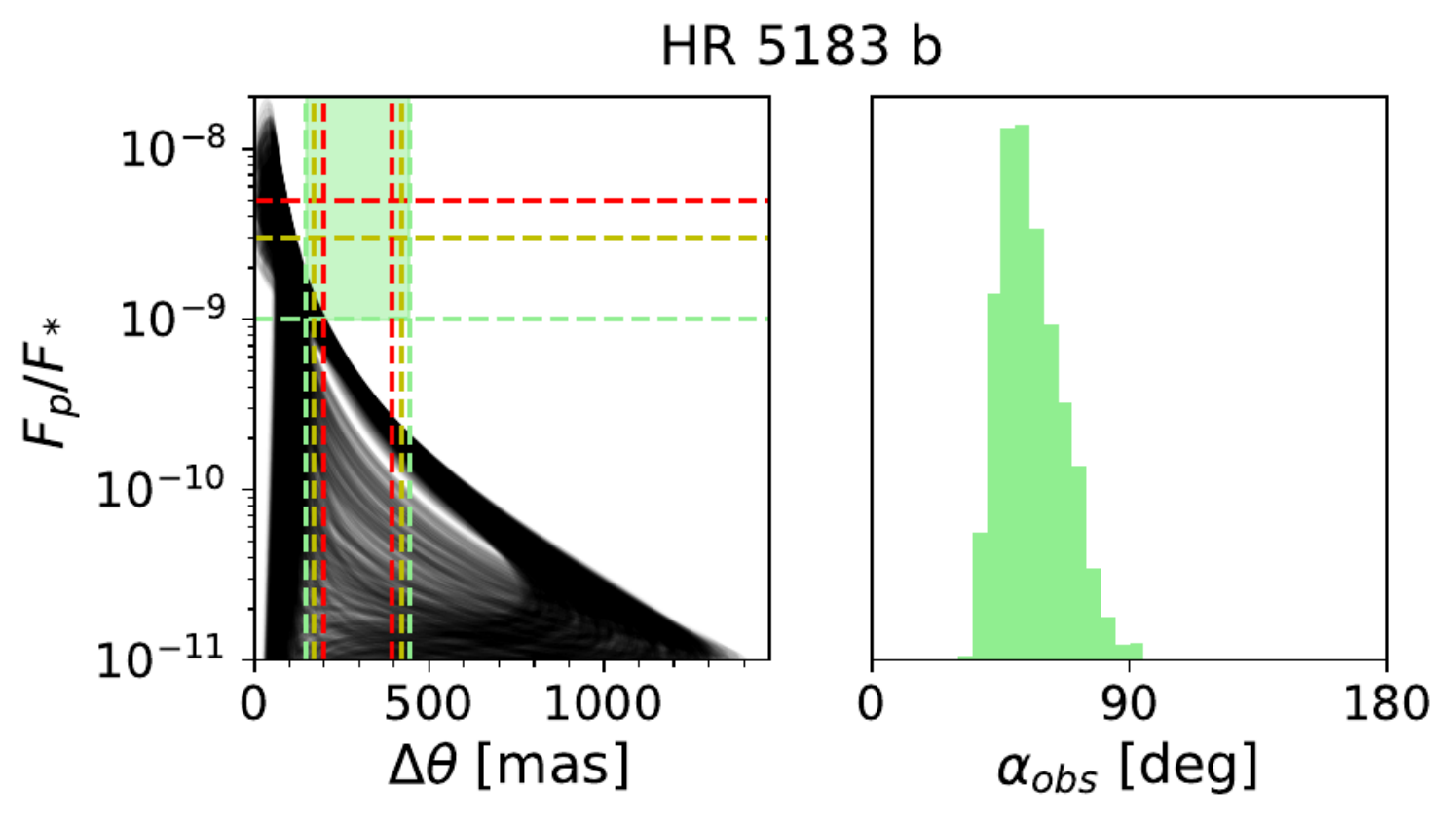} 
   \vspace{-0.1cm}
	   \\
	\includegraphics[width=6cm]{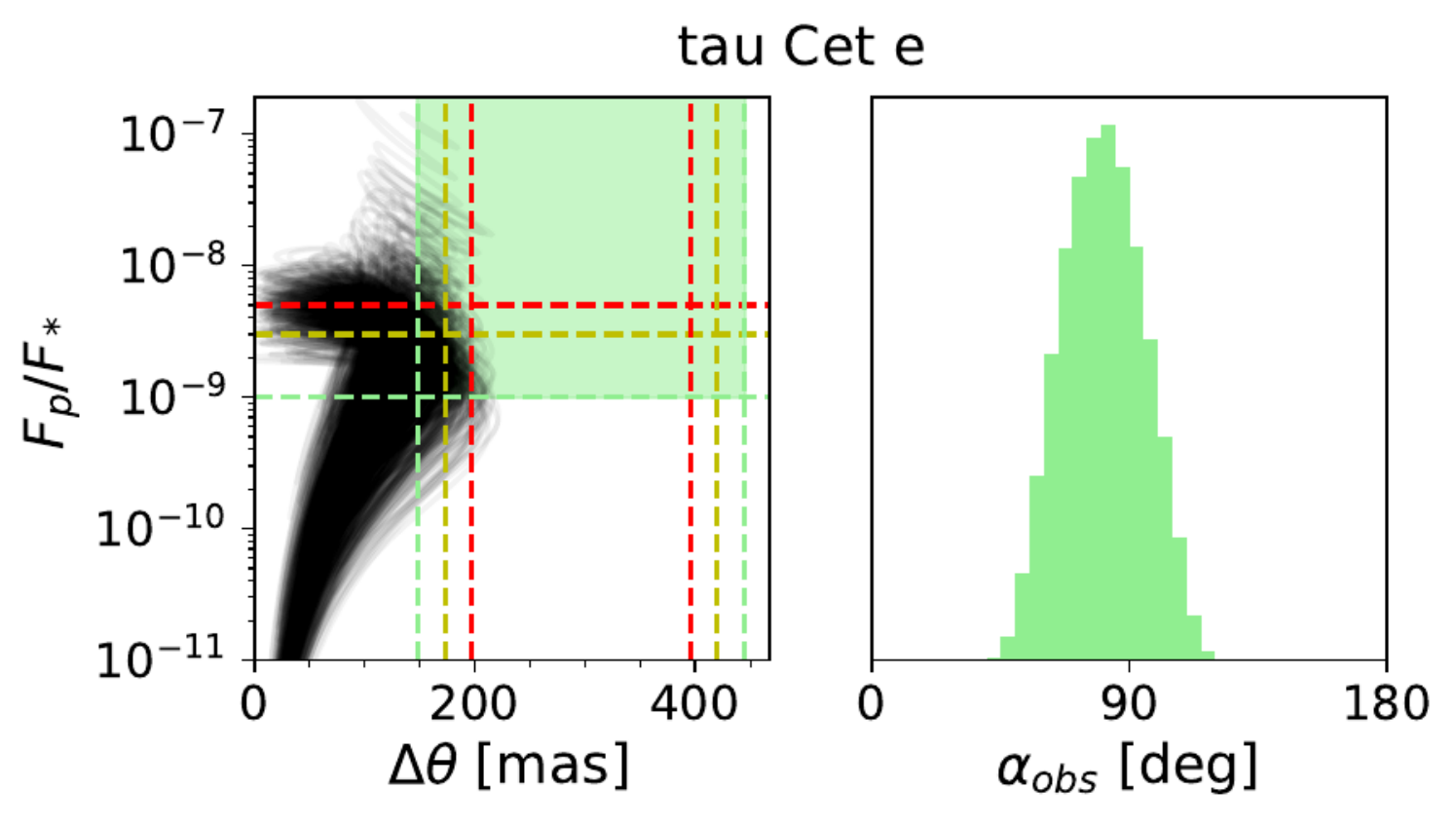} 
	\hfill
	\includegraphics[width=6cm]{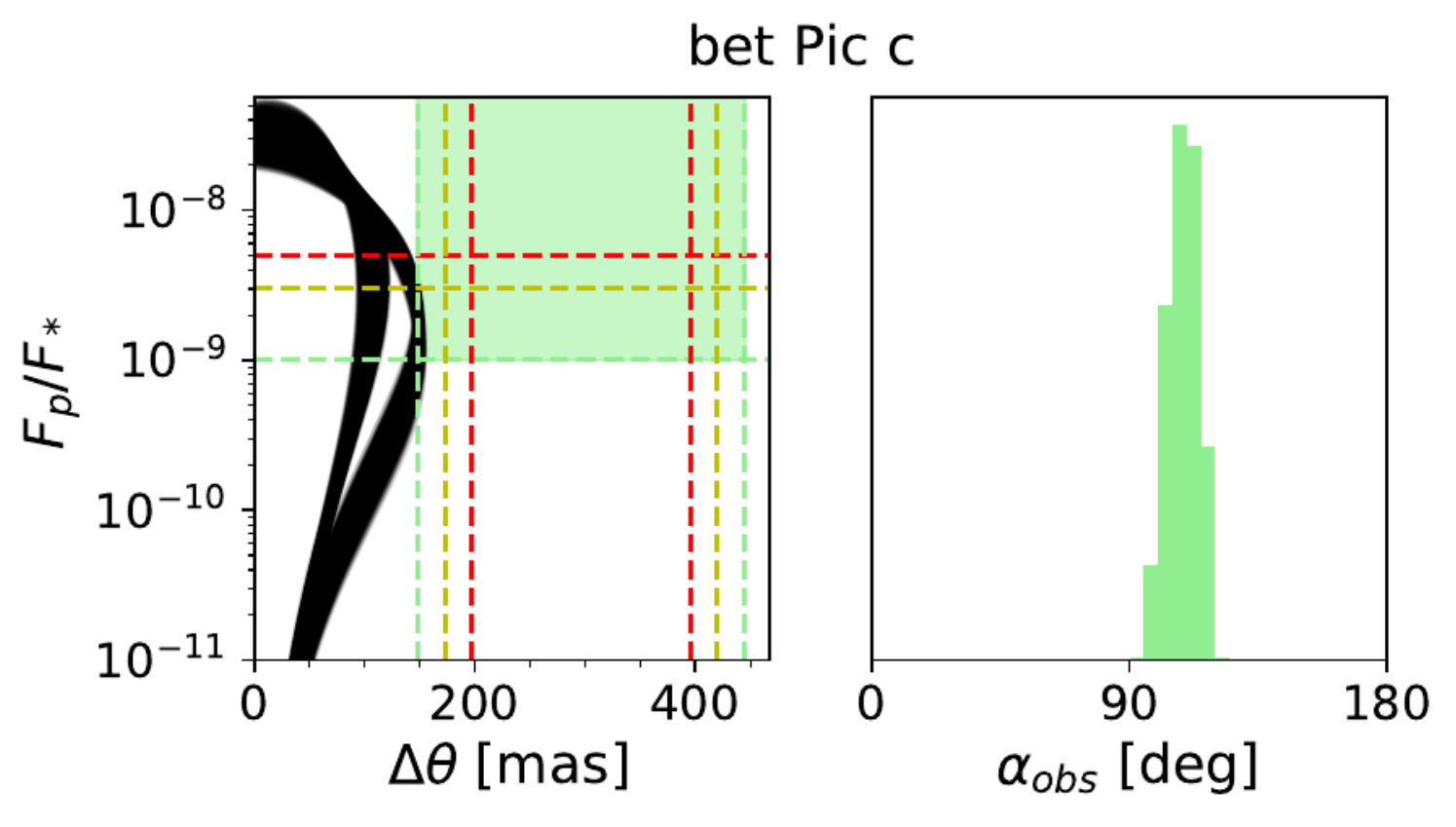} 
	\hfill
	\includegraphics[width=6cm]{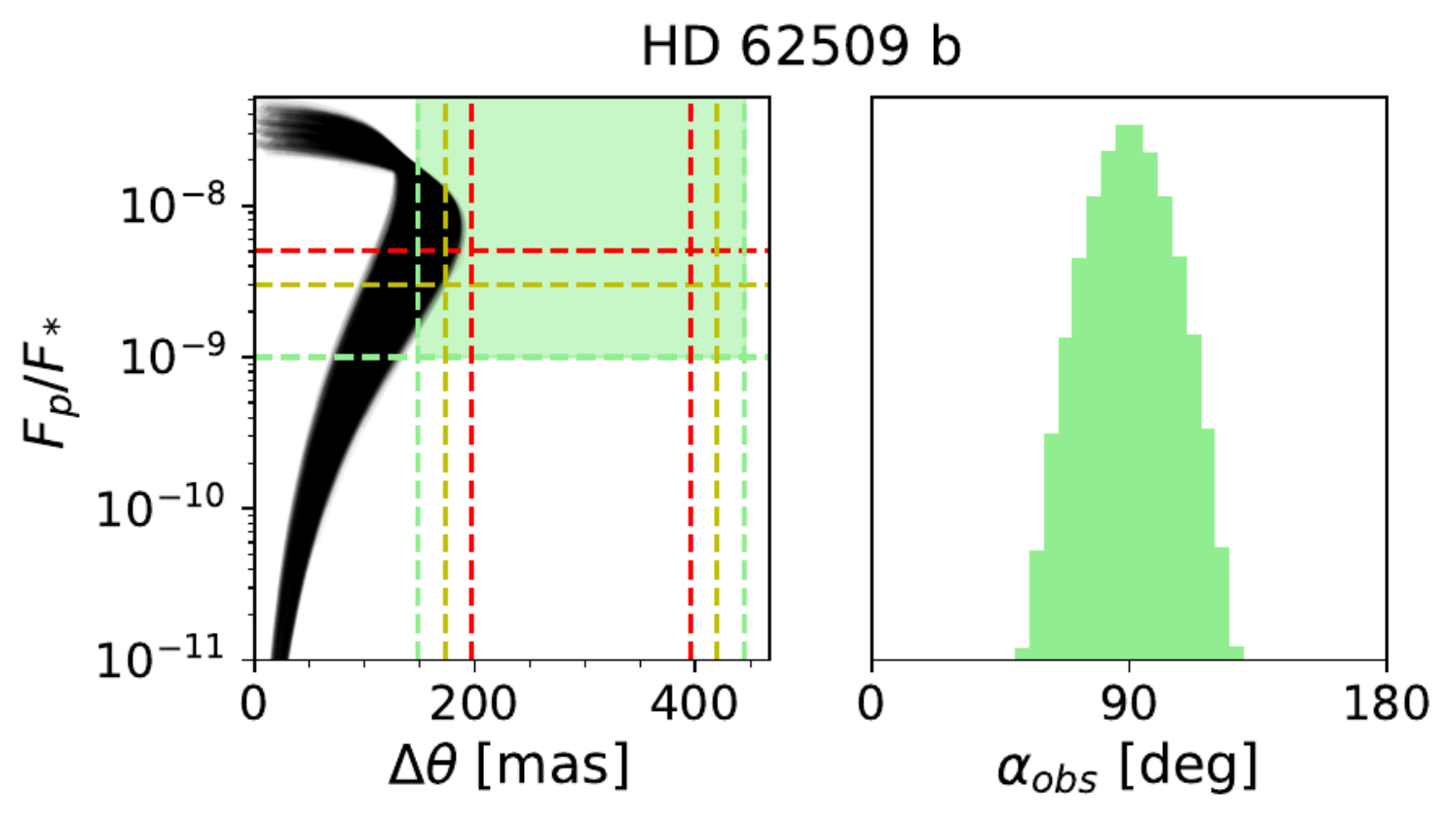} 
   \vspace{-0.1cm}
   \\
	\caption{Detectability conditions for the Roman-accessible exoplanets. In each left panel, solid black lines in the $F_p/F_\star$-$\Delta \theta$ diagram correspond to independent orbital realizations. 
	For the sake of clarity, only 1000 of the total 10000 realizations are shown.
	Horizontal dashed lines indicate the $C_{min}$ and vertical dashed lines, the IWA and OWA of the CGI at $\lambda$=575 nm for the optimistic (green), intermediate (yellow) and pessimistic (red) configurations (Table \ref{table:instrument_scenarios}). 
	Regions in green are the windows of detectability in the optimistic CGI configuration
	at this wavelength and
	the green histograms in the right panels show the posterior distributions of $\alpha_{obs}$ for this scenario.}
	\label{fig:unknownorbits_orbits&histograms}
	\end{figure*}

Some key findings ($P_{access}$, $\alpha_{obs}$, $t_{obs}$) on the detectability of the up to 26 Roman-accessible exoplanets with $P_{access}>25\%$ and $V$<7 mag are listed in Table \ref{table:results_detectability_multiwav} for all the CGI scenarios.
For reference, we also add the corresponding findings at $\lambda$=730 and 825 nm, the effective wavelengths of the two other commissioned filters for the coronagraph.
At these wavelengths, we assume an albedo of $A_g$=0.3 and account for the modified IWA and OWA.
The transit probability of these planets is listed in the output catalogue (Table \ref{table:output_catalogue}).
Figure \ref{fig:unknownorbits_orbits&histograms} (left panel of each diagram) shows the
tracks of contrast and angular separation of the random orbital realizations in our analysis. 
It also shows (right panel) the corresponding distributions of $\alpha_{obs}$ for the optimistic CGI scenario, which indicate the observable phase angles that occur more often. 
As we have discretised the orbits evenly in the true anomaly (rather than in time), these distributions do not translate directly into time spent at any given interval of phase angles.

At our reference wavelength $\lambda$=575 nm, the number of Roman-accessible exoplanets in the optimistic, intermediate and pessimistic CGI scenarios is 26, 10 and 3, respectively (Table \ref{table:results_detectability_multiwav}).
HD 219134 h, 47 UMa c and eps Eri b are the only planets that would be accessible in all three scenarios with $P_{access}$>25\%.
Generally, $P_{access}$ decreases at longer wavelengths because the IWA increases with $\lambda$, masking a larger region around the host star.
Particular cases like eps Eri b or HD 219134 h show an increase in $P_{access}$ at longer $\lambda$.
These are planets that reach large angular separations and, at $\lambda$=575 nm, orbit partly outside the OWA of the coronagraph (Fig. \ref{fig:unknownorbits_orbits&histograms}).
Hence, their $P_{access}$ increases at longer wavelengths because both the IWA and OWA move outwards.

\renewcommand{\thefigure}{\arabic{figure} (Cont.)}
\addtocounter{figure}{-1}
\begin{figure*}
	\centering
	\includegraphics[width=6cm]{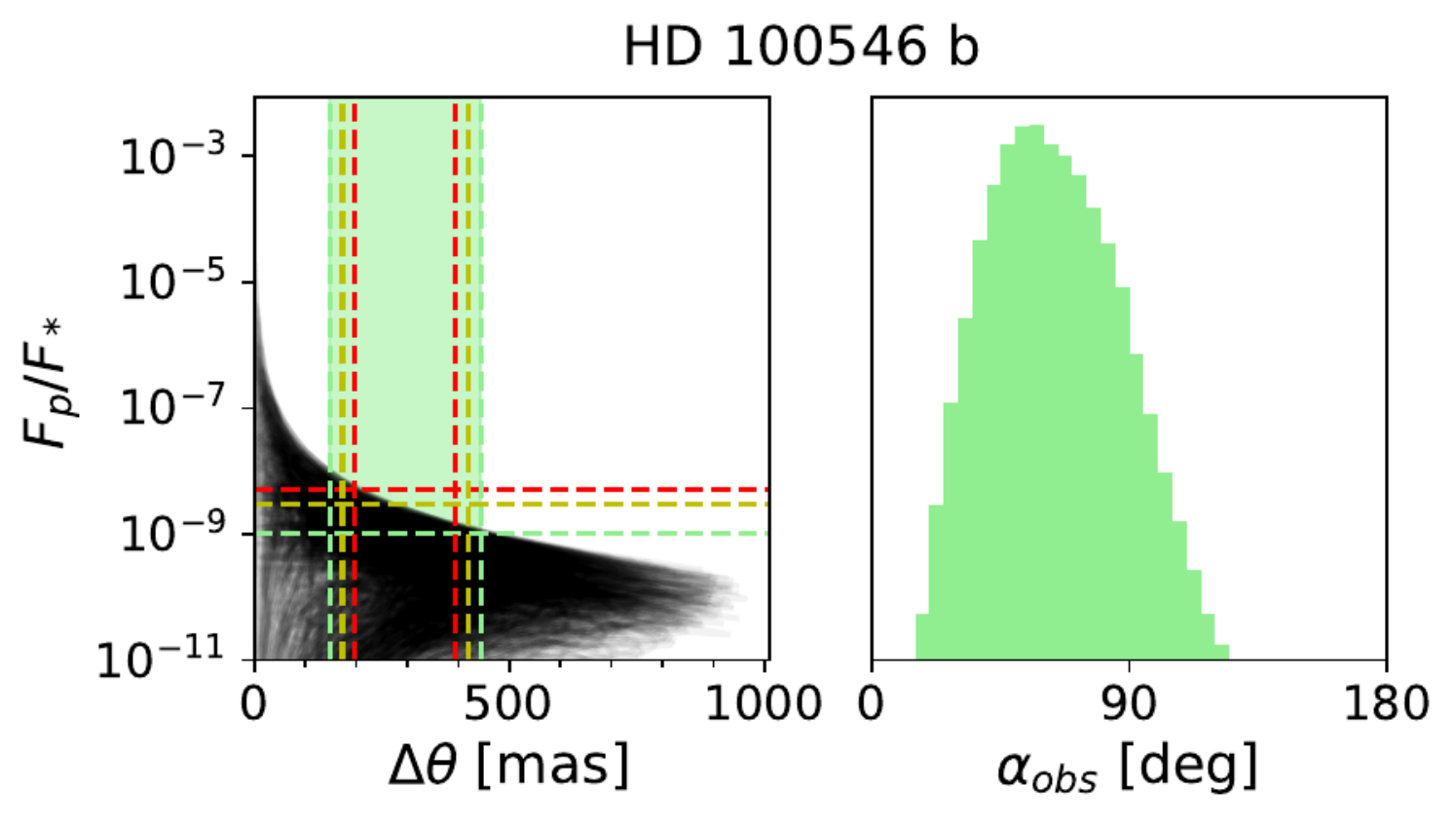} 
	\hfill
	\includegraphics[width=6cm]{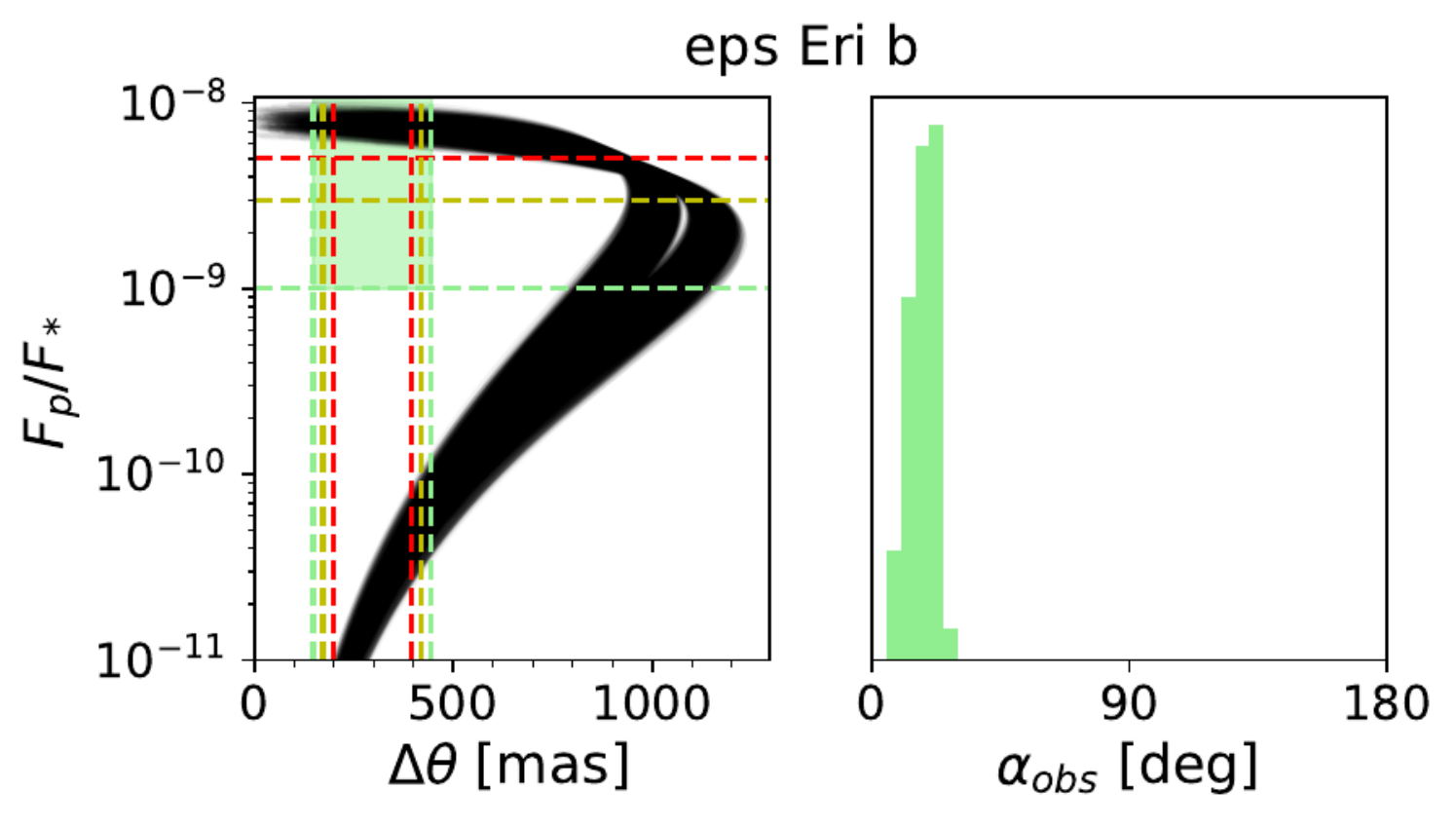} 
	\hfill
	\includegraphics[width=6cm]{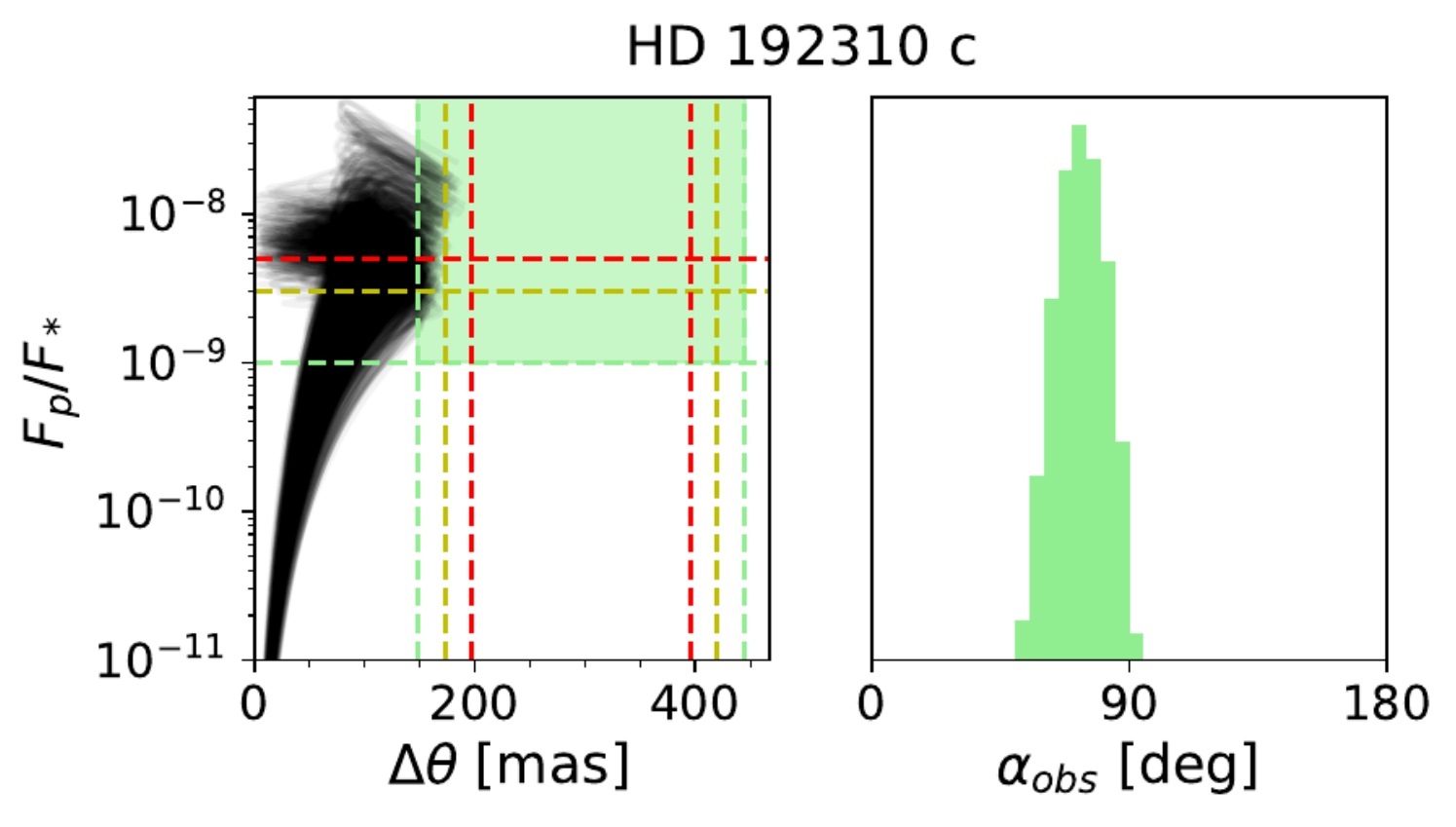} 
   \vspace{-0.1cm}
   \\
	\includegraphics[width=6cm]{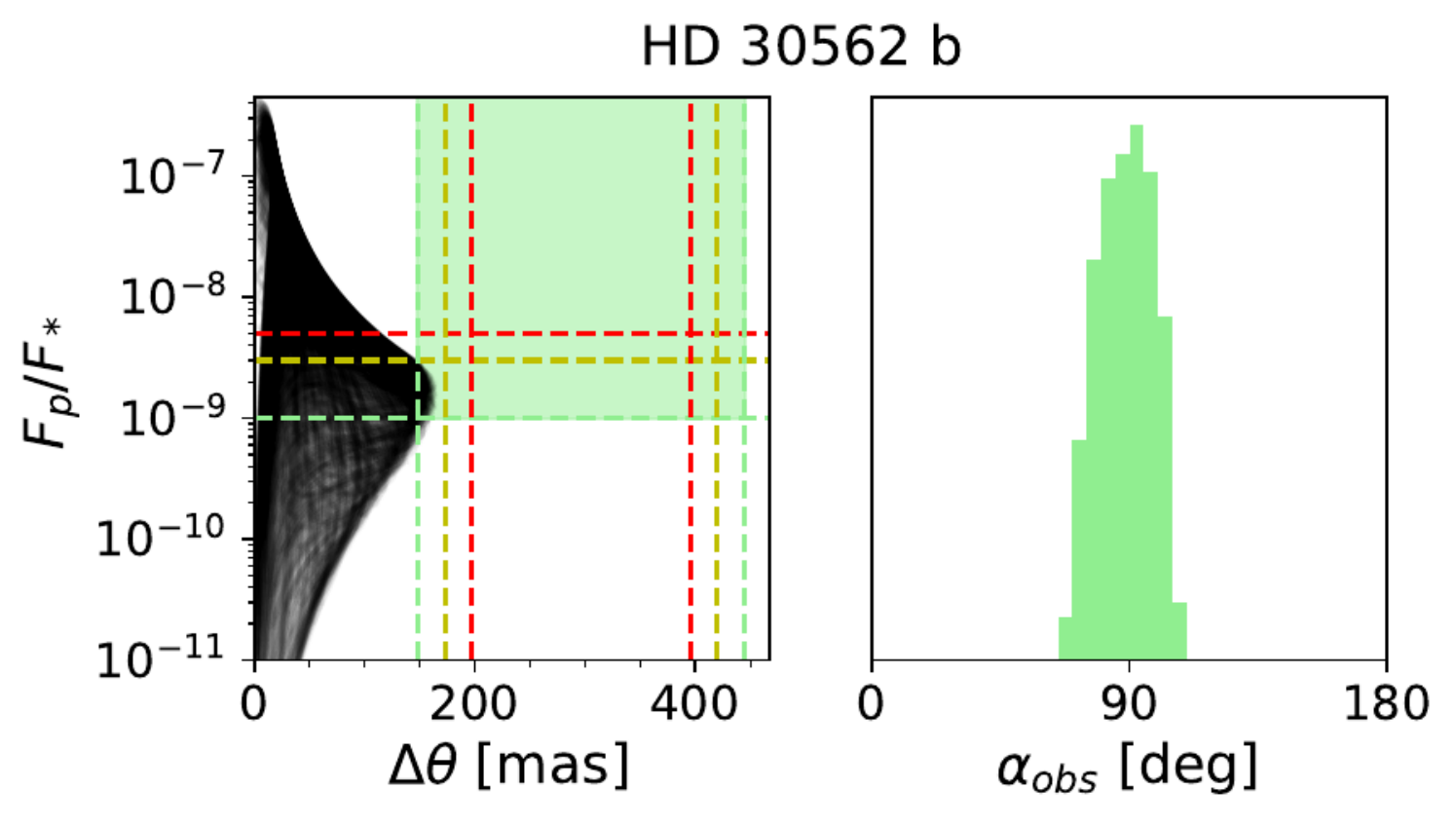} 
	\includegraphics[width=6cm]{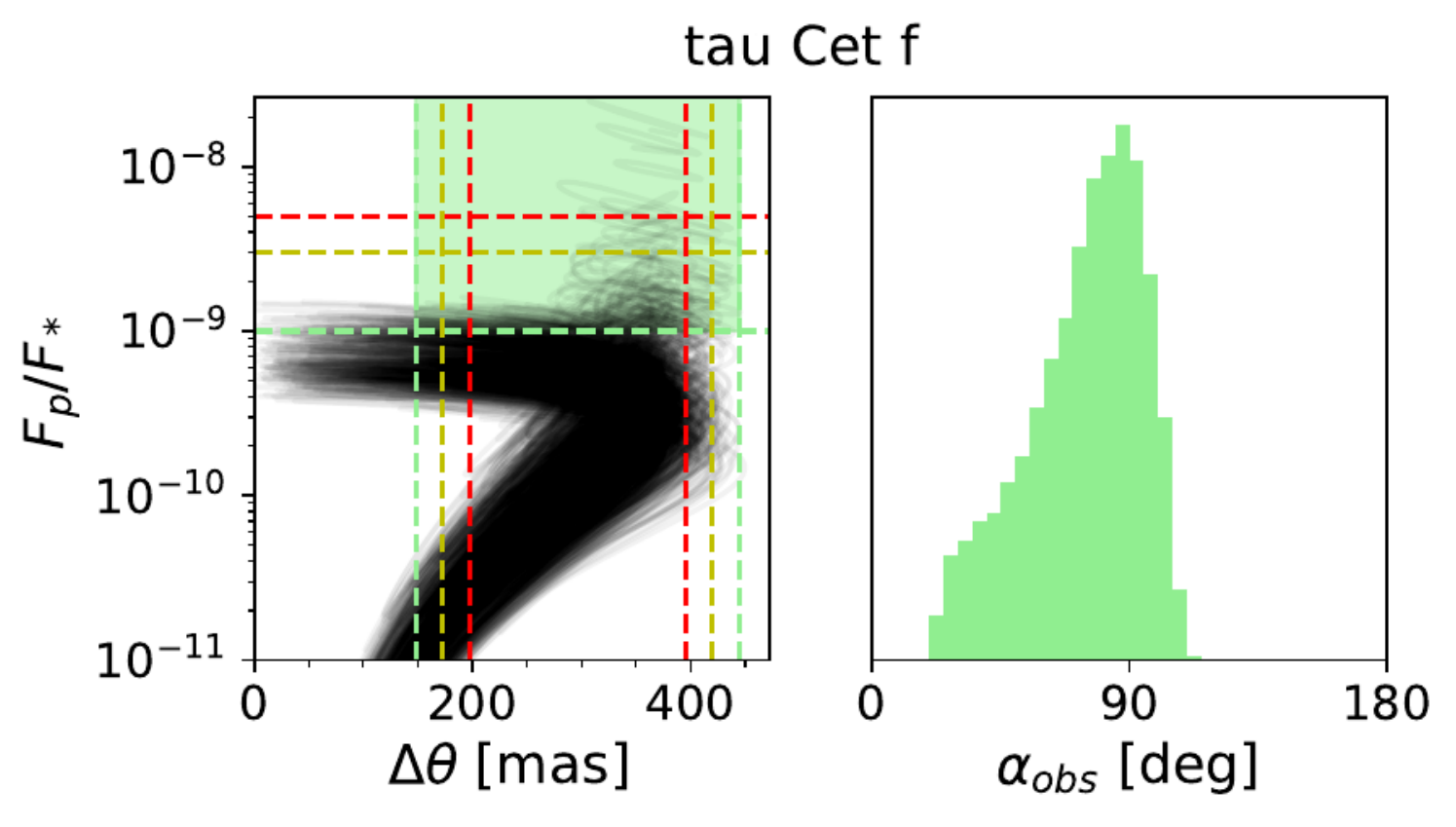} 
	\caption{}
\end{figure*}
\renewcommand{\thefigure}{\arabic{figure}}

The transit probability of the Roman-accessible exoplanets is low in all cases (Table \ref{table:output_catalogue}), with the maximum being $P_{tr}$=2.29\% for HD 62509 b.
This super-Jupiter ($M_p\, sin\,i$=2.3$M_J$) orbits the nearby ($d$=10.34 pc) K0 III giant Pollux.
With an orbital period of 589.6 days, HD 62509 b may require observations spanning multiple years to confirm its eventual transits. 
However, improving the orbital characterization with RV measurements could constrain the time of inferior conjunction and reduce the baseline needed.
This star was targeted for 27 days in TESS Sector 20, but its large optical magnitude ($V$=1.14)
poses a problem with photometric saturation. 
If this planet was found to transit and also imaged ($P_{access}$=73.84\% in the optimistic CGI scenario), it would be a unique opportunity to characterize its atmosphere by combining both techniques.
An astrometric determination of its inclination, which should be near 90$^{\circ}$ for the planet to transit, would help refine its transit probability.

In Fig. \ref{fig:unknownorbits_orbits&histograms}, those exoplanets with larger uncertainties in their orbital parameters (see Table \ref{table:NASA_database}) generally show larger scatter in their $F_p/F_\star$--$\Delta \theta$ tracks.
Figure \ref{fig:unknownorbits_orbits&histograms} also shows that planets in the sub-giant regime (i.e. those with $M_p<0.36M_J$) experience large increases of $F_p/F_\star$ in a small number of realizations (see e.g. tau Cet e, HD 192310 c, tau Cet f in Fig. \ref{fig:unknownorbits_orbits&histograms}).
This corresponds to orbital configurations with inclinations $i\approx0$ or 180$^\circ$ that result in large values of $M_p$ and in turn $R_p$ (Eq. \ref{eq:mass-radius_lowmass_volatiles}).
These unlikely configurations produce the outlying tracks in Fig. \ref{fig:unknownorbits_orbits&histograms}.

Generally, phase angles both before and after quadrature ($\alpha$<90$^\circ$ and $\alpha$>90$^\circ$, respectively) can be observed at $\lambda$=575 nm in the optimistic CGI configuration (Table \ref{table:results_detectability_multiwav} and Fig. \ref{fig:unknownorbits_orbits&histograms}).
This will be important to better constrain some of the optical properties of the atmosphere that may be more sensitive to the scattering angles \citep{carriongonzalezetal2020, damianoetal2020}.
The minimum value of $\alpha_{obs}$ is in most cases not smaller than about 30$^\circ$.
The main limitation to measure values of $\alpha$ closer to full phase is the IWA.
In this sense, eps Eri b is an outlier that can only be detected at small phase angles in the observing mode that we are considering here (see Sect. \ref{subsec:discussion_epsErib}).
Correspondingly, the maximum $\alpha_{obs}$ is not larger than 110$^\circ$ for most of these exoplanets.
Typically, at large phase angles, the planet is not bright enough and its contrast drops below the specified $C_{min}$.
Indeed, in the intermediate and pessimistic CGI scenarios, only phase angles smaller than quadrature are generally observed (Table \ref{table:results_detectability_multiwav}).
Therefore, both the IWA and $C_{min}$ are major factors limiting the windows of detectability.
This is also the reason why, typically, both $t_{obs}$ and the range of $\alpha_{obs}$ decrease at longer wavelengths (Table \ref{table:results_detectability_multiwav}).

\begin{table*}
\tiny
\caption{Detectability conditions for the Roman-accessible exoplanets with $P_{access}$>25\% and $V$<7 mag for each of the CGI scenarios. 
In addition to our reference wavelength $\lambda$=575 nm, we include the results at $\lambda$=730 and 825 nm.
For these cases we also assume a geometrical albedo $A_g$=0.3 to compute $F_p/F_\star$. }
\label{table:results_detectability_multiwav}
\begin{tabular}{c l c c c c c c c c c}
    \hline \hline
         &  & \multicolumn{3}{c}{575 nm}    & \multicolumn{3}{c}{730 nm}    & \multicolumn{3}{c}{825 nm} \\
    & Name	& $P_{access}$ \% & $t_{obs}$	[days]	& $\alpha_{obs}$ [deg] & $P_{access}$ \% & $t_{obs}$	[days]	& $\alpha_{obs}$ [deg] & $P_{access}$ \% & $t_{obs}$	[days]	& $\alpha_{obs}$ [deg]  \\ 
\hline
\multirow{26}{*}{\rotatebox{90}{Optimistic}} & HD 154345 b		& 100.00		& $1485^{+ 735 }_{- 242 }$		& [41$^{+ 16 }_{- 2 }$,107$^{+ 2 }_{- 2 }$]		& 100.00		& $1109^{+ 1033 }_{- 153 }$		& [55$^{+ 4 }_{- 3 }$,107$^{+ 2 }_{- 2 }$]		& 100.00		& $793^{+ 470 }_{- 131 }$		& [66$^{+ 4 }_{- 4 }$,106$^{+ 1 }_{- 2 }$] \\ 
 & pi Men b		& 100.00		& $330^{+ 32 }_{- 17 }$		& [69$^{+ 7 }_{- 2 }$,95$^{+ 1 }_{- 1 }$]		& 100.00		& $227^{+ 16 }_{- 15 }$		& [80$^{+ 3 }_{- 1 }$,95$^{+ 1 }_{- 1 }$]		& 99.98		& $139^{+ 16 }_{- 15 }$		& [87$^{+ 1 }_{- 1 }$,95$^{+ 1 }_{- 2 }$] \\ 
 & 55 Cnc d		& 100.00		& $2117^{+ 125 }_{- 318 }$		& [30$^{+ 20 }_{- 10 }$,84$^{+ 2 }_{- 2 }$]		& 100.00		& $2052^{+ 169 }_{- 424 }$		& [30$^{+ 20 }_{- 5 }$,84$^{+ 2 }_{- 2 }$]		& 100.00		& $1985^{+ 218 }_{- 462 }$		& [30$^{+ 20 }_{- 1 }$,84$^{+ 2 }_{- 2 }$] \\ 
 & HD 114613 b		& 100.00		& $750^{+ 271 }_{- 168 }$		& [42$^{+ 17 }_{- 5 }$,106$^{+ 10 }_{- 33 }$]		& 100.00		& $514^{+ 75 }_{- 55 }$		& [49$^{+ 12 }_{- 5 }$,76$^{+ 4 }_{- 4 }$]		& 100.00		& $431^{+ 57 }_{- 49 }$		& [54$^{+ 10 }_{- 5 }$,76$^{+ 3 }_{- 4 }$] \\ 
 & ups And d		& 100.00		& $394^{+ 224 }_{- 44 }$		& [69$^{+ 5 }_{- 1 }$,124$^{+ 1 }_{- 5 }$]		& 100.00		& $114^{+ 190 }_{- 51 }$		& [98$^{+ 2 }_{- 3 }$,117$^{+ 2 }_{- 3 }$]		& 14.26		& $456^{+ 73 }_{- 346 }$		& [102$^{+ 6 }_{- 6 }$,112$^{+ 3 }_{- 9 }$] \\ 
 & HD 217107 c		& 100.00		& $522^{+ 96 }_{- 60 }$		& [43$^{+ 14 }_{- 4 }$,72$^{+ 30 }_{- 4 }$]		& 100.00		& $391^{+ 40 }_{- 38 }$		& [51$^{+ 10 }_{- 4 }$,72$^{+ 4 }_{- 4 }$]		& 99.99		& $313^{+ 33 }_{- 33 }$		& [56$^{+ 9 }_{- 4 }$,72$^{+ 4 }_{- 4 }$] \\ 
 & 14 Her b		& 100.00		& $523^{+ 391 }_{- 252 }$		& [58$^{+ 22 }_{- 14 }$,107$^{+ 7 }_{- 7 }$]		& 72.63		& $390^{+ 209 }_{- 208 }$		& [69$^{+ 20 }_{- 9 }$,104$^{+ 3 }_{- 13 }$]		& 40.13		& $245^{+ 106 }_{- 110 }$		& [78$^{+ 12 }_{- 6 }$,100$^{+ 4 }_{- 12 }$] \\ 
 & 47 UMa c		& 100.00		& $1288^{+ 539 }_{- 227 }$		& [37$^{+ 20 }_{- 3 }$,116$^{+ 3 }_{- 3 }$]		& 100.00		& $1055^{+ 769 }_{- 155 }$		& [47$^{+ 9 }_{- 3 }$,116$^{+ 3 }_{- 3 }$]		& 100.00		& $898^{+ 770 }_{- 127 }$		& [55$^{+ 6 }_{- 4 }$,116$^{+ 2 }_{- 3 }$] \\ 
 & 47 UMa b		& 100.00		& $153^{+ 95 }_{- 32 }$		& [73$^{+ 2 }_{- 2 }$,109$^{+ 2 }_{- 2 }$]		& 0.00		& $-$		& $-$		& 0.00		& $-$		& $-$ \\ 
 & HD 190360 b		& 100.00		& $1371^{+ 845 }_{- 529 }$		& [44$^{+ 14 }_{- 14 }$,119$^{+ 3 }_{- 9 }$]		& 100.00		& $833^{+ 701 }_{- 272 }$		& [51$^{+ 15 }_{- 13 }$,106$^{+ 7 }_{- 13 }$]		& 100.00		& $698^{+ 423 }_{- 296 }$		& [55$^{+ 18 }_{- 12 }$,99$^{+ 9 }_{- 7 }$] \\ 
 & psi 1 Dra B b		& 100.00		& $1530^{+ 960 }_{- 366 }$		& [34$^{+ 24 }_{- 2 }$,109$^{+ 3 }_{- 5 }$]		& 100.00		& $908^{+ 899 }_{- 229 }$		& [45$^{+ 13 }_{- 2 }$,96$^{+ 4 }_{- 4 }$]		& 94.02		& $589^{+ 695 }_{- 265 }$		& [54$^{+ 8 }_{- 4 }$,86$^{+ 4 }_{- 7 }$] \\ 
 & HD 219077 b		& 99.84		& $216^{+ 103 }_{- 66 }$		& [75$^{+ 4 }_{- 30 }$,89$^{+ 0 }_{- 1 }$]		& 37.66		& $20^{+ 23 }_{- 20 }$		& [75$^{+ 2 }_{- 13 }$,76$^{+ 1 }_{- 14 }$]		& 0.00		& $-$		& $-$ \\ 
 & HD 134987 c		& 99.26		& $1309^{+ 642 }_{- 230 }$		& [43$^{+ 13 }_{- 4 }$,93$^{+ 4 }_{- 4 }$]		& 99.24		& $677^{+ 333 }_{- 169 }$		& [56$^{+ 8 }_{- 6 }$,90$^{+ 4 }_{- 7 }$]		& 97.70		& $257^{+ 85 }_{- 78 }$		& [66$^{+ 8 }_{- 7 }$,79$^{+ 5 }_{- 6 }$] \\ 
 & HD 160691 c		& 98.84		& $2008^{+ 420 }_{- 374 }$		& [30$^{+ 26 }_{- 6 }$,94$^{+ 4 }_{- 3 }$]		& 98.84		& $1840^{+ 507 }_{- 412 }$		& [33$^{+ 23 }_{- 2 }$,94$^{+ 4 }_{- 3 }$]		& 98.84		& $1671^{+ 607 }_{- 377 }$		& [37$^{+ 19 }_{- 2 }$,94$^{+ 4 }_{- 3 }$] \\ 
 & HD 219134 h		& 97.93		& $917^{+ 152 }_{- 159 }$		& [30$^{+ 25 }_{- 11 }$,123$^{+ 2 }_{- 2 }$]		& 100.00		& $1559^{+ 413 }_{- 325 }$		& [30$^{+ 26 }_{- 6 }$,123$^{+ 2 }_{- 2 }$]		& 100.00		& $1547^{+ 425 }_{- 356 }$		& [30$^{+ 26 }_{- 3 }$,123$^{+ 2 }_{- 2 }$] \\ 
 & HD 142 c		& 97.55		& $865^{+ 418 }_{- 150 }$		& [44$^{+ 11 }_{- 4 }$,77$^{+ 5 }_{- 5 }$]		& 97.51		& $425^{+ 155 }_{- 80 }$		& [56$^{+ 7 }_{- 5 }$,77$^{+ 5 }_{- 5 }$]		& 90.88		& $94^{+ 66 }_{- 48 }$		& [64$^{+ 6 }_{- 5 }$,73$^{+ 4 }_{- 5 }$] \\ 
 & gam Cep b		& 97.28		& $135^{+ 94 }_{- 69 }$		& [73$^{+ 5 }_{- 4 }$,101$^{+ 4 }_{- 5 }$]		& 0.00		& $-$		& $-$		& 0.00		& $-$		& $-$ \\ 
 & HR 5183 b		& 94.24		& $145^{+ 87 }_{- 83 }$		& [47$^{+ 15 }_{- 7 }$,52$^{+ 13 }_{- 10 }$]		& 19.19		& $0^{+ 15 }_{- 0 }$		& [58$^{+ 6 }_{- 4 }$,58$^{+ 6 }_{- 4 }$]		& 0.00		& $-$		& $-$ \\ 
 & tau Cet e		& 87.75		& $34^{+ 36 }_{- 14 }$		& [61$^{+ 10 }_{- 8 }$,100$^{+ 9 }_{- 10 }$]		& 16.04		& $19^{+ 20 }_{- 11 }$		& [75$^{+ 8 }_{- 8 }$,92$^{+ 8 }_{- 9 }$]		& 1.02		& $12^{+ 11 }_{- 7 }$		& [82$^{+ 4 }_{- 6 }$,90$^{+ 7 }_{- 5 }$] \\ 
 & bet Pic c		& 78.81		& $64^{+ 22 }_{- 27 }$		& [102$^{+ 5 }_{- 4 }$,116$^{+ 2 }_{- 4 }$]		& 0.00		& $-$		& $-$		& 0.00		& $-$		& $-$ \\ 
 & HD 62509 b		& 73.84		& $208^{+ 136 }_{- 106 }$		& [64$^{+ 11 }_{- 8 }$,116$^{+ 8 }_{- 12 }$]		& 1.06		& $35^{+ 21 }_{- 18 }$		& [84$^{+ 3 }_{- 4 }$,97$^{+ 4 }_{- 4 }$]		& 0.00		& $-$		& $-$ \\ 
 & HD 100546 b		& 73.54		& $8771^{+ 10693 }_{- 5996 }$		& [41$^{+ 27 }_{- 17 }$,90$^{+ 20 }_{- 29 }$]		& 70.31		& $7192^{+ 10936 }_{- 5356 }$		& [45$^{+ 26 }_{- 17 }$,84$^{+ 20 }_{- 26 }$]		& 65.50		& $6415^{+ 10842 }_{- 4827 }$		& [46$^{+ 25 }_{- 17 }$,82$^{+ 20 }_{- 24 }$] \\ 
 & eps Eri b		& 57.99		& $252^{+ 60 }_{- 59 }$		& [12$^{+ 8 }_{- 4 }$,24$^{+ 1 }_{- 1 }$]		& 74.58		& $336^{+ 74 }_{- 77 }$		& [16$^{+ 11 }_{- 6 }$,32$^{+ 2 }_{- 1 }$]		& 86.29		& $390^{+ 86 }_{- 92 }$		& [18$^{+ 12 }_{- 7 }$,37$^{+ 2 }_{- 2 }$] \\ 
 & HD 192310 c		& 49.36		& $99^{+ 123 }_{- 53 }$		& [62$^{+ 8 }_{- 6 }$,85$^{+ 5 }_{- 8 }$]		& 0.15		& $39^{+ 27 }_{- 7 }$		& [82$^{+ 4 }_{- 2 }$,82$^{+ 4 }_{- 2 }$]		& 0.00		& $-$		& $-$ \\ 
 & HD 30562 b		& 33.83		& $235^{+ 84 }_{- 113 }$		& [83$^{+ 13 }_{- 10 }$,95$^{+ 10 }_{- 14 }$]		& 0.00		& $-$		& $-$		& 0.00		& $-$		& $-$ \\ 
 & tau Cet f		& 26.74		& $181^{+ 307 }_{- 114 }$		& [53$^{+ 20 }_{- 29 }$,74$^{+ 26 }_{- 28 }$]		& 25.88		& $184^{+ 334 }_{- 130 }$		& [54$^{+ 19 }_{- 23 }$,75$^{+ 25 }_{- 26 }$]		& 25.12		& $189^{+ 366 }_{- 142 }$		& [55$^{+ 19 }_{- 20 }$,76$^{+ 24 }_{- 25 }$] \\ 

\hline

\multirow{10}{*}{\rotatebox{90}{Intermediate}} & ups And d		& 100.00		& $60^{+ 20 }_{- 7 }$		& [84$^{+ 2 }_{- 1 }$,99$^{+ 2 }_{- 1 }$]		& 0.00		& $-$		& $-$		& 0.00		& $-$	& $-$ \\ 
 & 47 UMa c		& 97.95		& $599^{+ 303 }_{- 99 }$		& [43$^{+ 12 }_{- 3 }$,85$^{+ 4 }_{- 4 }$]		& 97.95		& $336^{+ 184 }_{- 64 }$		& [57$^{+ 5 }_{- 4 }$,85$^{+ 4 }_{- 4 }$]		& 96.41		& $97^{+ 46 }_{- 35 }$		& [69$^{+ 6 }_{- 6 }$,79$^{+ 4 }_{- 6 }$] \\ 
 & HD 190360 b		& 92.97		& $232^{+ 159 }_{- 71 }$		& [49$^{+ 13 }_{- 12 }$,89$^{+ 5 }_{- 20 }$]		& 40.62		& $17^{+ 19 }_{- 17 }$		& [61$^{+ 8 }_{- 7 }$,65$^{+ 7 }_{- 9 }$]		& 0.00		& $-$		& $-$ \\ 
 & HD 219134 h		& 92.15		& $577^{+ 118 }_{- 89 }$		& [27$^{+ 23 }_{- 6 }$,68$^{+ 5 }_{- 3 }$]		& 99.87		& $1039^{+ 110 }_{- 230 }$		& [31$^{+ 26 }_{- 3 }$,96$^{+ 2 }_{- 2 }$]		& 99.87		& $954^{+ 191 }_{- 196 }$		& [34$^{+ 23 }_{- 2 }$,96$^{+ 2 }_{- 2 }$] \\ 
 & HD 154345 b		& 87.68		& $329^{+ 172 }_{- 58 }$		& [49$^{+ 3 }_{- 2 }$,66$^{+ 4 }_{- 4 }$]		& 0.00		& $-$		& $-$		& 0.00		& $-$		& $-$ \\ 
 & 14 Her b		& 67.24		& $70^{+ 82 }_{- 41 }$		& [60$^{+ 14 }_{- 7 }$,69$^{+ 10 }_{- 8 }$]		& 0.00		& $-$		& $-$		& 0.00		& $-$		& $-$ \\ 
 & HD 114613 b		& 65.57		& $32^{+ 29 }_{- 22 }$		& [49$^{+ 12 }_{- 4 }$,51$^{+ 12 }_{- 6 }$]		& 0.00		& $-$		& $-$		& 0.00		& $-$		& $-$ \\ 
 & eps Eri b		& 54.45		& $214^{+ 64 }_{- 35 }$		& [11$^{+ 8 }_{- 2 }$,23$^{+ 1 }_{- 1 }$]		& 70.21		& $285^{+ 79 }_{- 48 }$		& [15$^{+ 10 }_{- 3 }$,30$^{+ 2 }_{- 1 }$]		& 80.68		& $330^{+ 93 }_{- 55 }$		& [17$^{+ 11 }_{- 4 }$,35$^{+ 2 }_{- 1 }$] \\ 
 & pi Men b		& 53.24		& $0^{+ 9 }_{- 0 }$		& [74$^{+ 1 }_{- 1 }$,75$^{+ 1 }_{- 1 }$]		& 0.00		& $-$		& $-$		& 0.00		& $-$		& $-$ \\ 
 & HD 62509 b		& 26.75		& $100^{+ 62 }_{- 54 }$		& [75$^{+ 7 }_{- 5 }$,105$^{+ 5 }_{- 7 }$]		& 0.00		& $-$		& $-$		& 0.00		& $-$		& $-$ \\  
\hline

\multirow{3}{*}{\rotatebox{90}{Pessim.}} & HD 219134 h		& 86.59		& $444^{+ 143 }_{- 68 }$		& [27$^{+ 19 }_{- 2 }$,61$^{+ 3 }_{- 3 }$]		& 95.43		& $591^{+ 199 }_{- 91 }$		& [34$^{+ 19 }_{- 2 }$,78$^{+ 3 }_{- 3 }$]		& 95.43		& $522^{+ 232 }_{- 81 }$		& [38$^{+ 15 }_{- 2 }$,78$^{+ 3 }_{- 3 }$] \\ 
 & 47 UMa c		& 82.84		& $87^{+ 45 }_{- 33 }$		& [50$^{+ 4 }_{- 3 }$,62$^{+ 6 }_{- 6 }$]		& 0.00		& $-$		& $-$		& 0.00		& $-$		& $-$ \\ 
 & eps Eri b		& 51.29		& $172^{+ 67 }_{- 28 }$		& [11$^{+ 7 }_{- 1 }$,21$^{+ 1 }_{- 1 }$]		& 65.57		& $232^{+ 84 }_{- 38 }$		& [15$^{+ 9 }_{- 1 }$,28$^{+ 2 }_{- 1 }$]		& 74.99		& $268^{+ 100 }_{- 42 }$		& [17$^{+ 10 }_{- 1 }$,32$^{+ 2 }_{- 1 }$] \\

 \hline
\end{tabular}
\end{table*}

\begin{table}
\tiny
\caption{Exoplanets with the widest ranges of $\alpha_{obs}$ at $\lambda$=575 nm for each of the CGI configurations.}
\label{table:results_unknownorbits_largealphaobs}
\centering 
\begin{tabular}{c l c c c}
\hline \hline
 & Planet      & $t_{obs}$		& $\alpha_{obs}$    & $\Delta \alpha_{obs}$ \\
 &             & [days]        & [deg]             & [deg]        \\ 
\hline
\multirow{4}{*}{\rotatebox{90}{Optimist.}} & HD 219134 h 		 & 917$^{+ 152 }_{- 159 }$ 		 & [30$^{+ 25 }_{- 11 }$,123$^{+ 2 }_{- 2 }$] 		 & 94$^{+ 11 }_{- 27 }$ \\ 
 & 47 UMa c 		 & 1288$^{+ 539 }_{- 227 }$ 		 & [37$^{+ 20 }_{- 3 }$,116$^{+ 3 }_{- 3 }$] 		 & 79$^{+ 5 }_{- 21 }$ \\ 
 & HD 190360 b 		 & 1371$^{+ 845 }_{- 529 }$ 		 & [44$^{+ 14 }_{- 14 }$,119$^{+ 3 }_{- 9 }$] 		 & 75$^{+ 15 }_{- 20 }$ \\ 
 & psi 1 Dra B b 		 & 1530$^{+ 960 }_{- 366 }$ 		 & [34$^{+ 24 }_{- 2 }$,109$^{+ 3 }_{- 5 }$] 		 & 73$^{+ 6 }_{- 22 }$ \\

\hline
\multirow{4}{*}{\rotatebox{90}{Interm.}} & 47 UMa c 		 & 599$^{+ 303 }_{- 99 }$ 		 & [43$^{+ 12 }_{- 3 }$,85$^{+ 4 }_{- 4 }$] 		 & 41$^{+ 7 }_{- 12 }$ \\ 
 & HD 219134 h 		 & 577$^{+ 118 }_{- 89 }$ 		 & [27$^{+ 23 }_{- 6 }$,68$^{+ 5 }_{- 3 }$] 		 & 41$^{+ 7 }_{- 22 }$ \\ 
 & HD 190360 b 		 & 232$^{+ 159 }_{- 71 }$ 		 & [49$^{+ 13 }_{- 12 }$,89$^{+ 5 }_{- 20 }$] 		 & 36$^{+ 15 }_{- 24 }$ \\ 
 & HD 62509 b 		 & 100$^{+ 62 }_{- 54 }$ 		 & [75$^{+ 7 }_{- 5 }$,105$^{+ 5 }_{- 7 }$] 		 & 30$^{+ 9 }_{- 13 }$ \\ 
 
\hline
\multirow{3}{*}{\rotatebox{90}{Pessim.}} & HD 219134 h 		 & 444$^{+ 143 }_{- 68 }$ 		 & [27$^{+ 19 }_{- 2 }$,61$^{+ 3 }_{- 3 }$] 		 & 33$^{+ 4 }_{- 18 }$ \\ 
 & 47 UMa c 		 & 87$^{+ 45 }_{- 33 }$ 		 & [50$^{+ 4 }_{- 3 }$,62$^{+ 6 }_{- 6 }$] 		 & 10$^{+ 7 }_{- 4 }$ \\ 
 & eps Eri b 		 & 172$^{+ 67 }_{- 28 }$ 		 & [11$^{+ 7 }_{- 1 }$,21$^{+ 1 }_{- 1 }$] 		 & 10$^{+ 1 }_{- 7 }$ \\ 
\hline

\end{tabular}
\end{table}

\begin{figure}
   \includegraphics[width=9.cm]{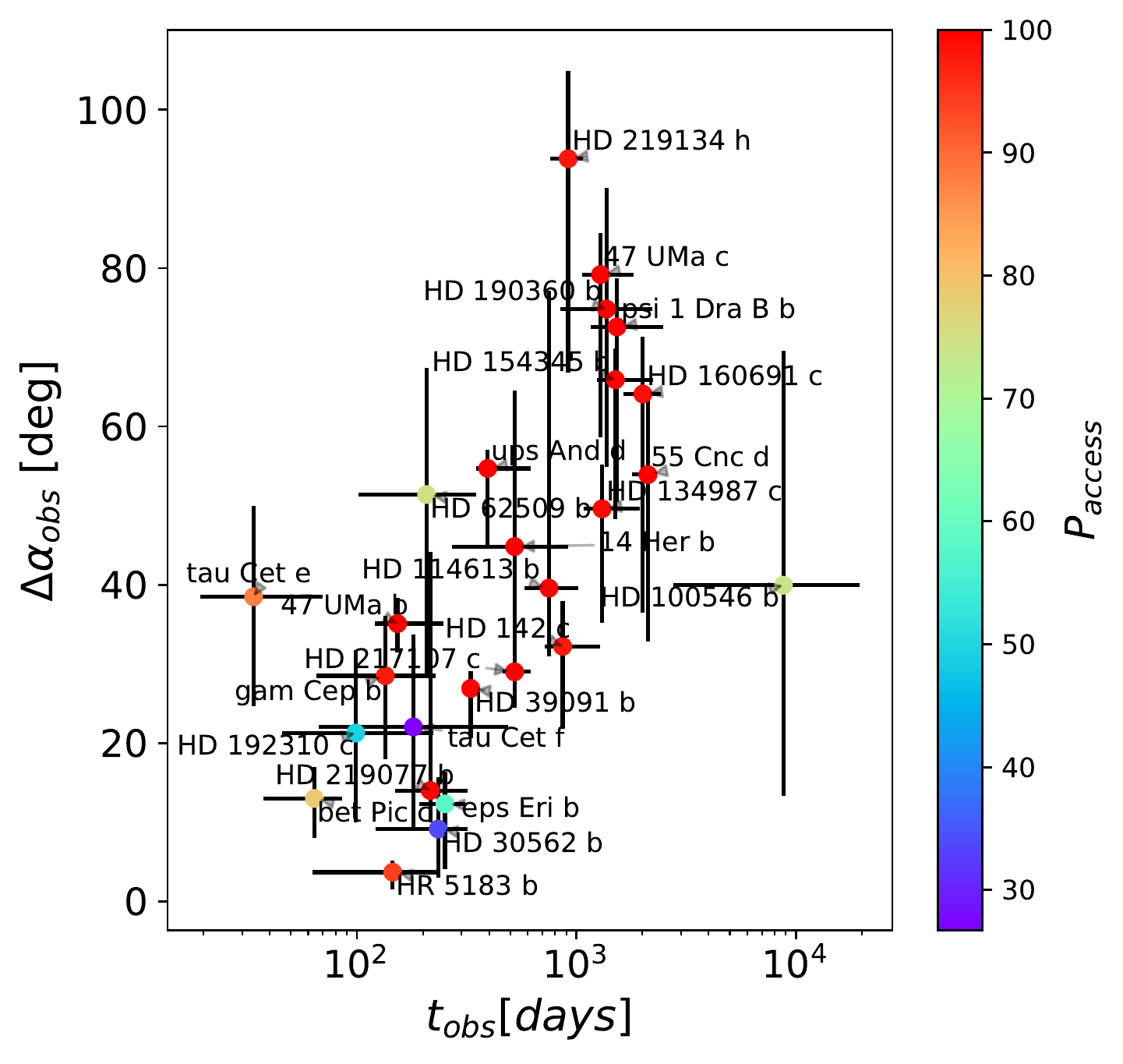}
   \caption{
   \label{fig:delta_alphaobs_VS_timeobs} 
   Range of observable phase angles against the time the planet is accessible per orbit at $\lambda$=575 nm in the optimistic CGI configuration.
   The colour of the markers indicates the $P_{access}$ of the exoplanet.
   Horizontal and vertical errorbars correspond to the upper and lower uncertainties of $t_{obs}$ and $\Delta \alpha_{obs}$, respectively.
   }
\end{figure}

We define the interval of observable phase angles as $\Delta \alpha_{obs} = \alpha_{obs(max)} - \alpha_{obs(min)}$ and compute the corresponding upper and lower uncertainties.
Table \ref{table:results_unknownorbits_largealphaobs} shows the planets with the largest $\Delta \alpha_{obs}$ at our reference $\lambda$=575 nm, which a priori might become prime targets for phase-curve measurements in each CGI scenario. 
Figure \ref{fig:delta_alphaobs_VS_timeobs} shows, for the optimistic CGI configuration, the computed ranges of 
$\Delta \alpha_{obs}$ for each exoplanet against the total time they are observable, $t_{obs}$.
This information is potentially relevant to find optimal targets for phase-curve measurements.
For instance, HD 219134 h shows a large variation of $\alpha$ in the optimistic configuration ($\Delta \alpha_{obs}$=94$^{+ 11 }_{- 27 }$) taking place in a detectability window of 2.5 years ($t_{obs}=$917$^{+ 152 }_{- 159 }$ days), the shortest value of $t_{obs}$ among the planets of Table \ref{table:results_unknownorbits_largealphaobs}.
Furthermore, this planet has particularly large intervals of $\alpha_{obs}$ in the intermediate and pessimistic scenarios ($\Delta \alpha_{obs}$=41$^{+ 7 }_{- 22 }$ and 33$^{+ 4 }_{- 18 }$ deg, respectively).

\subsubsection{Multiplanetary systems}   \label{subsubsec:results_unknown_MultiplanetarySystems}

\begin{table*}
\caption{Multiplanetary systems that are Roman-accessible in each CGI configuration.
We also quote under the "Technique" column the observing techniques with which the planets have been detected and the number of planets detected with each of these techniques. For each CGI configuration, only those exoplanets with $P_{access}$>25\% are shown.}
\label{table:results_multiplanet}
\centering 
\begin{tabular}{l c c c c c}
\hline \hline
System  	& Planets	&  Techniques  & Roman-access. &  Roman-access. &  Roman-access.            \\ 
        	& Total	&                  &  (Optimistic) &  (Intermediate) & (Pessimistic)            \\ 
\hline                                                                                                                                     
47 UMa 	 & 3 	& RV (3)   & 2 (c: 100.00\%; b: 100.00\%) 	 &    1 (c: 97.95\%)     &      1 (c: 82.84\%)      \\ 
tau Cet 	 & 4 	& RV (4)  & 2 (e: 87.75\%; f: 26.74\%) 	 &    0                  &      0                   \\ 
pi Men 	 & 2 	& RV (2); Transit (1)  & 1 (b: 100.00\%) 	             &    1 (b: 53.24\%)     &      0                   \\ 
55 Cnc 	 & 5 	& RV (5); Transit (1) & 1 (d: 100.00\%) 	             &    0                  &      0                   \\ 
ups And 	 & 3 	& RV (3) & 1 (d: 100.00\%) 	             &    1 (d: 100.00\%)    &      0                   \\ 
HD 217107 	 & 2 	& RV (2) & 1 (c: 100.00\%) 	             &    0                  &      0                   \\ 
HD 190360 	 & 2 	& RV (2) & 1 (b: 100.00\%) 	             &    1 (b: 92.97\%)     &      0                   \\ 
HD 134987 	 & 2 	& RV (2) & 1 (c: 99.26\%) 	             &    0                  &      0                    \\ 
HD 160691 	 & 4 	& RV (4) & 1 (c: 98.84\%) 	             &    0                  &      0                    \\ HD 219134 	 & 6 	& RV (6); Transit (2) & 1 (h: 97.93\%) 	             &    1 (h: 92.15\%)     &      1 (h: 86.59\%)       \\ 
HD 142 	 & 2 	& RV (2) & 1 (c: 97.55\%) 	             &    0                  &      0                    \\ 
bet Pic 	 & 2 	& RV (1); Astrometry (1); Imaging (1)    & 1 (c: 78.81\%) 	             &    0                  &      0                    \\ 
HD 192310 	 & 2 	& RV (2) & 1 (c: 49.36\%) 	             &    0                  &      0                   \\ 
\hline
\end{tabular}
\end{table*}

Among the optimistic 26 Roman-accessible exoplanets, 13 of them are part of stellar systems with other confirmed planetary companions.
Table \ref{table:results_multiplanet} lists these multiplanetary systems, with the number of exoplanets that they host as well as the number of them that are Roman-accessible in each CGI scenario.
Three of these exoplanets are also among those with a larger $\Delta \alpha_{obs}$ in Table \ref{table:results_unknownorbits_largealphaobs}: HD 219134 h, 47 UMa c and HD 190360 b.

We find that, in the optimistic CGI scenario, the systems 47 UMa and tau Cet have more than one Roman-accessible exoplanet.
In the case of 47 UMa, planets $b$ and $c$ are accessible with $P_{access}$=100\%.
We note that 47 UMa $d$ also has a marginal $P_{access}$=9.41\% in this scenario (Table \ref{table:output_catalogue}).
The system tau Cet stands out because planets $e$ and $f$ ($M_p\, sin\,i\sim 4M_\oplus$) are Roman-accessible ($P_{access}$=87.75 and 26.74\%, resp.).
In Sect. \ref{subsec:discussion_tauCet} we discuss more thoroughly the prospects to observe tau Cet e and f.

Table \ref{table:results_multiplanet} also shows three systems for which a transiting, inner exoplanet is known to exist. This offers the possibility of studying both the outer planet in direct imaging and the inner planet with transmission spectroscopy.
Such scenarios are potentially valuable to gain insight into the system as a whole, and the processes that may have led to the final arrangements. 
In the optimistic scenario, this is the case of 55 Cnc d, with the transiting ultra-short-period planet e, pi Men b, with a transiting super-Earth (planet c) and HD 219134 h, with two transiting super-Earths (b and c). 
These systems will be discussed in more detail in Sect. \ref{subsec:discussion_outercompanions}.

\subsection{Equilibrium temperatures of the Roman-accessible planets} \label{subsec:results_Teq}
In order to facilitate future atmospheric modeling of the Roman-accessible exoplanets, we computed their $T_{eq}$ at each orbital position by means of Eq. (\ref{eq:Teq}).
In our output catalogue (Table \ref{table:output_catalogue}) we quote the range of $T_{eq}$, and the corresponding uncertainties, computed for each planet in the 10000 orbital realizations (whether detectable or not).
In addition, for some planets we report under $T_{eq(obs)}$ the range of equilibrium temperatures that correspond only to those orbital positions that are Roman-accessible (Table \ref{table:results_largeTeqobs}).
This provides a first estimate of the possible variations that the planetary atmosphere might undergo during the time that it remains accessible.

\begin{figure*}
	\centering
	\includegraphics[width=3cm]{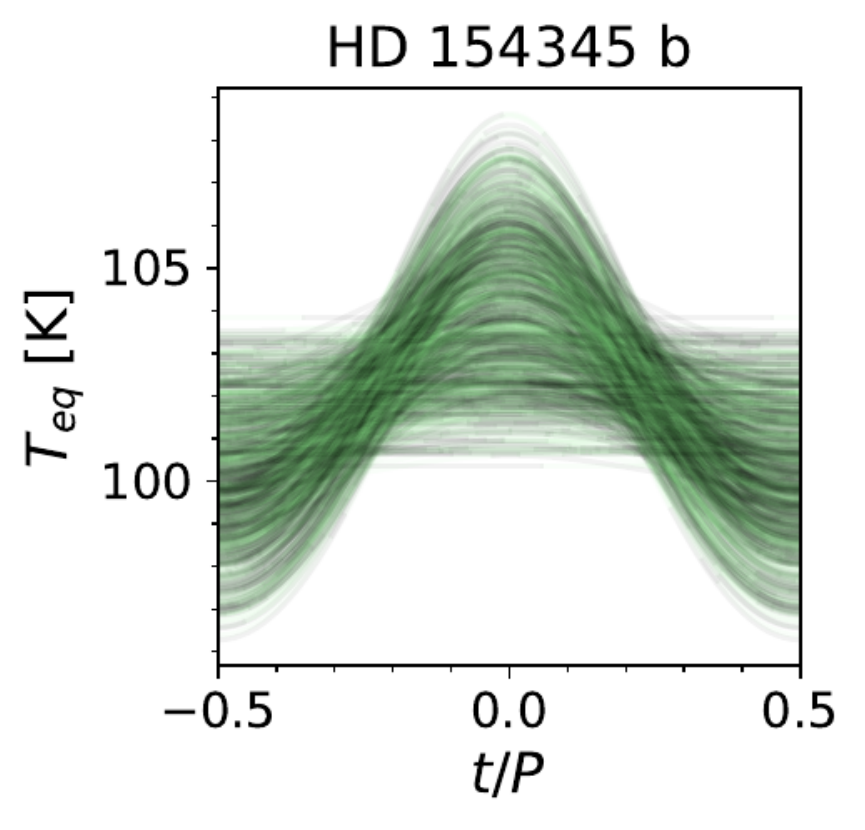} 
	\hspace{-0.4cm}
	\includegraphics[width=3cm]{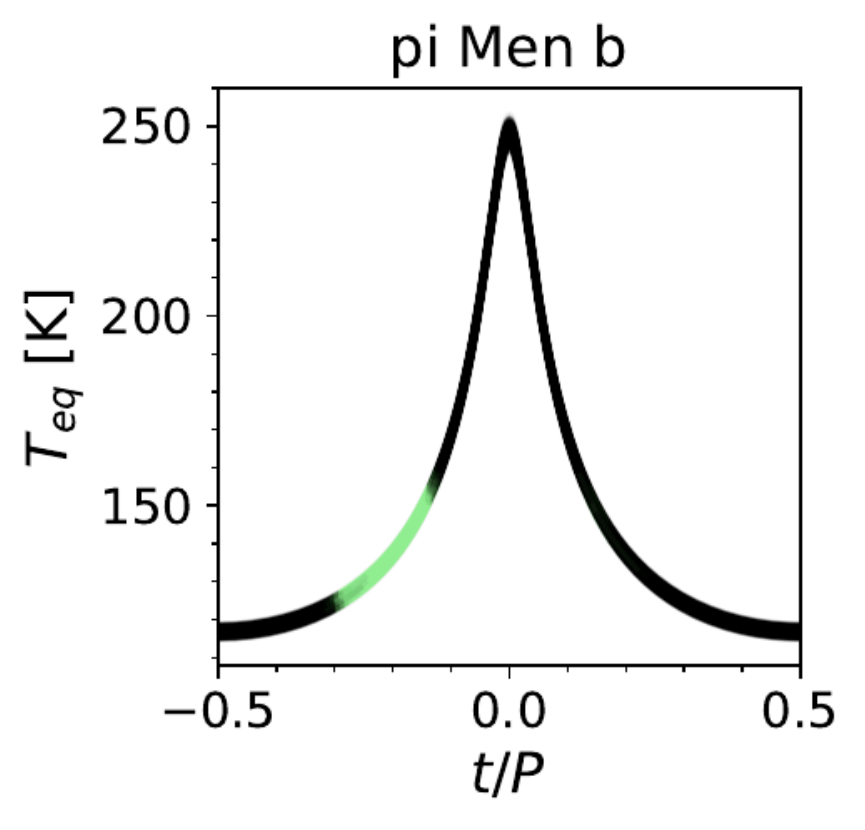} 
	\hspace{-0.4cm}
	\includegraphics[width=3cm]{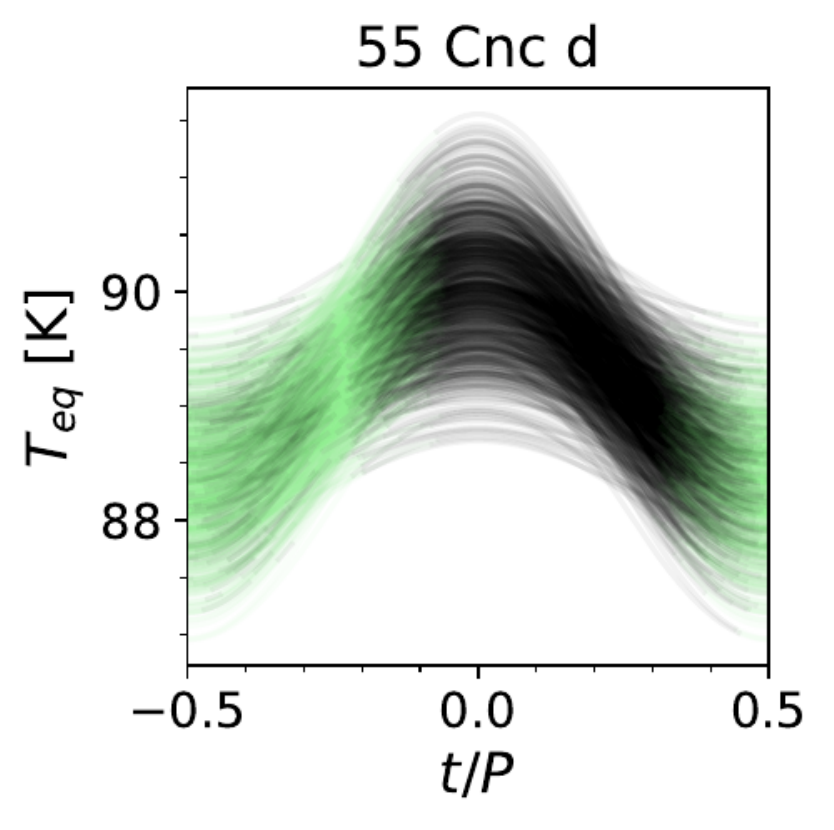}
	\hspace{-0.4cm}
	\includegraphics[width=3cm]{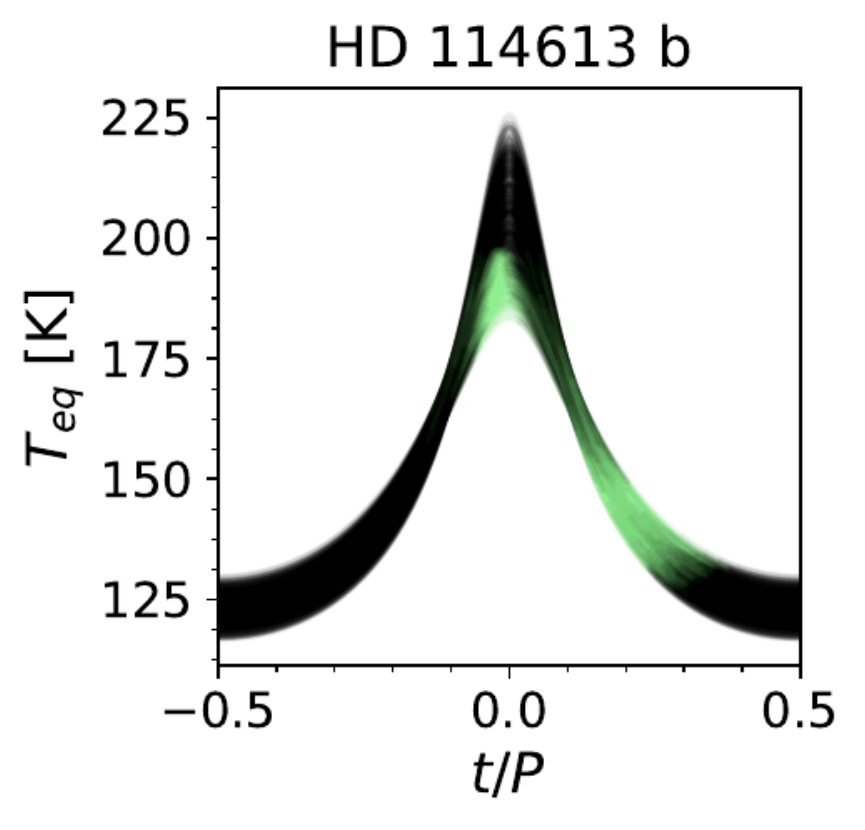} 
	\hspace{-0.4cm}
	\includegraphics[width=3cm]{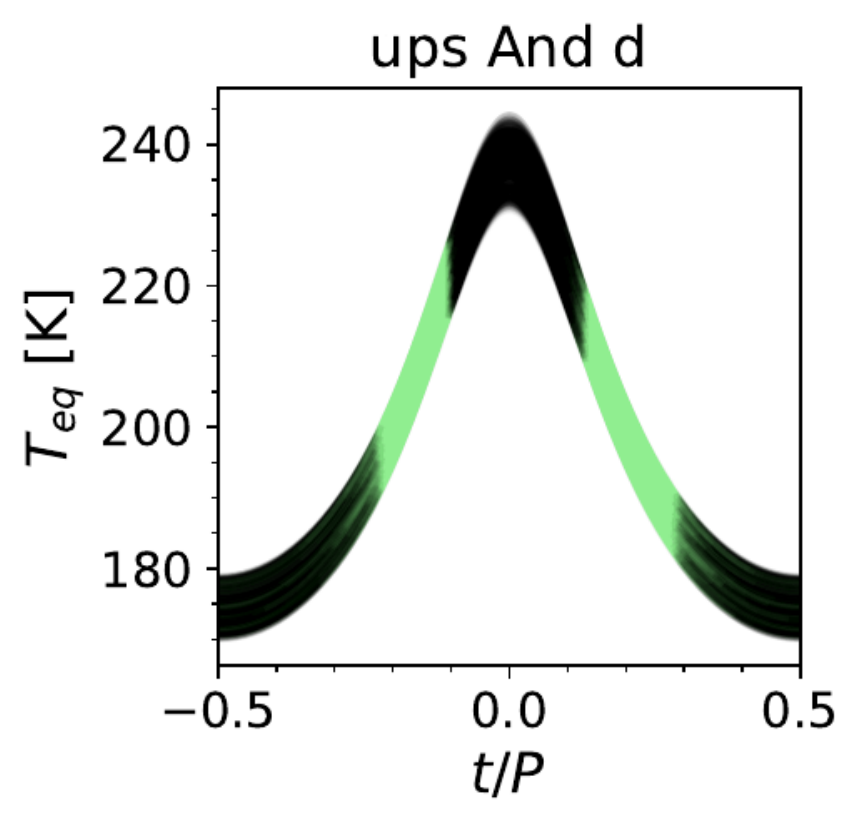} 
	\hspace{-0.4cm}
	\includegraphics[width=3cm]{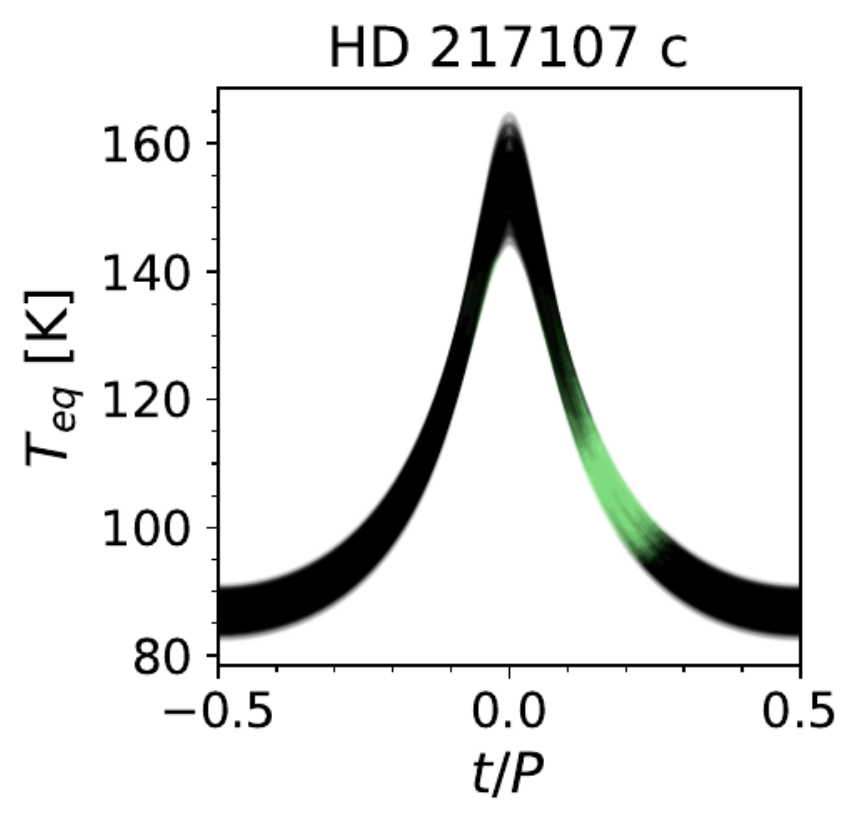} 
	\vspace{-0.1cm}
	   \\
	\includegraphics[width=3cm]{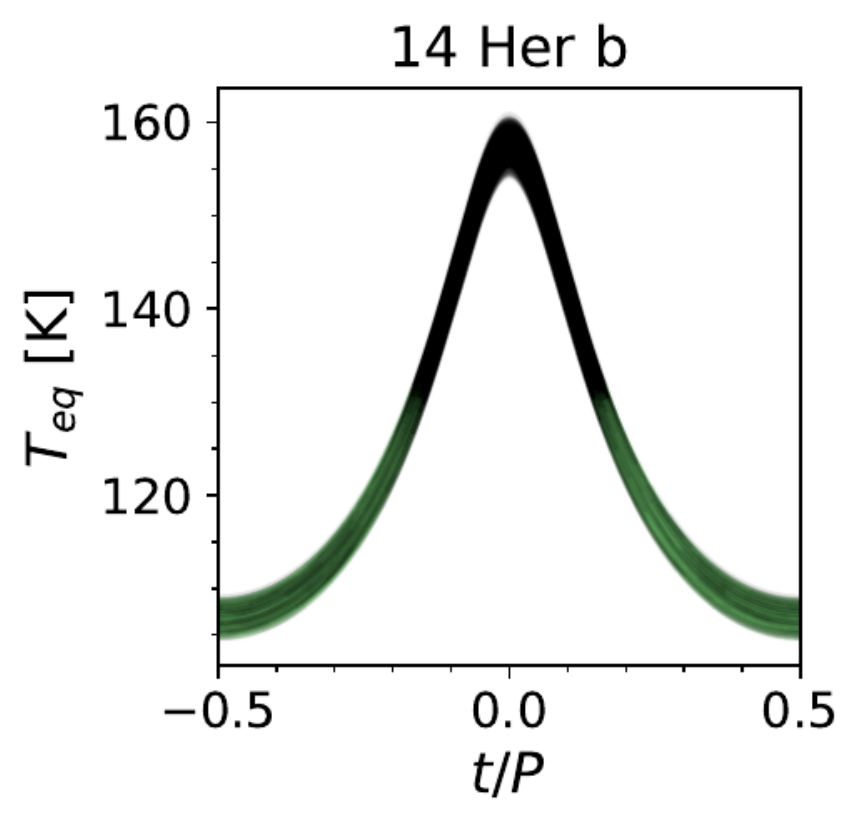} 
	\hspace{-0.4cm}
	\includegraphics[width=3cm]{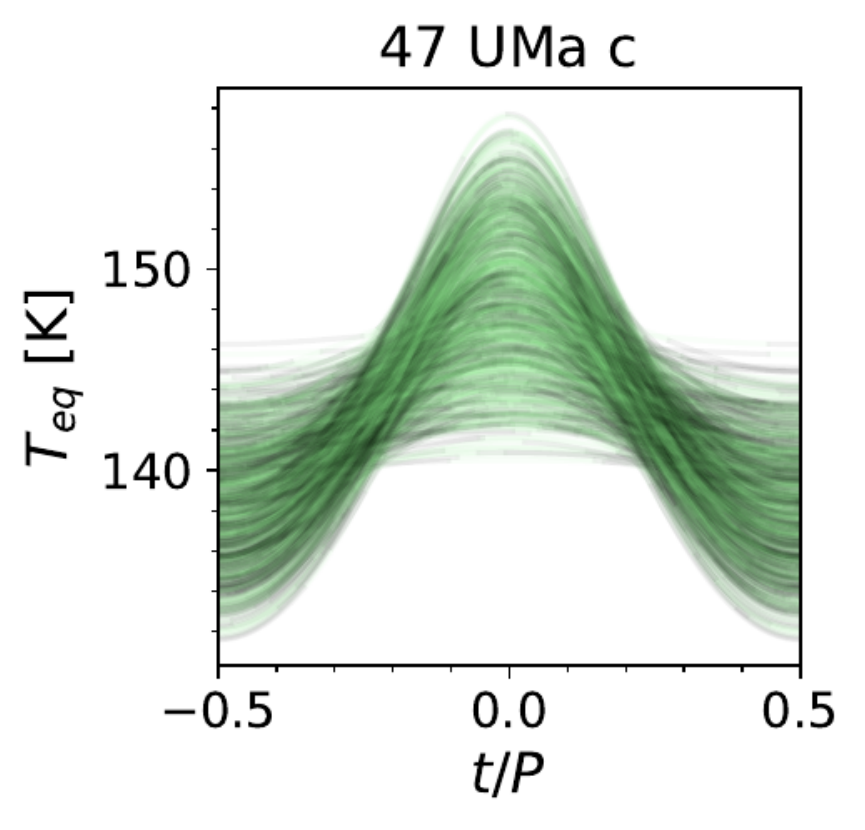} 
	\hspace{-0.4cm}
	\includegraphics[width=3cm]{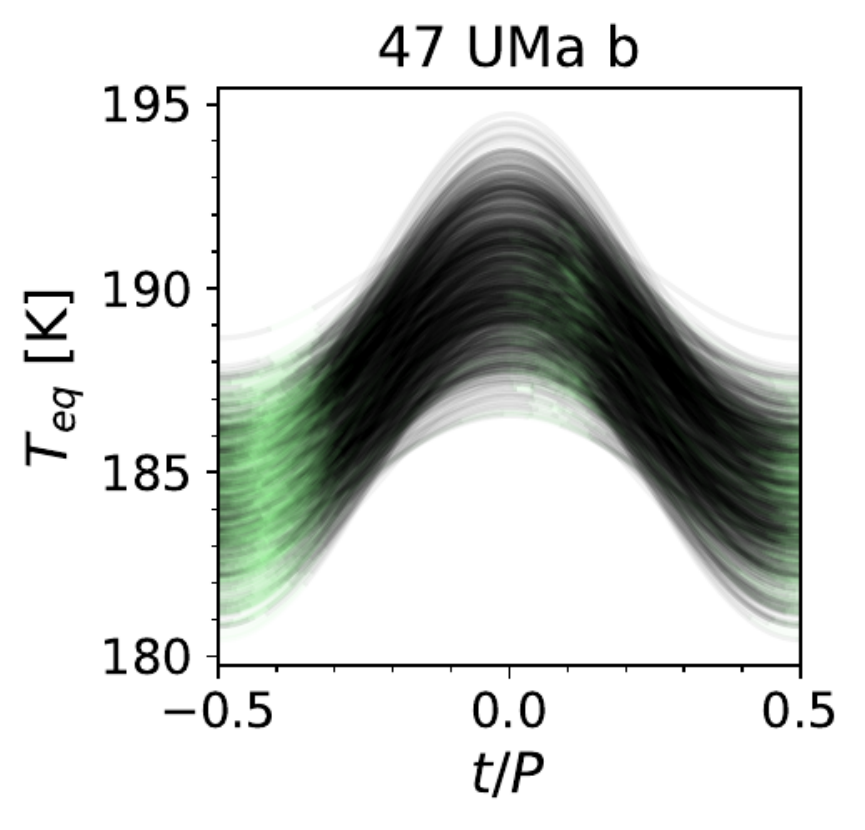} 
	\hspace{-0.4cm}
	\includegraphics[width=3cm]{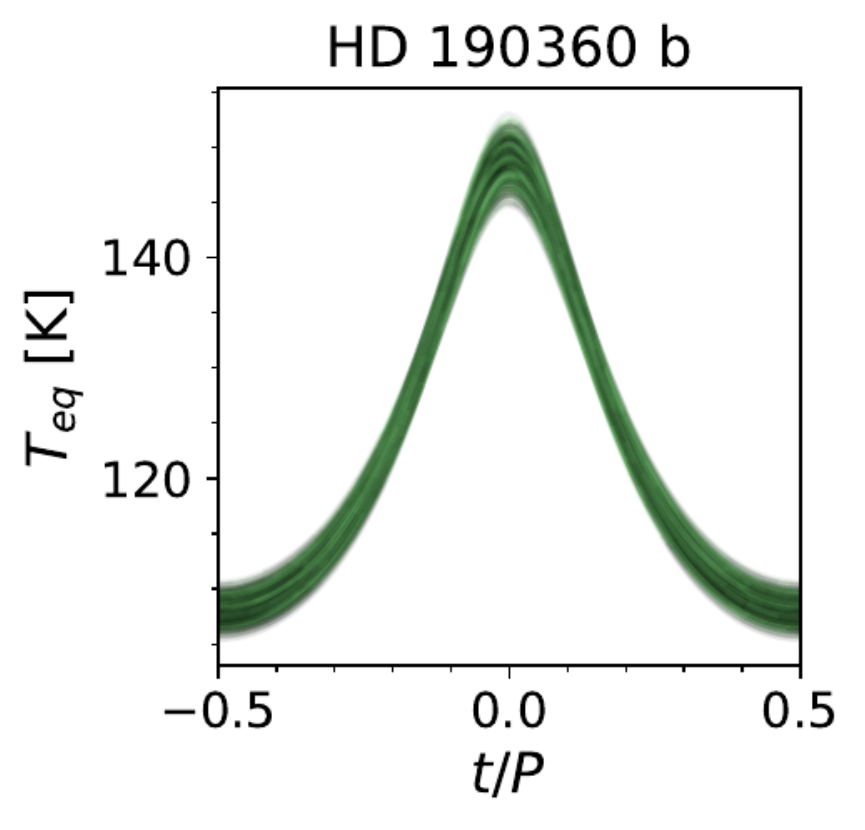}
	\hspace{-0.4cm}
	\includegraphics[width=3cm]{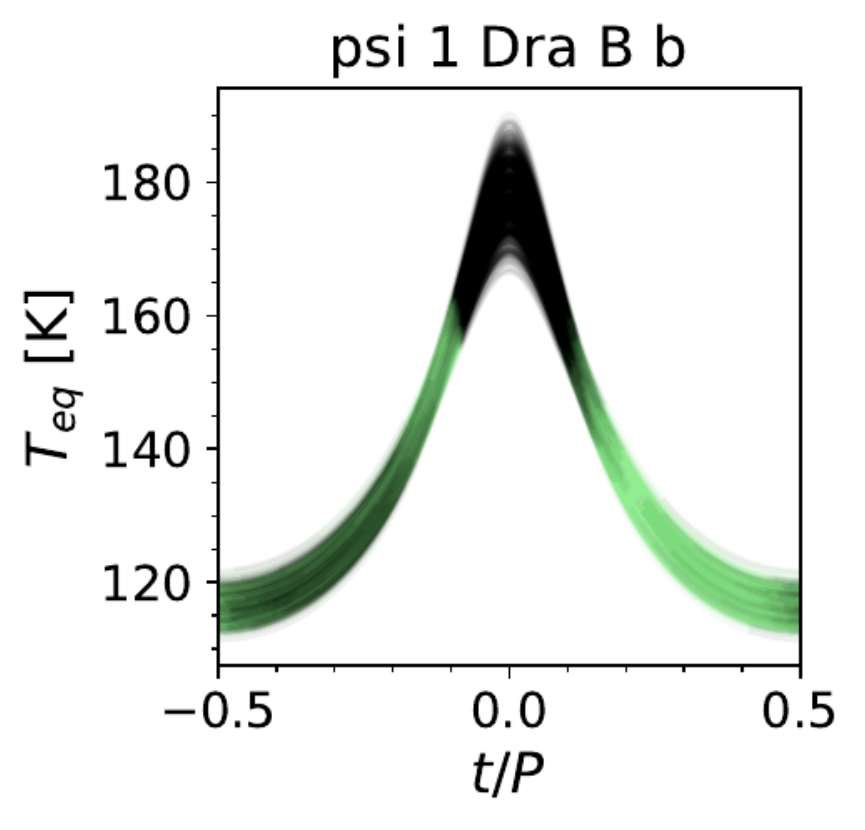} 
	\hspace{-0.4cm}
	\includegraphics[width=3cm]{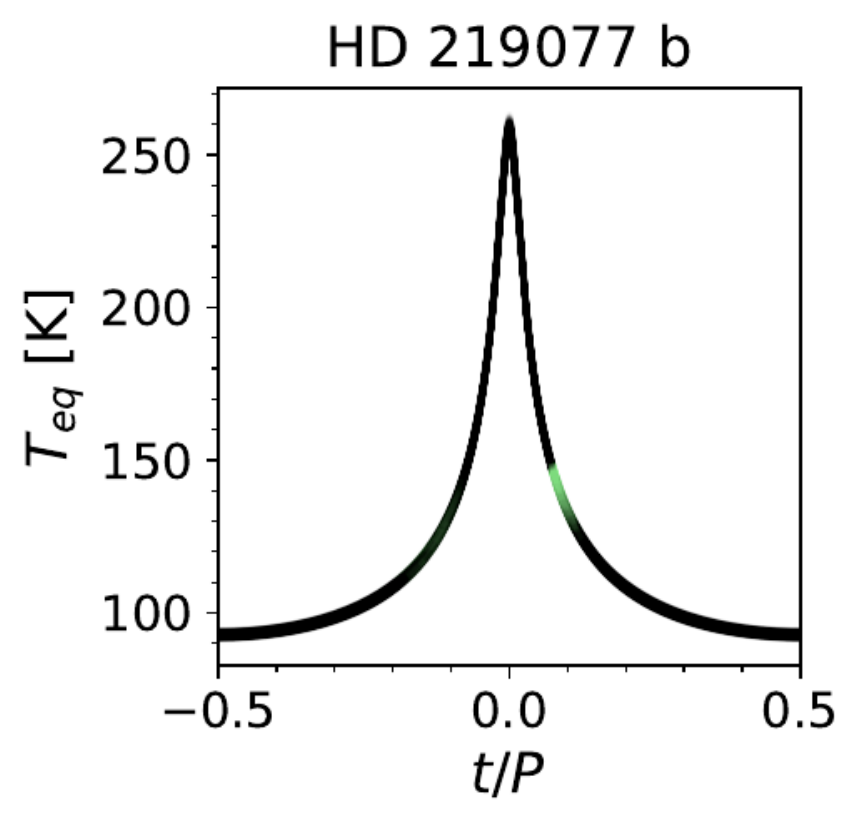} 
	\vspace{-0.1cm}
	   \\
	\includegraphics[width=3cm]{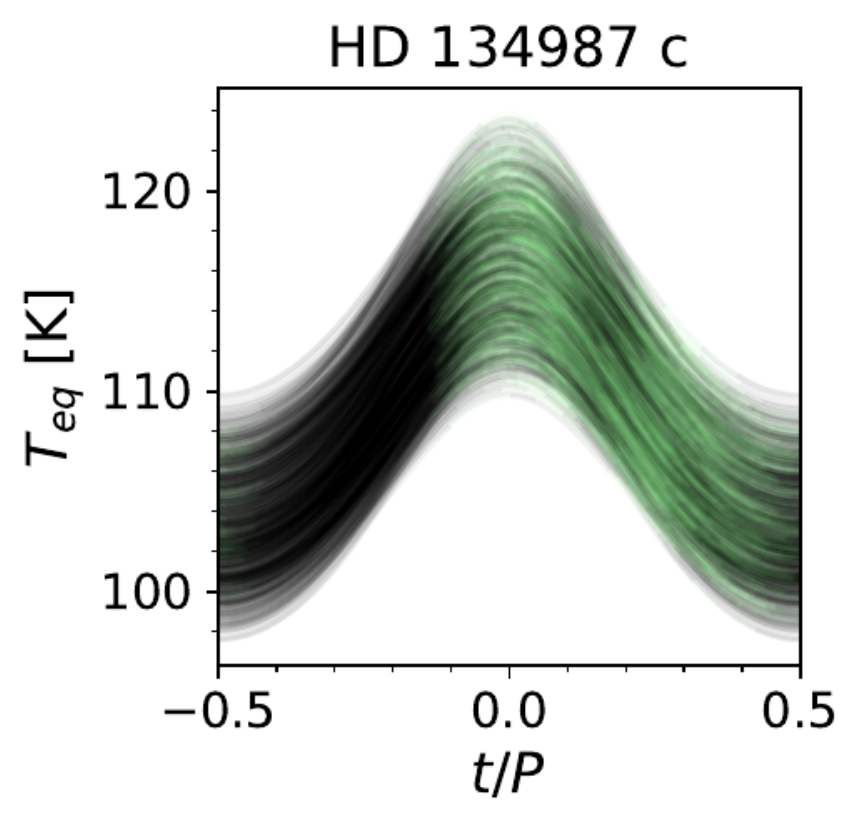} 
	\hspace{-0.4cm}
	\includegraphics[width=3cm]{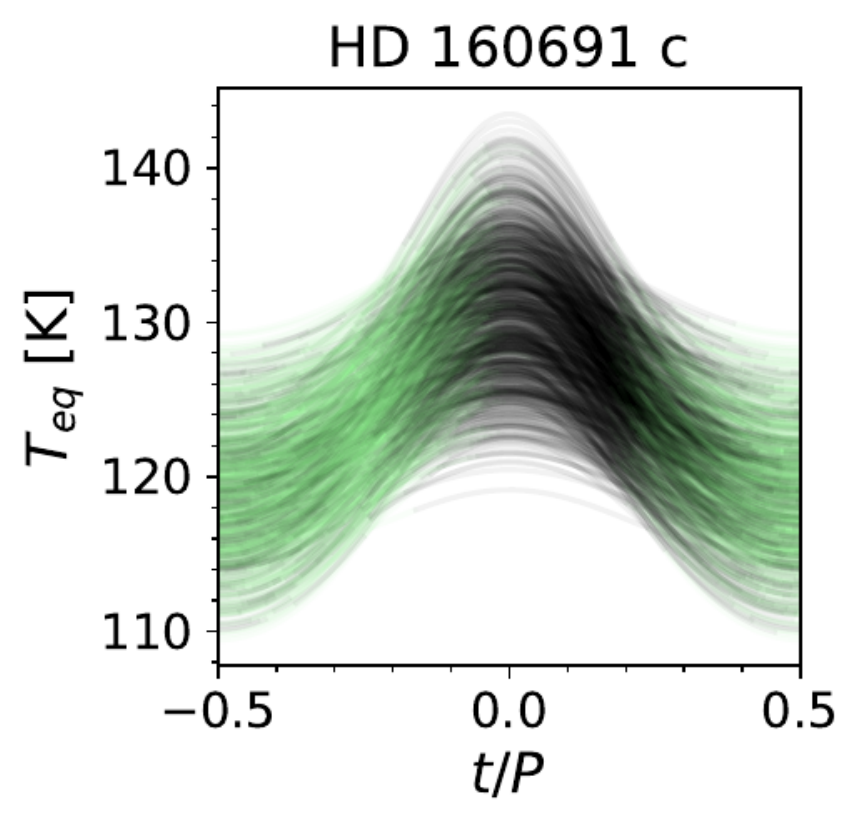} 
	\hspace{-0.4cm}
	\includegraphics[width=3cm]{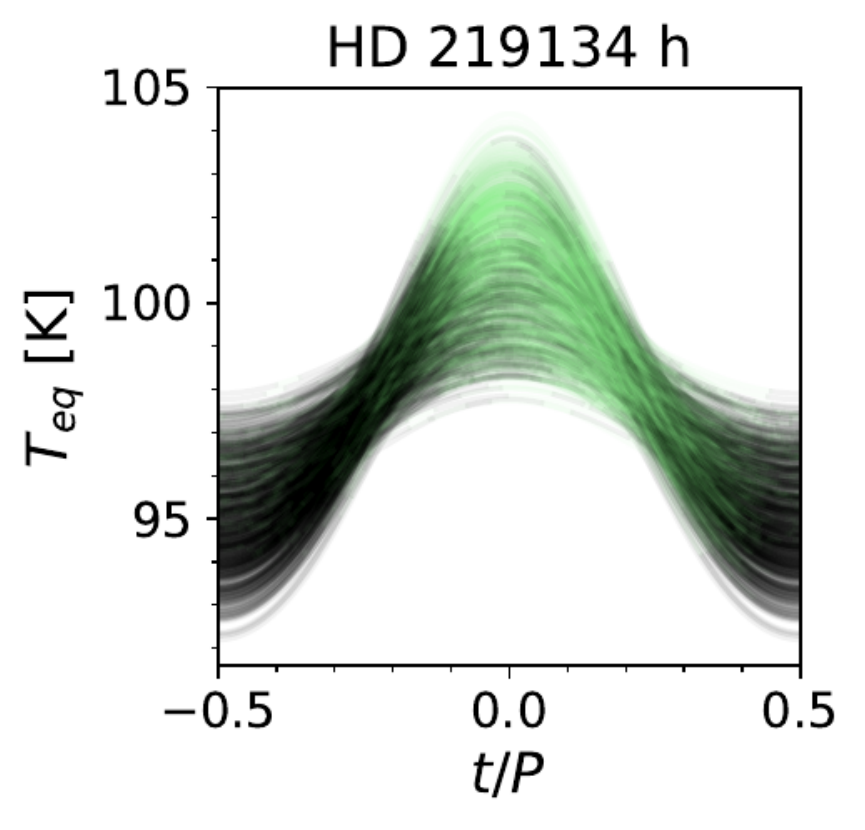} 
	\hspace{-0.4cm}
	\includegraphics[width=3cm]{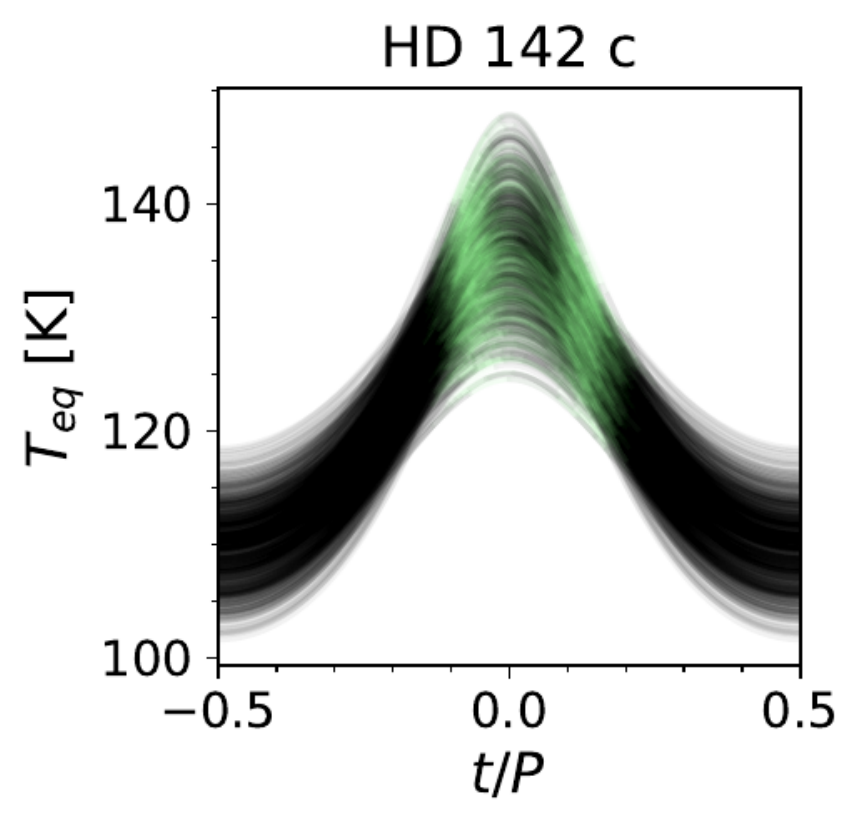} 
	\hspace{-0.4cm}
	\includegraphics[width=3cm]{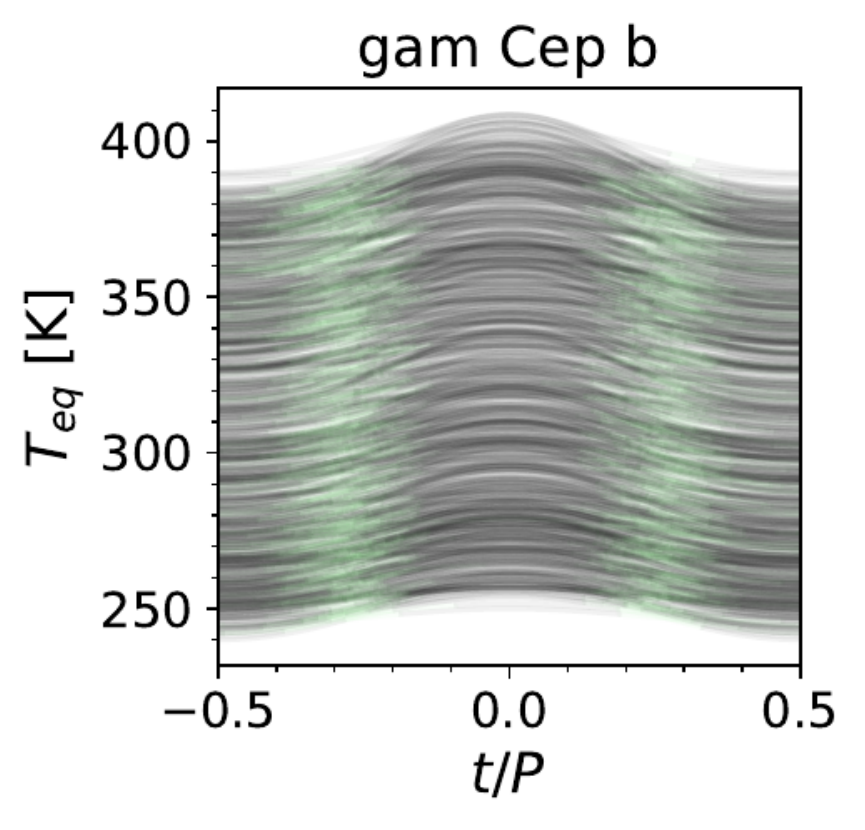} 
	\hspace{-0.4cm}
	\includegraphics[width=3cm]{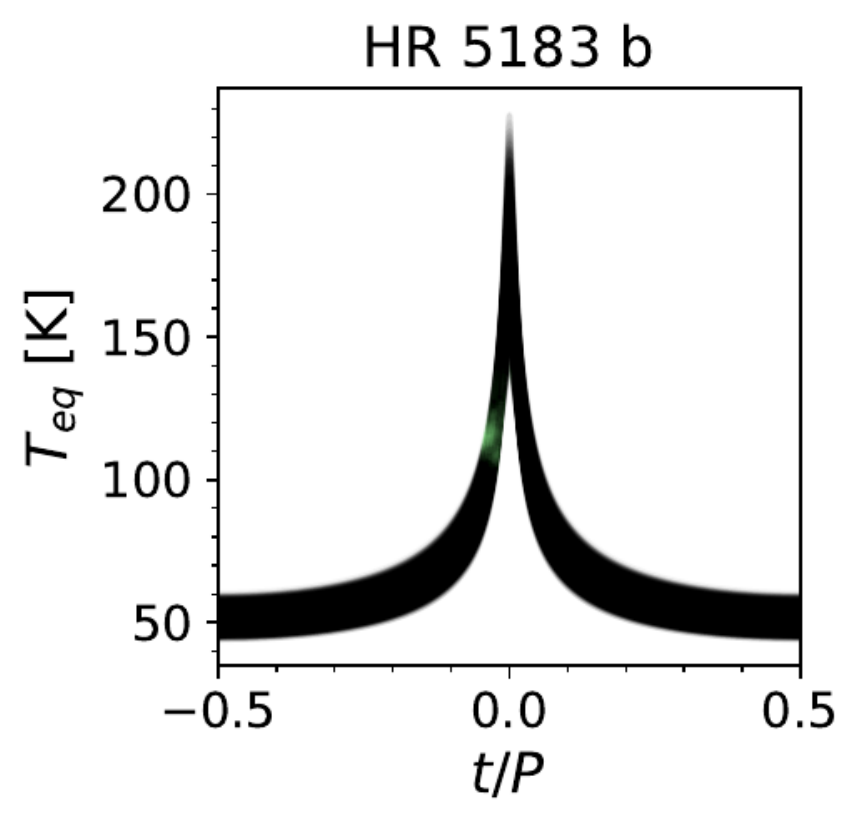} 
	\vspace{-0.1cm}
	   \\
	\includegraphics[width=3cm]{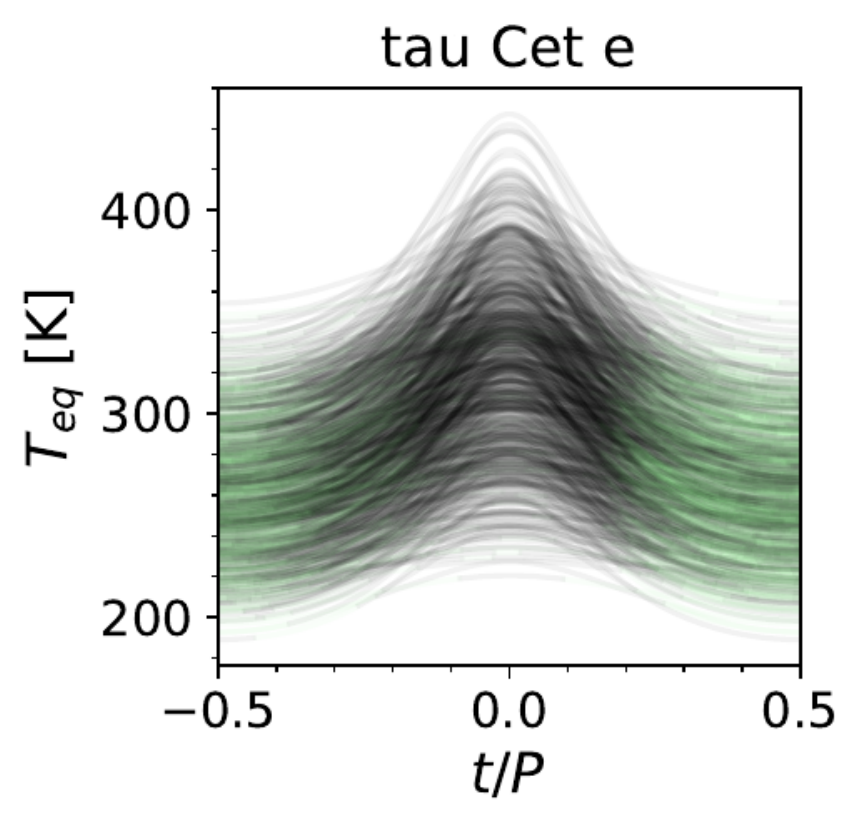} 
	\hspace{-0.4cm}
	\includegraphics[width=3cm]{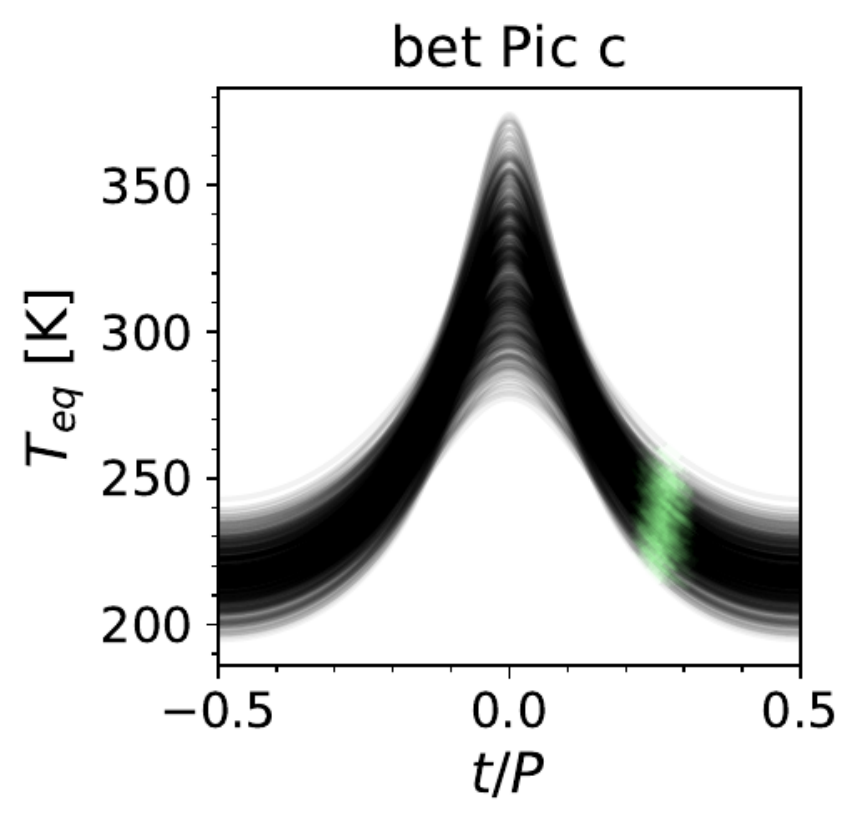} 
	\hspace{-0.4cm}
	\includegraphics[width=3cm]{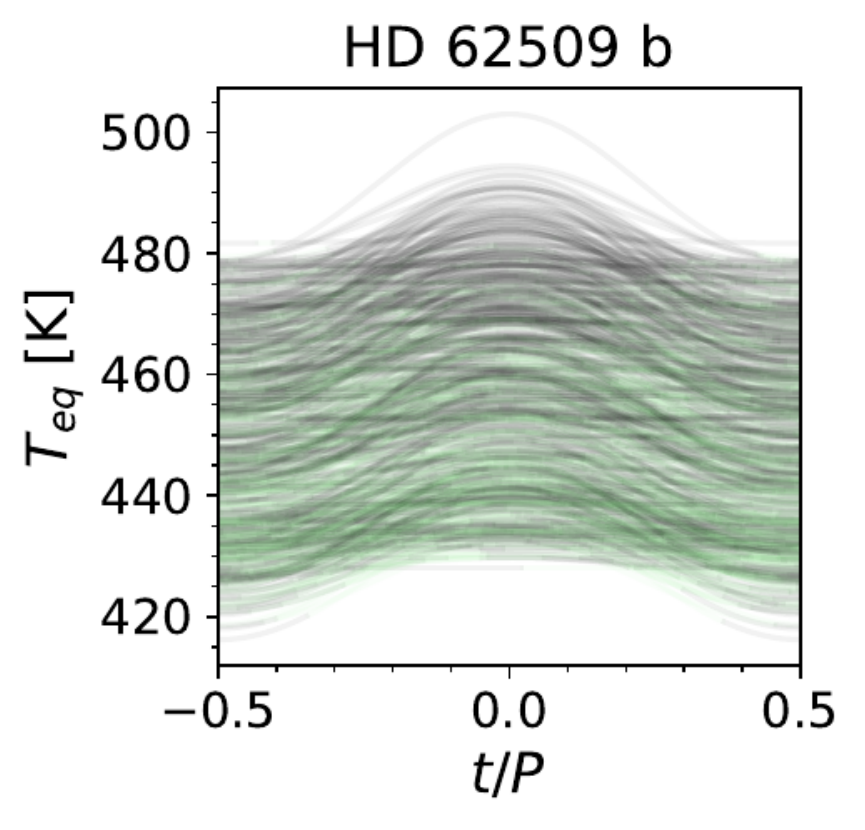} 
	\hspace{-0.4cm}
	\includegraphics[width=3cm]{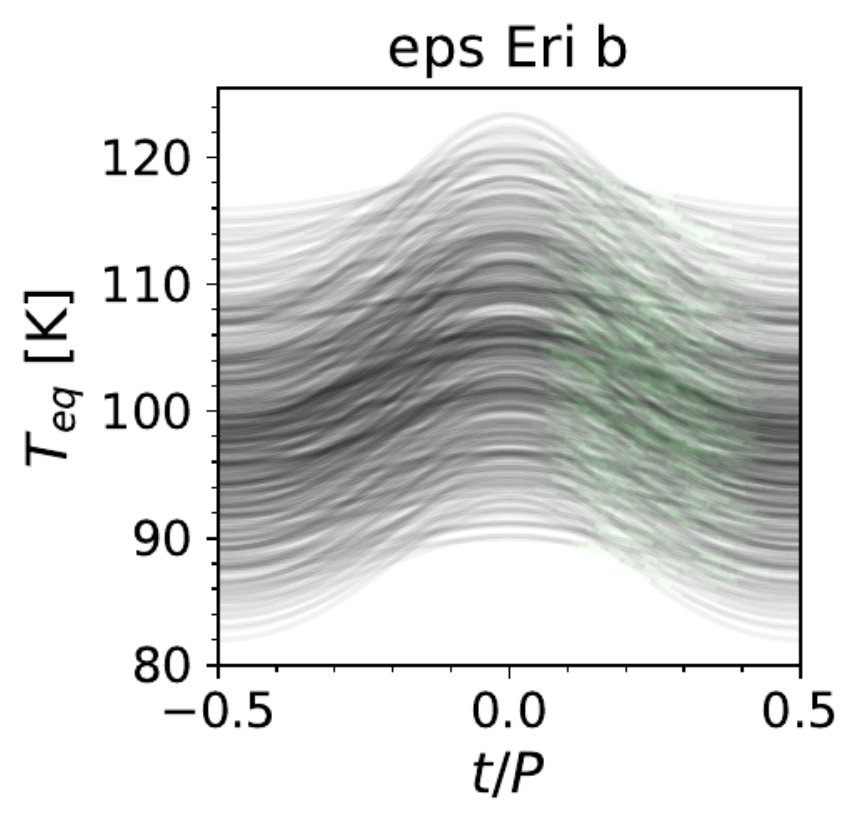} 
	\hspace{-0.4cm}
	\includegraphics[width=3cm]{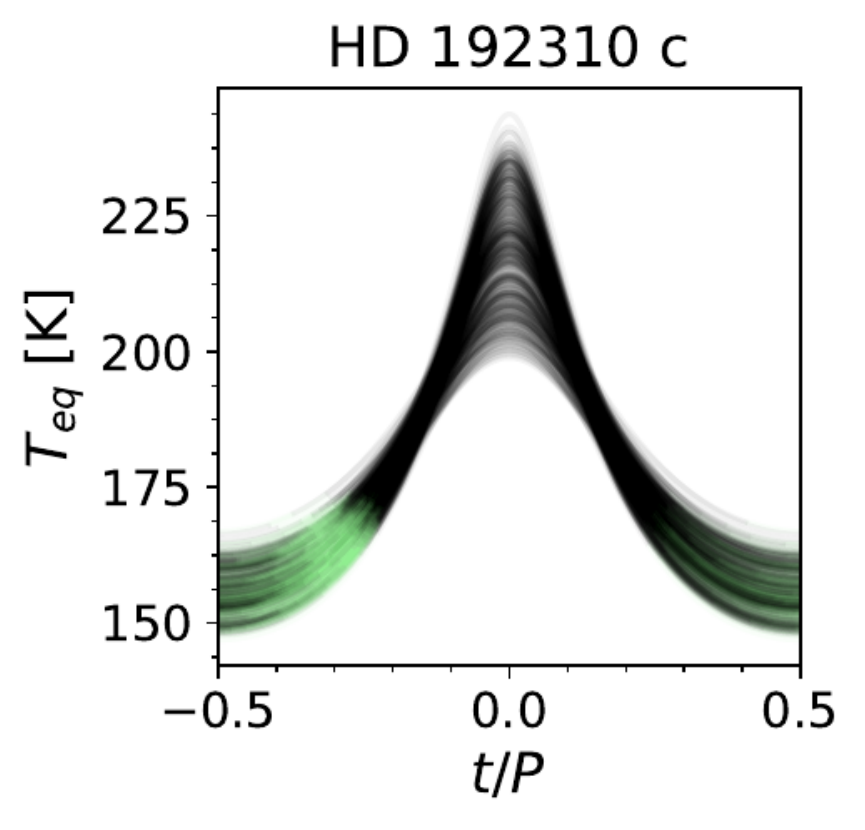} 
	\hspace{-0.4cm}
	\includegraphics[width=3cm]{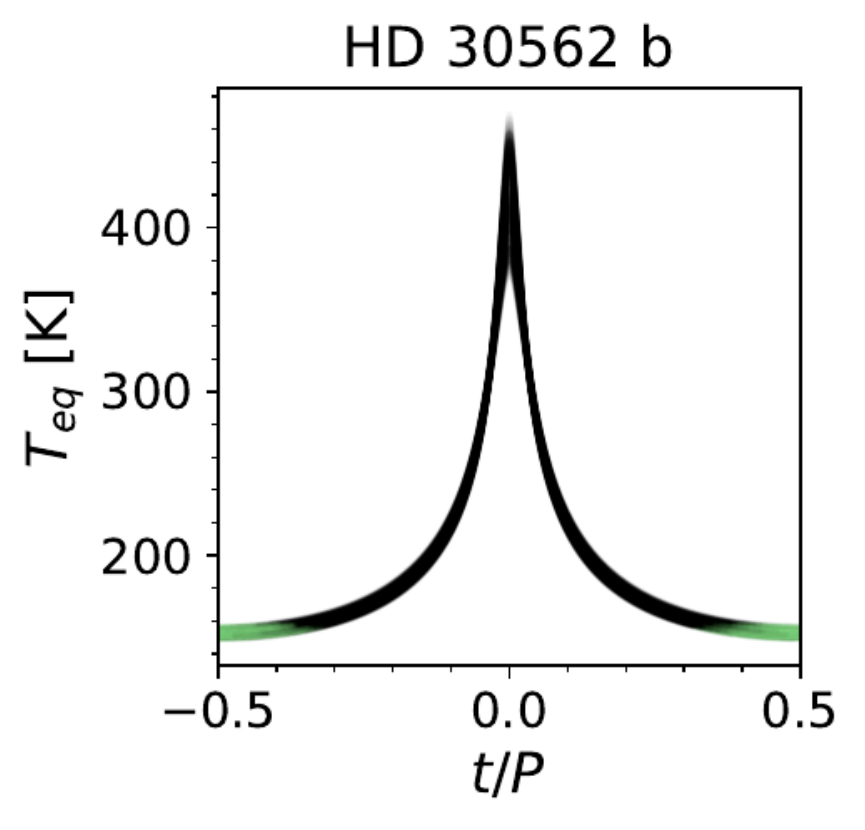} 
	\vspace{-0.1cm}
	   \\
	\includegraphics[width=3cm]{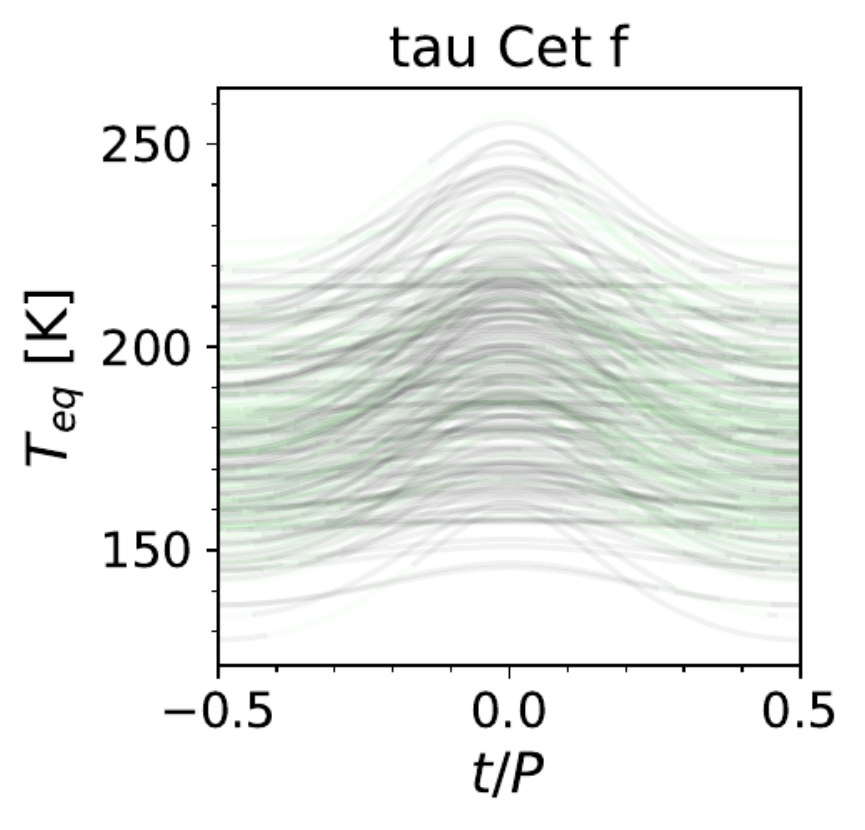} 
	\hspace{-0.4cm}
	\caption{
	\label{fig:results_Teqobs_VS_time}
	Evolution of $T_{eq}$ with time for the accessible orbits of the Roman-accessible exoplanets with a constrained value of $e$ (Table \ref{table:NASA_database}). Green colour indicates the orbital positions which are accessible in the optimistic CGI scenario. For the sake of clarity, only 1 of each 10 orbital realizations are shown.}
\end{figure*}

Figure \ref{fig:results_Teqobs_VS_time} shows the evolution of $T_{eq}$ with time for the accessible orbits of those planets that have an estimate of $e$ in the NASA Archive (all but HD 100546 b). Planets in eccentric orbits experience large changes of $T_{eq(obs)}$ and therefore are prime targets to search for atmospheric variability.
On the other hand, this would complicate an eventual atmospheric characterization by multiple-phase observations.

The planets with the largest changes in $T_{eq(obs)}$ ($\Delta T_{eq(obs)}$) for each CGI configuration are listed in Table \ref{table:results_largeTeqobs}.
In the optimistic scenario, ups And d and pi Men b experience changes of about 30 K during the time that they remain accessible, which is about a year in both cases.
HD 114613 b remains observable for about two years and we find that it undergoes a $\Delta T_{eq(obs)}=53^{+ 9 }_{- 28 }$ K.
Both psi 1 Dra B b and HD 190360 b have a $t_{obs}$ of about four years and, in this time, they show variations in $T_{eq}$ of about 40 K.
Such variations in $T_{eq}$ during the time that they are observable will likely trigger variability in the cloud coverage of their atmospheres \citep{sanchezlavegaetal2004}.
These five planets have $P_{access}=100\%$ and hence they appear as suitable targets to search for atmospheric variability with the Roman Telescope.
In more conservative CGI scenarios, however, the observable variability of $T_{eq}$ is significantly reduced.
In these cases, only HD 190360 b in the intermediate CGI scenario shows a noteworthy $\Delta T_{eq(obs)}$ (17$^{+ 13 }_{- 10 }$ K).
The rest of planets in the intermediate or pessimistic scenario have $\Delta T_{eq(obs)}$ smaller than 10 K, which is likely unable to trigger atmospheric variability during the time that they are observable.

\begin{table}
\tiny
\caption{Exoplanets with the widest ranges of $T_{eq(obs)}$ at $\lambda$=575 nm for each of the CGI configurations. Only planets with an estimate of $e$ in the NASA Archive were considered.}
\label{table:results_largeTeqobs}
\centering 
\begin{tabular}{c l c c c}
\hline \hline
 & Planet        & $t_{obs}$		& $T_{eq(obs)}$    & $\Delta T_{eq(obs)}$ \\
  &          & [days]        & [K]             & [K]        \\ 
\hline
\multirow{7}{*}{\rotatebox{90}{Optimistic}} & HD 114613 b 		 & 750$^{+ 271 }_{- 168 }$ 		 & [132$^{ +3 }_{ -3 }$,188$^{ +8 }_{ -31 }$] 		 & 53$^{+ 9 }_{- 28 }$ \\ 
 & psi 1 Dra B b 		 & 1530$^{+ 960 }_{- 366 }$ 		 & [117$^{ +2 }_{ -2 }$,158$^{ +3 }_{ -4 }$] 		 & 42$^{+ 2 }_{- 4 }$ \\ 
 & HD 190360 b 		 & 1371$^{+ 845 }_{- 529 }$ 		 & [109$^{ +7 }_{ -1 }$,148$^{ +2 }_{ -2 }$] 		 & 39$^{+ 3 }_{- 7 }$ \\ 
 & ups And d 		 & 394$^{+ 224 }_{- 44 }$ 		 & [183$^{ +4 }_{ -6 }$,221$^{ +4 }_{ -4 }$] 		 & 36$^{+ 8 }_{- 2 }$ \\ 
 & pi Men b 		 & 330$^{+ 32 }_{- 17 }$ 		 & [125$^{ +1 }_{ -1 }$,154$^{ +3 }_{ -2 }$] 		 & 29$^{+ 3 }_{- 2 }$ \\ 
 & HD 217107 c 		 & 522$^{+ 96 }_{- 60 }$ 		 & [98$^{ +3 }_{ -2 }$,118$^{ +20 }_{ -5 }$] 		 & 20$^{+ 14 }_{- 3 }$ \\ 
 & HD 219077 b 		 & 216$^{+ 103 }_{- 66 }$ 		 & [127$^{ +5 }_{ -13 }$,147$^{ +1 }_{ -1 }$] 		 & 20$^{+ 13 }_{- 4 }$ \\ 
\hline

\multirow{4}{*}{\rotatebox{90}{Interm.}}  & HD 190360 b 		 & 232$^{+ 159 }_{- 71 }$ 		 & [121$^{ +9 }_{ -8 }$,144$^{ +3 }_{ -16 }$] 		 & 17$^{+ 13 }_{- 10 }$ \\ 
 & ups And d 		 & 60$^{+ 20 }_{- 7 }$ 		 & [197$^{ +3 }_{ -3 }$,206$^{ +3 }_{ -3 }$] 		 & 9$^{+ 1 }_{- 1 }$ \\ 
 & 47 UMa c 		 & 599$^{+ 303 }_{- 99 }$ 		 & [141$^{ +3 }_{ -4 }$,148$^{ +4 }_{ -4 }$] 		 & 7$^{+ 7 }_{- 5 }$ \\ 
 & HD 219134 h 		 & 577$^{+ 118 }_{- 89 }$ 		 & [97$^{ +1 }_{ -2 }$,101$^{ +2 }_{ -1 }$] 		 & 3$^{+ 2 }_{- 2 }$ \\ 
\hline

\multirow{3}{*}{\rotatebox{90}{Pessim.}} & 47 UMa c 		 & 87$^{+ 45 }_{- 33 }$ 		 & [143$^{ +3 }_{ -3 }$,148$^{ +4 }_{ -3 }$] 		 & 3$^{+ 5 }_{- 2 }$ \\ 
 & HD 219134 h 		 & 444$^{+ 143 }_{- 68 }$ 		 & [98$^{ +2 }_{ -2 }$,101$^{ +2 }_{ -1 }$] 		 & 3$^{+ 2 }_{- 1 }$ \\ 
 & eps Eri b 		 & 172$^{+ 67 }_{- 28 }$ 		 & [101$^{ +7 }_{ -7 }$,103$^{ +7 }_{ -7 }$] 		 & 2$^{+ 1 }_{- 1 }$ \\ 

\hline
\end{tabular}
\end{table}

\begin{figure}
   \includegraphics[width=9.cm]{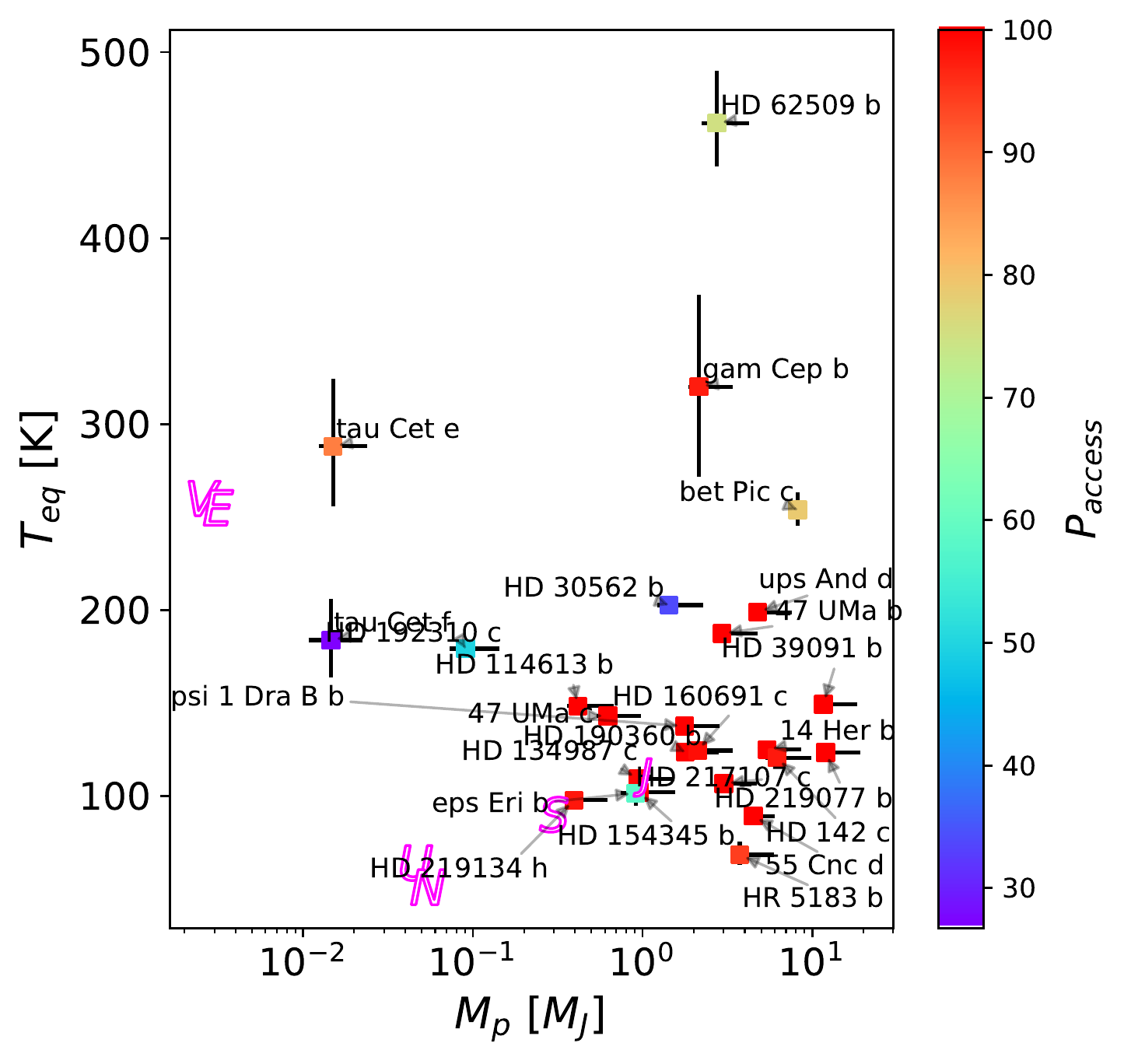}
   \caption{
   \label{fig:Mp_VS_Teq_output} 
   Median $T_{eq}$ against the median $M_p$ for each Roman-accessible planet in the optimistic CGI configuration, as computed in our 10000 orbital realizations.
   The colour of the markers indicates the $P_{access}$ of the exoplanet.
   Horizontal and vertical errorbars correspond to the upper and lower uncertainties of $M_p$ and $T_{eq}$, respectively.
   Magenta letters in the diagram indicate the Solar System planets: Venus (V), Earth (E), Jupiter (J), Saturn (S), Uranus (U) and Neptune (N).
   }
\end{figure}

Figure \ref{fig:Mp_VS_Teq_output} shows, for the optimistic CGI scenario, the median value of the computed $T_{eq}$ distributions against the median value of the $M_p$ resulting from our statistical exercise (Table \ref{table:output_catalogue}).
It shows that the population of exoplanets probed with the Roman Telescope will be remarkably different from the one that has been explored with previous techniques, with Jupiter and Saturn analogues amenable to characterization.
On the other hand, analogues of Uranus and Neptune are still out of reach for the Roman Telescope.
We note that, although some planets in this range of $T_{eq}$ and $M_p$ can be found in our output catalogue (Table \ref{table:output_catalogue}), they orbit stars fainter than $V$=7 mag and are thus excluded from our analysis.
Interestingly, we find Roman-accessible planets with $T_{eq}$ comparable to that of the Earth such as the super-Earth tau Cet e, the giant planet gam Cep b or the super-Jupiter bet Pic c. 
bet Pic c is a young exoplanet in a system of about 18.5 Myr \citep{miretroigetal2020} and thus Eq. (\ref{eq:Teq}) used here will severely underestimate its effective atmospheric temperatures.
On the other hand, both gam Cep b and tau Cet e are mature systems with ages of 6.6 Gyr \citep{torres2006} and 5.8 Gyr \citep{tuomietal2013}, respectively.

\section{Discussion on selected targets} \label{sec:discussion_selectedtargets}
We next elaborate on eight targets that showcase new study cases in exoplanet science.
The exercise explores possibilities for their characterization in reflected starlight, but also limitations arising from, for example, uncertainties in their orbital solutions or their host stars' brightness.
First, we focus on the two super-Earths tau Cet e and f, which orbit near their star's habitable zone (HZ).
Then, we study the cases of pi Men b, 55 Cnc d and HD 219134 h, planets in multi-planetary systems whose
known innermost companions are accessible to atmospheric characterization through transit spectroscopy. 
We also analyse the gas giant eps Eri b, whose orbital solution remains somewhat controversial, demonstrating the potential of the Roman Telescope to characterize its orbit.
Finally, we discuss the candidate super-Earths Proxima Centauri c (hereon, Proxima c) and Barnard's Star b (hereon, Barnard b) as key targets for the next generation of directly-imaged exoplanets.

In addition, estimates or reasonable guesses of the orbital inclination are available for most of these exoplanets.
This affects their prospects for direct imaging.
For such cases, we compare their detectability against the scenario in which $i$ is unconstrained.
This way we show the relevance of multi-technique strategies for exoplanet characterization, an approach that will become more common with upcoming Gaia data releases.

\subsection{Two super-Earths near the habitable zone} \label{subsec:discussion_tauCet}
tau Cet is a nearby G8 V star with an effective temperature $T_\star$=5344 K \citep{santosetal2004}.
It  hosts four super-Earths with minimum masses in the range $1.75-3.93\,M_\oplus$ \citep{tuomietal2013, fengetal2017}. 
Based on Hipparcos astrometry, \citet{kervellaetal2019} report an anomaly in the star's tangential velocity attributable to a possible outer giant companion. 
We find that the two outermost confirmed exoplanets, tau Cet e and f, are Roman-accessible in the optimistic CGI scenario with $P_{access}$ of about 88\% and 27\%, respectively.
In the intermediate and pessimistic CGI scenarios, the probabilities drop below 13\% for both planets (Table \ref{table:output_catalogue}).

For a planet with a mass of $5M_\oplus$, \citet{fengetal2017} estimated a conservative HZ between 0.68 and 1.26 AU and an optimistic HZ between 0.55 and 1.32 AU.
This mass is consistent with the $M_p\,\approx\,4.84\,M_\oplus$
obtained from our statistical method for tau Cet e and f (Table \ref{table:output_catalogue}).
This, together with our obtained $R_p\,\approx\,1.87\,R_\oplus$, places them in the super-Earth regime if defined as $R_p<2\,R_\oplus$ and $M_p<10\,M_\oplus$.\footnote{\url{https://exoplanets.nasa.gov/what-is-an-exoplanet/planet-types/super-earth/}}
Accounting for the uncertainties in the values of $a$ (Table \ref{table:NASA_database}), tau Cet e and f orbit within the optimistic HZ and slightly outside the conservative HZ.
We note that, if additional planets in this mass range were found inside the HZ of the system as suggested by e.g. \citet{dietrich-apai2021}, they would likely fall in the accessible region of the Roman Telescope too.
The possibility of characterizing the atmospheres of these planets represents a remarkable step toward a better understanding of habitability beyond the Earth, making these targets quite unique.

\begin{figure}
   \centering
   \includegraphics[width=7cm]{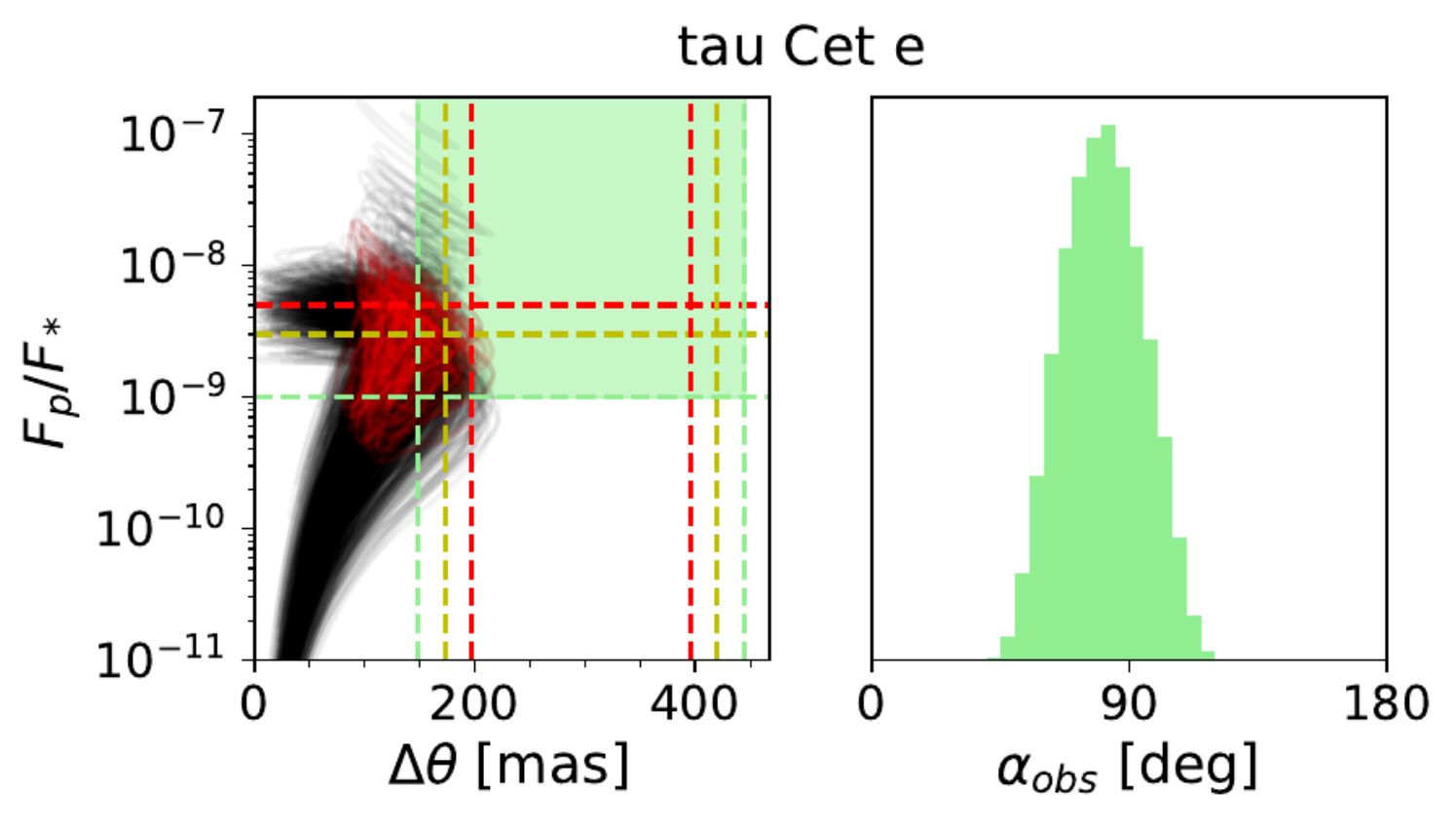} 
   \\
   \includegraphics[width=7cm]{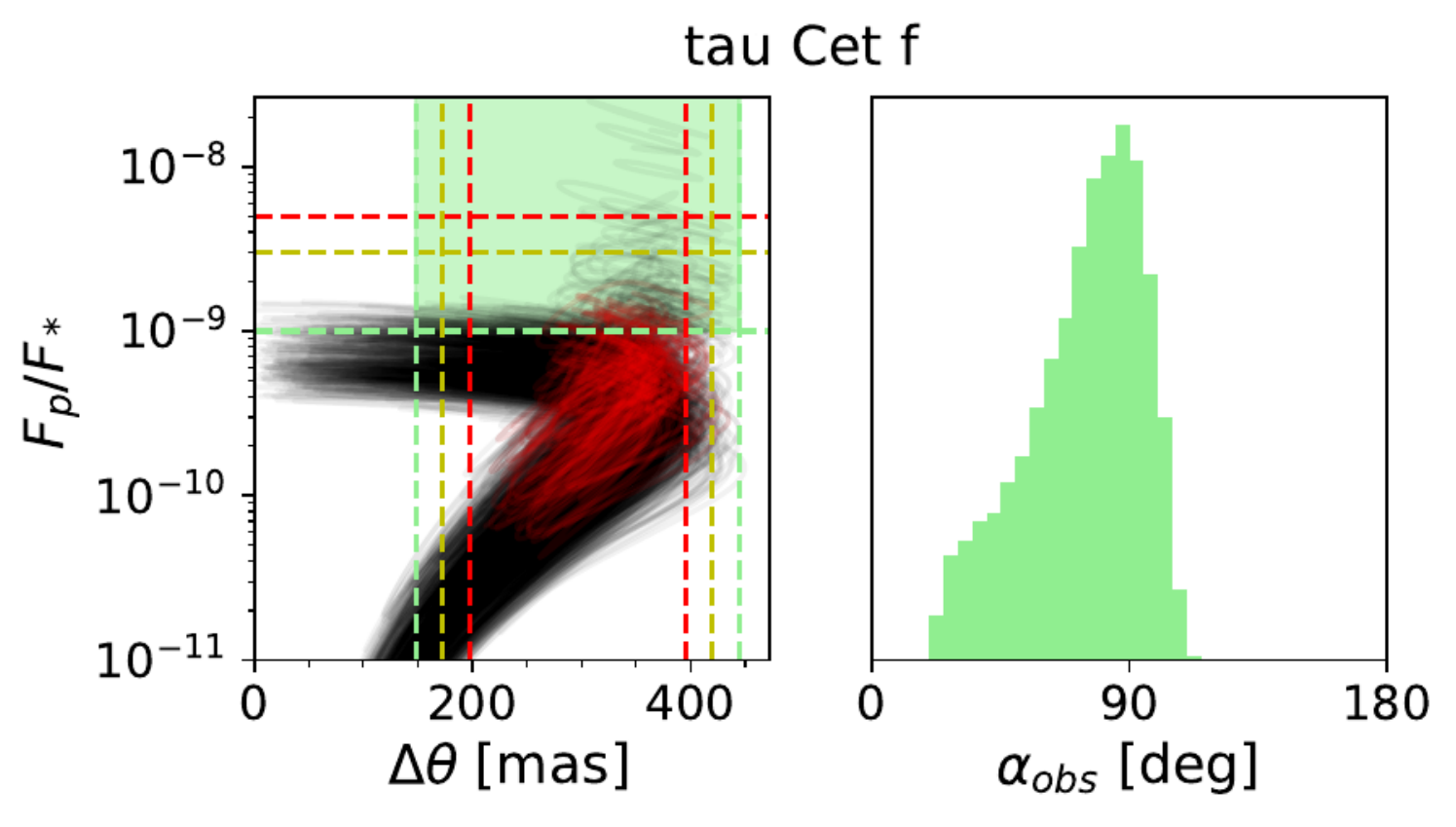} 
  \caption{\label{fig:discussion_tauCete-f} 
   Detectability of tau Cet e and f in each CGI configuration, following the same colour code as in Fig. \ref{fig:unknownorbits_orbits&histograms}. 
   Left: Black lines correspond to orbital realizations without an inclination constraint.
   Solid red lines correspond to orbital configurations with $25^\circ<i<45^\circ$, coplanar with the debris disc of the system \citep{lawleretal2014}. 
   For the latter case, the inclination is sampled from a uniform distribution within the quoted limits.
   Right: the histograms show the posterior distributions of $\alpha_{obs}$.
   }
   \end{figure}

tau Cet hosts a debris disc with a total mass of about $1\,M_\oplus$ \citep{greavesetal2004} that might potentially hinder the direct imaging of the system's planets.
Based on Herschel images, \citet{lawleretal2014} find that the disc is inclined by $i=35^\circ \pm 10^\circ$ from face-on.
They also find that the disc's inner edge is most likely located between 2 and 3 AU (555 and 833 milliarcseconds, respectively)
 although cannot rule out solutions between 1 and 10 AU.
The disc's outer edge is at about 55 AU.
\citet{macgregoretal2016} observed this system with the Atacama Large Millimeter/submillimeter Array (ALMA) and estimated an inner edge of the disc at $6.2^{+9.8}_{-4.6}$ AU.
This is consistent with recent findings by \citet{hunzikeretal2020} based on observations in the $600-900$ nm range with the SPHERE/ZIMPOL instrument at the Very Large Telescope (VLT).
Based on their non-detection of extended sources around tau Cet, \citet{hunzikeretal2020} concluded that either the disc is too faint or its inner edge is at a distance farther than about 6 AU.
Overall, the ALMA and SPHERE/ZIMPOL observations suggest that the debris disc will not interfere with the prospective imaging of the exoplanets but further measurements are needed to confirm it.
Indeed, we find that if the disc's inner edge is at 2 AU, it remains outside the optimistic OWA of the Roman Telescope for $\lambda$=575 nm but it could be detected at $\lambda$=730 nm and 825 nm (see Table \ref{table:instrument}).
The disc is not detectable in any of the exoplanet-devoted CGI filters of any CGI scenario considered here if the inner edge is further out than 2.3 AU.

The debris disc may negatively affect the habitability of these planets if they are frequently subject to large impacts.
On the other hand, the existence of abundant debris from such impacts may have favoured the formation of exomoons, which could be searched for in direct imaging \citep{cabrera-schneider2007}.
Furthermore, the disc can be used for a first guess on the planets' inclinations because systems hosting debris discs and multiple planets are frequently coplanar \citep{watsonetal2011, greavesetal2014}. 
Figure \ref{fig:discussion_tauCete-f} shows, in black, the orbital realizations for tau Cet e and f following our general methodology for planets without a constraint on inclination and, in red, those configurations with $i$ coincident with the disc's orientation. 
Table \ref{table:discussion_detectability_tauCet} compares the detectability results for tau Cet e and f in all CGI scenarios if no prior knowledge of the inclination is assumed and if the orbits of the planets are assumed coplanar with the disc.
In the optimistic CGI configuration, an estimate of $i=35^{+10}_{-10}$ deg results for both planets in statistically larger values of $P_{access}$.
This corresponds to an increase in $t_{obs}$, while the ranges of $\alpha_{obs}$ remain similar in both cases.
Similar conclusions are found for tau Cet e in the intermediate CGI configuration, whereas in this configuration tau Cet f reduces its small $P_{access}$ to zero when $i$ is constrained (Table \ref{table:discussion_detectability_tauCet}).
In fact tau Cet f remains inaccessible for any CGI configuration out of the best-case, optimistic scenario.
This reduction of $P_{access}$ when $i=35^{+10}_{-10}$, in comparison to the case of unconstrained $i$, also happens for both planets in the pessimistic CGI configuration.
In the case of tau Cet f, the reason is that if $C_{min}$ increases, only those orbital realizations with $i$ close to 0 or 180$^\circ$ (and hence very large $M_p$ and $R_p$) would be accessible.
If $i$ is constrained within 25 and 45$^\circ$, these orbital realizations will not reach the $C_{min}$ threshold.
For tau Cet e, the large IWA in the pessimistic CGI scenario is the main limitation for the detectability of the planet.

In order to determine the orbital parameters that have a larger impact on the detectability of these planets, we carried out a sensitivity study included in Appendix \ref{sec:appendix_sensitivity_study}. 
There, we fix all orbital parameters $a$, $e$, $i$ and $\omega_p$ except for one at a time and check how $P_{access}$ and $\alpha_{obs}$ change. 
For tau Cet e, we find (Fig. \ref{fig:appendix_sensitivity_study_tauCete}) that $P_{access}$ does not change significantly, with the largest effect being due to variations in $\omega_p$.
In the case of tau Cet f, $i$ and $\omega_p$ are the main parameters affecting the detectability (Fig. \ref{fig:appendix_sensitivity_study_tauCetf}).
This sensitivity study shows the relative effects of each orbital parameter on $P_{access}$ and $\alpha_{obs}$, but the correct values of these parameters are those reported in Table \ref{table:results_detectability_multiwav}, where all uncertainties are accounted for simultaneously.

\begin{table}
\tiny
\caption{Detectability of tau Cet e and f at $\lambda$=575 nm in each CGI scenario, both without prior knowledge on the orbital inclination and assuming $25^\circ<i<45^\circ$.}
\label{table:discussion_detectability_tauCet}
\centering
    \begin{tabular}{l c c c c }
    \hline \hline
    Name	& $i$  & $P_{access}$ [\%] & $P_{access}$ [\%] & $P_{access}$ [\%] \\ 
    	    & [deg] & (Optimistic)   &  (Intermediate) & (Pessimistic) \\ 
    \hline
tau Cet e	& $-$	&	87.75		& 12.99     &  0.93 \\  
tau Cet e	& $35^{+10}_{-10}$	& 90.57	& 28.91 &   0.16	 \\  
tau Cet f	& $-$	&  26.74		& 1.82         &  0.90 \\
tau Cet f	& $35^{+10}_{-10}$	& 41.84	& 0     &  0	 \\
\hline
\end{tabular}
\end{table}

Contamination from exo-zodiacal dust (exozodi) might also limit the detectability of tau Cet e and f.
\citet{erteletal2020} found that exozodiacal dust levels in the HZ around nearby early- and solar-type stars are generally about 3 times that of the Solar System.
They conclude that these levels are low enough that would not impede the spectral characterization of HZ rocky planets with current direct-imaging mission concepts such as the WFIRST Starshade Rendezvous \citep{seageretal2019}, HabEx or LUVOIR.
\citet{erteletal2020} did not detect exozodi around tau Cet and set an upper limit of 120 exozodis (that is, 120 times that of the Solar System).
Follow-up observations should help determine the actual amount of exozodi, which could also be constrained by the Roman Telescope in its observing mode devoted to disc measurements \citep{mennessonetal2019}.
\citet{erteletal2020} suggest multi-epoch observations as a path to distinguishing between the signal from the exoplanets and that from exozodi dust clumps, given their different phase functions.
In this respect, we find that even in the optimistic CGI scenario, only a modest phase coverage could be achieved for tau Cet e and f ($\alpha \in$ [61$^{+ 10 }_{- 8 }$,100$^{+ 9 }_{- 10 }$] and [53$^{+ 20 }_{- 29 }$,74$^{+ 26 }_{- 28 }$], respectively).

\subsection{Outer companions of transiting exoplanets} \label{subsec:discussion_outercompanions}

The Roman Telescope will be able to characterize several exoplanets in multi-planetary systems, some of them with inner companions accessible to transmission or occultation spectroscopy. 
This provides unprecedented possibilities for understanding their bulk atmospheric compositions, histories and the connection between formation, migration and current-time architecture.
Here we discuss the cases of pi Men b and 55 Cnc d, as representatives of this type of exoplanets.

\subsubsection{pi Men b} \label{subsubsec:discussion_selectedtargets_piMenb}

\begin{figure}
   \centering
   \includegraphics[width=7cm]{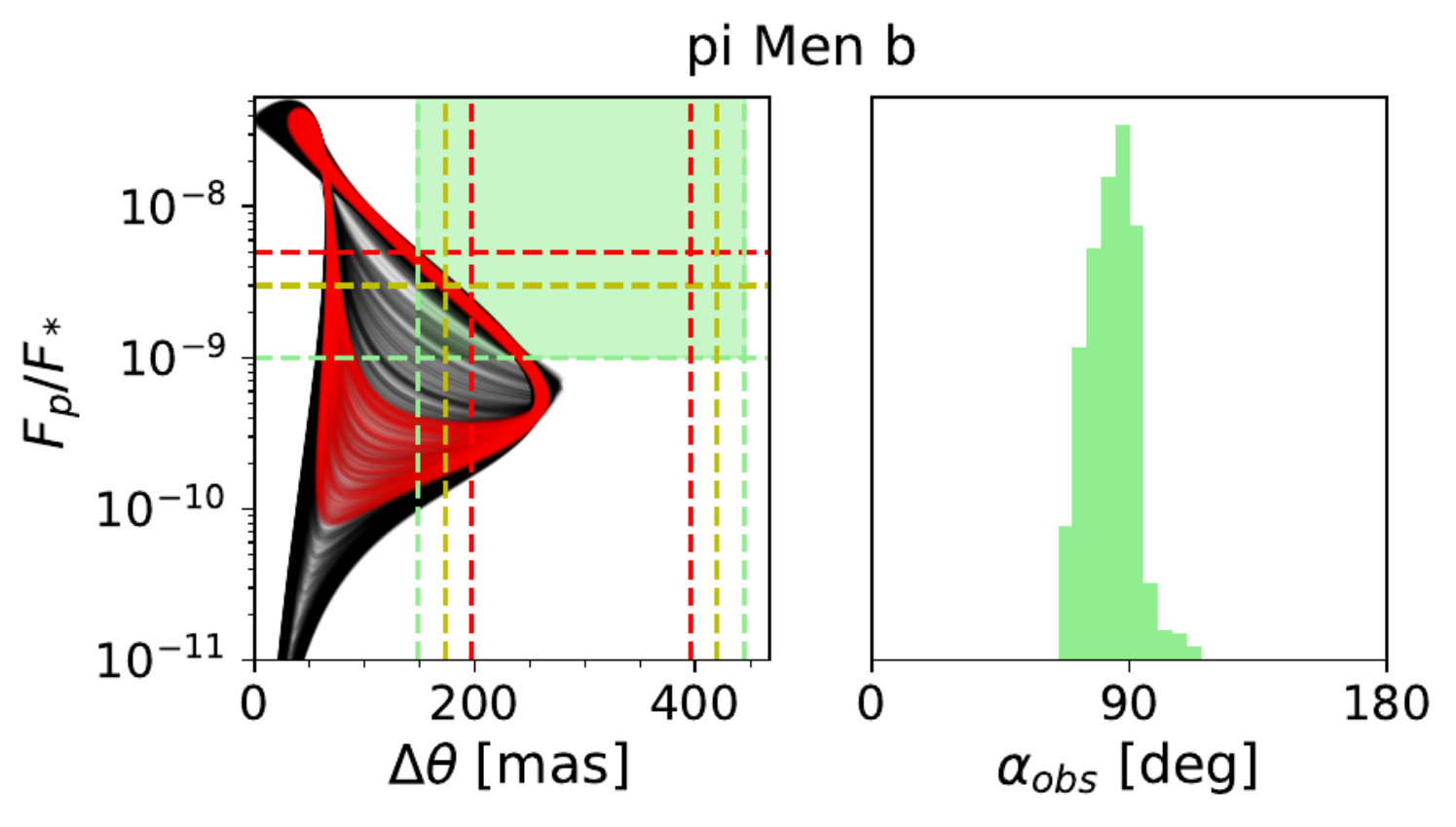} 
  \caption{\label{fig:discussion_piMenb} 
   As Fig. \ref{fig:discussion_tauCete-f}, but for the case of pi Men b.
   Solid red lines correspond to orbital configurations with $i=128.8^{+9.8}_{-14.1}$ deg, in accordance with the findings in \citet{xuan-wyatt2020}. 
   }
   \end{figure}

Planetary systems that contain a far-out Jupiter and a close-in super-Earth appear to be relatively common \citep{bryanetal2019}. The mechanisms that result in such architectures remain unclear but are  potentially important for understanding the origin and evolution of super-Earths. pi Men ($V$=5.67 mag) is one of such systems. It hosts a far-out Jupiter discovered with RV \citep{jonesetal2002} and a close-in transiting super-Earth discovered with photometry and RV \citep{gandolfietal2018, huangetal2018}.

The outer planet, pi Men b, has also been detected in joint Hipparcos and Gaia astrometry \citep{xuan-wyatt2020, derosaetal2020, damassoetal2020pimen} thereby providing constraints on its sky-projected inclination and the mutual inclination between both planets in the system. Constraints of this kind will become more usual with future releases of Gaia astrometric data. Pi Men b is now known to follow an eccentric orbit that is most likely not coplanar with the orbit of the inner planet.

The super-Earth in the system, pi Men c, is amenable to in-transit atmospheric characterization \citep{garciamunozetal2020, garciamunozetal2021}. It has been proposed that its atmosphere may not be hydrogen/helium-dominated but rather contains large amounts of heavy gases. Rossiter-McLaughlin measurements during the transit of pi Men c have revealed that its orbital plane is misaligned with the stellar spin axis \citep{kunovachodzicetal2021}.

Interestingly, the eccentricity and inclination of the outer planet and the orbital misalignment of the inner one support a formation scenario in which the super-Earth is formed far from the star and migrated into its current orbit following high-eccentricity migration \citep{kunovachodzicetal2021}. The possibility of obtaining detailed orbital information of both planets and atmospheric information of the inner one make the pi Men system quite unique. Of interest here, pi Men b is amenable to direct imaging with the Roman Telescope. This will help further constrain its orbit, especially if multi-phase measurements are made. It will also enable the spectroscopic investigation of its atmosphere, which should set valuable constraints on its chemical composition \citep[e.g.][]{lupuetal2016, nayaketal2017, carriongonzalezetal2020}.

To explore the detectability of pi Men b, we compare the orbital solution given in the NASA Archive \citep{huangetal2018}, which has no estimate of $i$, and the scenario in which $i$ is constrained.
We use an inclination of $128.8^{+9.8}_{-14.1}$ deg that results from translating the inclination angle defined in \citet{xuan-wyatt2020} to our own definition in Figure \ref{fig:sketch_orbit_XYZ}. The inclination is such that the angular momentum vector of pi Men b’s orbit points toward the observer (Xuan, private communication).

Figure \ref{fig:discussion_piMenb} compares the $F_p/F_\star$-$\Delta \theta$ diagrams if the inclination is constrained and if it is not.
In case $i$ is constrained, $t_{obs}=334^{+ 15 }_{- 15 }$ days and $\alpha_{obs}$=[70$^{+ 2 }_{- 1 }$,95$^{+ 1 }_{- 1 }$] in the optimistic CGI scenario.
This does not differ substantially from the results for the analysis with unconstrained inclination (Table \ref{table:results_detectability_multiwav}), in which pi Men b is accessible over $330^{+ 32 }_{- 17 }$ days of its 2093-day orbital period and phase angles $\alpha \in$ [69$^{+ 7 }_{- 2 }$,95$^{+ 1 }_{- 1 }$].
Similarly, if $i$ is constrained the conclusions for the other CGI scenarios are comparable to those in Table \ref{table:results_detectability_multiwav}, finding that the planet is only marginally accessible in the intermediate scenario and not accessible in the pessimistic one.
The sensitivity study in Fig. \ref{fig:appendix_sensitivity_study_piMenb} shows that the detectability of the planet does not change much if the orbital parameters vary within the uncertainties reported in the input catalogue (Table \ref{table:NASA_database}).
For comparison, we note that a shift of $180^\circ$ in the value of $\omega_p$ (as if $\omega_{\star}$ was mistaken for $\omega_p$) would yield a significantly larger range of observable phase angles $\alpha \in$ [42$^{+ 16 }_{- 3 }$,111$^{+ 2 }_{- 7 }$] (see Appendix \ref{sec:appendix_effectomega}).

\subsubsection{55 Cnc d} \label{subsubsec:discussion_selectedtargets_55Cncd}

A total of five planets have been confirmed around 55 Cnc ($V$=5.96 mag) to date \citep{butleretal1997, marcyetal2002, fischeretal2008, winnetal2011}.
The super-Earth 55 Cnc e is the only one found to transit, which allowed to constrain the inclination of its orbit.

\citet{nelsonetal2014} carried out dynamical simulations and determined that the inclination of planets $b$, $c$, $d$ and $f$, assumed coplanar, likely coincides with that of planet $e$. 
They also found that the system becomes unstable if the mutual inclination between planet $e$ and the others is between $60^\circ$ and $125^\circ$. 
\citet{baluev2015} considered this an optimistic estimate and concluded that the inclination of the outer planets could not be below $30^\circ$.
The NASA Exoplanet Archive quotes $i=90^\circ$, with no upper or lower uncertainties, for 55 Cnc $b$, $c$, $d$ and $f$.
We manually set the inclination of these planets to $i=90 \pm 60^\circ$, more in accordance with the conservative scenario in \citet{baluev2015}.
Hence, the values of $M_p$ quoted in the NASA Archive for these planets become their minimum masses. 
In our exploration, we determine the planet masses according to the sampling of $i$ in each realization (see Sect. \ref{subsec:dataprocessing_bootstraping}).

We find that the only planet observable by the Roman Telescope in this system is 55 Cnc d, with $P_{access}$=100\% in the optimistic CGI scenario.
This is the outermost and a priori most massive planet ($M_p\,sin\,i$=3.878 $M_J$) in the system, which appears to be a frequent architecture in multiplanetary systems (e.g. ups And, pi Men, HD 160691, HD 219134).
The detectability window spans over $2117^{+ 125 }_{- 318 }$ days, with a range of observable phase angles $\alpha \in [30^{+ 20 }_{- 10 },84^{+ 2 }_{- 2 }]$.
One of the limitations in the detectability of 55 Cnc d is the IWA, which affects mainly the smaller phase angles.
The value of $C_{min}$ prevents the detection of the planet as it orbits from quadrature to inferior conjunction and $\alpha$ increases, reducing $F_p/F_\star$.
In the intermediate and pessimistic CGI scenarios, the planet is below the $C_{min}$ and therefore it is not Roman-accessible.

From the sensitivity study for this planet (Fig. \ref{fig:appendix_sensitivity_study_55Cncd}) we conclude that the uncertainties in the orbital parameters have no significant effect on $P_{access}$ and that $i$ is the main parameter affecting the range of $\alpha_{obs}$.
Detecting 55 Cnc d in reflected starlight will set constraints on its atmospheric structure and composition. 
This may help understand the possible evolution of the system and the dynamical processes that have brought 55 Cnc e to its ultra-short-period orbit of $P$=0.74 days \citep{winnetal2011}.

\subsubsection{HD 219134 h} \label{subsubsec:discussion_selectedtargets_HD219134h}

The K3 V star HD 219134 ($V$=5.570 mag) hosts a multi-planetary system with up to six exoplanets \citep{motalebietal2015, vogtetal2015, gillonetal2017}.
The two innermost of them, super-Earths b ($M_p=4.74\pm0.19\, M_\oplus$) and c ($M_p=4.36\pm0.22\, M_\oplus$), have been observed in transit \citep{motalebietal2015, gillonetal2017}.
The system also includes three mini-Neptunes (planets d, f and g) and an outer Saturn-mass planet (h), all discovered in RV.
Given the different nomenclatures used in literature, we adopt here that of the NASA Archive.
\citet{johnsonetal2016} proposed that the signal attributed to planet f may be a false positive due to stellar rotation  and this planet is indeed marked as controversial in the NASA Archive.
HD 219134 h, on the other hand, has been suggested to be real despite its reported orbital period of about half the 12-year stellar activity cycle \citep{johnsonetal2016}.

In this work, we have found that HD 219134 h is one of the only three exoplanets that are Roman-accessible in all the CGI configurations considered.
In all scenarios, it is also the exoplanet that shows the largest interval of $\alpha_{obs}$ and therefore the most favourable target to perform phase-curve measurements on (Table \ref{table:results_unknownorbits_largealphaobs}).
Phase angles near quadrature are however less likely to be observed because those orbital positions tend to fall outside the OWA (see Fig. \ref{fig:unknownorbits_orbits&histograms}).
An observing mode reaching larger angular separations, such as the CGI mode devoted to disc measurements (Sect. \ref{sec:directimagingtechnical}), may complement the observations in that region of $\Delta \theta$.
In the intermediate CGI scenario and even the pessimistic one, HD 219134 h would remain accessible for about 577 and 444 days, respectively. This could facilitate higher S/N observations being obtained.
We also find that this planet is suitable to be observed with a broad wavelength coverage.
Remarkably, its $P_{access}$ is about 90\% or higher for the three CGI filters (575, 730 and 825 nm) in the optimistic, intermediate as well as in the pessimistic CGI scenario (Table \ref{table:results_detectability_multiwav}).

There have been recent investigations of the evolution and current composition of HD 219134 b and c \citep{vidottoetal2018, nikolaouetal2019}.
The broad phase and wavelength coverage achievable for HD 219134 h makes it a promising target for atmospheric characterization \citep{damianoetal2020}.
Furthermore, it can be considered one of the most reliable targets for the Roman Telescope given its great detectability prospects in all CGI scenarios and wavelenghts.
The orbital parameters reported in the NASA Archive for this planet correspond to those in the discovery papers, which have not been further updated.
Planning for direct-imaging observations will require a refined orbital characterization, for which additional RV campaigns are strongly needed.
Such follow-up RV measurements would also help clarify which of the reported signals in the system correspond to actual planets and which are caused by stellar activity.

\subsection{Prospects to confirm controversial exoplanets: eps Eri b} \label{subsec:discussion_epsErib}

\begin{figure}
   \centering
   \includegraphics[width=9.cm]{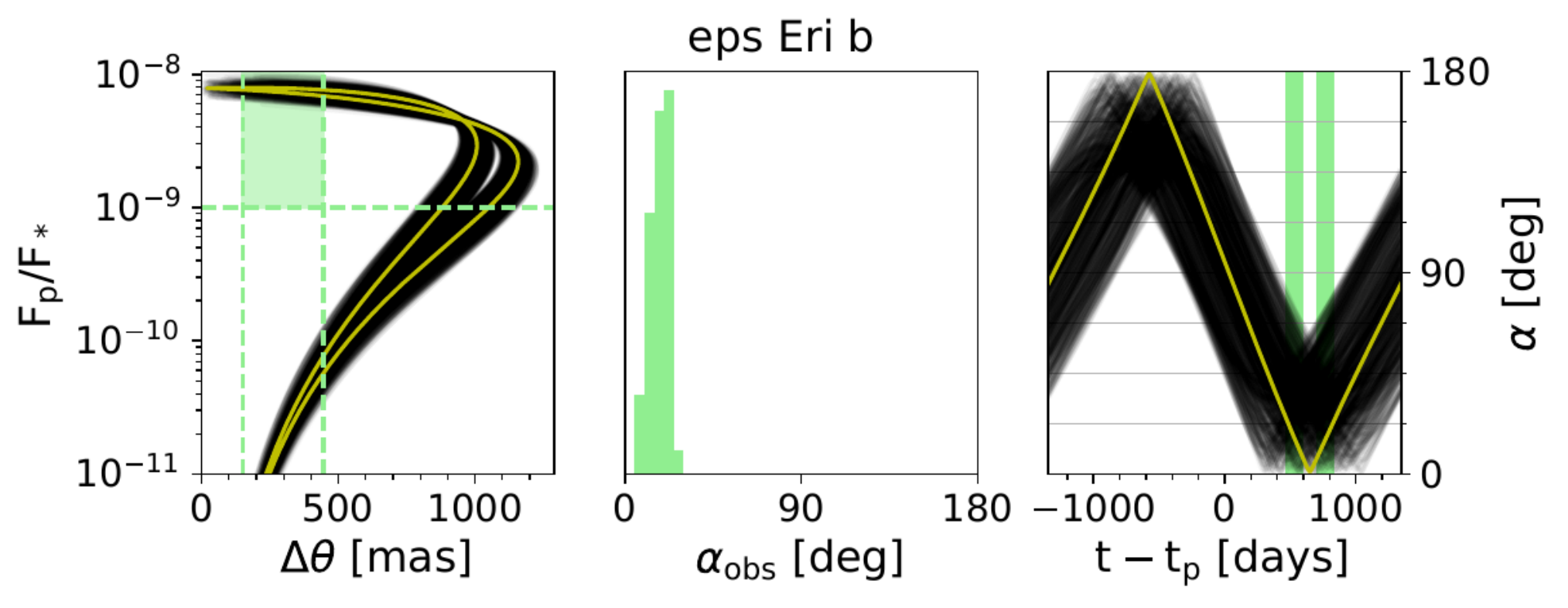}
   \\
   \includegraphics[width=9.cm]{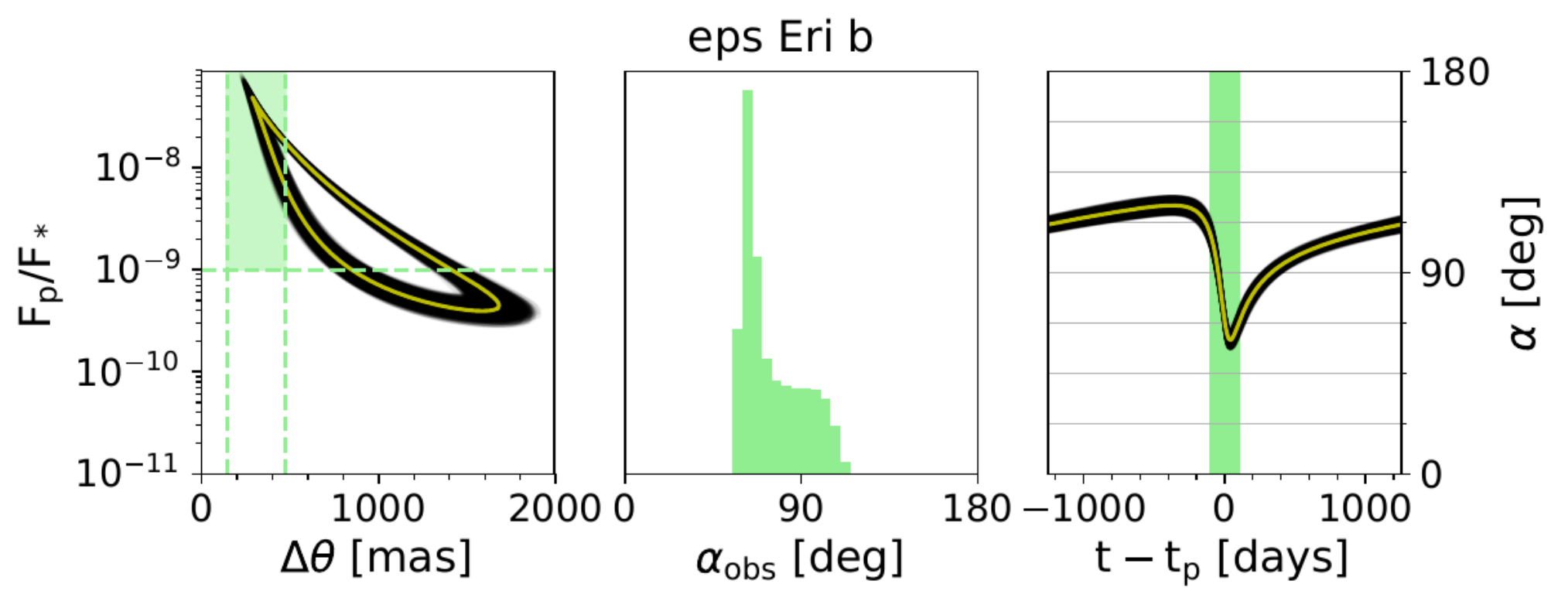}
      \caption{\label{fig:discussion_epsErib} 
      Detectability of eps Eri b in the optimistic CGI scenario, with the orbital parameters from \citet{mawetetal2019} (top panel) and \citet{benedictetal2006} (bottom panel).
      Left column: $F_p/F_\star$-$\Delta \theta$ diagram. 
      Yellow lines are specific to the maximum-likelihood orbital configuration provided in the corresponding reference.
      Middle: posterior distributions of $\alpha_{\rm{obs}}$.
      Right: variation of $\alpha$ with time for each orbital realization.
      In this panel, green regions correspond to detectability windows for the maximum-likelihood orbit (yellow line).
      All orbital realizations are shown for reference in the $\alpha$-$t$ diagram (black lines), but their corresponding detectability windows are omitted. 
      }
   \end{figure}

eps Eri b is a giant planet orbiting a young K2 V nearby star ($d$=3.22 pc) with a period of about 7 years, discovered in RV data by \citet{hatzesetal2000}.
\citet{benedictetal2006} combined RV and astrometry, and found an orbital solution with $i=30.1^\circ\pm3.8^\circ$ and $e=0.70^{+0.04}_{-0.04}$.
It has since been a promising target for direct-imaging given its predicted large angular separation of up to 1600 mas \citep{kaneetal2018}
and the interest in the atmospheric processes that could take place on a planet with such an eccentric orbit \citep{sanchezlavegaetal2003}.
However, the orbital solution of this planet has remained controversial \citep[e.g.][]{hollisetal2012} and, furthermore, the existence of the planet has also been questioned \citep{angladaescude-butler2012}.
\citet{mawetetal2019} combined RV data with high-contrast direct imaging observations at 4.67 $\mu$m, finding a RV signal consistent with a planet in a 7-year orbit but no thermal emission.
They inferred a minimum age of 800 Myr, an orbital inclination $i=89^\circ\pm42^\circ$ and an eccentricity of $e=0.07^{+0.06}_{-0.05}$, an order of magnitude smaller than the previous reference adopted as default in the NASA Exoplanet Archive.
They find this solution marginally compatible with the planet being co-planar with the outer debris disc in the system, which has $i=34\pm2^\circ$ \citep{boothetal2017}.

The NASA Exoplanet Archive updated on 2020-09-03 the information on eps Eri b from that provided by \citet{benedictetal2006} to that by \citet{mawetetal2019}.
The scope of our work is not to determine which one of the orbital solutions is more reliable.
This said, and as shown here, the update dramatically changes the prospects for detecting the planet, and demonstrates the importance of follow-up measurements, preferably with multiple techniques.
Focusing on the optimistic CGI scenario, we compare both solutions in Fig. \ref{fig:discussion_epsErib} and find that the one in \citet{benedictetal2006} is accessible in all of our realizations ($P_{access}$=100\%) and produce $\alpha_{obs}$=$[60^{+ 3 }_{- 3 }, 107^{+ 4 }_{- 5 }]$ whereas the orbital solution proposed by \citet{mawetetal2019} yields $P_{access}$=57.99\% and $\alpha_{obs}$=[12$^{+ 8 }_{- 4 }$,24$^{+ 1 }_{- 1 }$].
These obvious differences, which are also observable in the itermediate and pessimistic CGI scenarios, have potential implications on the prospects to characterize the exoplanet's atmosphere. 
In a more positive note, given that the ranges of $\alpha_{obs}$ do not overlap, reflected-starlight observations of the planet may help determine the actual orbital solution.
In both cases, we find that the OWA of the Roman Telescope is a major limitation to observe the planet. 
Observing modes with larger OWAs or telescope architectures more flexible in this regard \citep[e.g.][]{seageretal2019, lifeteam2021} will facilitate the detection of this planet and increase the interval of $\alpha_{obs}$.
In our sensitivity study for the orbital solution given by \citet{mawetetal2019}, we find that $i$ is the key factor affecting the detectability of this planet (Fig. \ref{fig:appendix_sensitivity_study_epsErib}).
Fig. \ref{fig:appendix_sensitivity_study_epsErib} shows that orbital realizations with $i$ of about 50 or 130$^\circ$ would remain outside the OWA for the whole orbital period, but those close to edge-on reach smaller $\Delta \theta$ making the planet accessible.

The abundant exo-zodiacal dust in the system \citep{erteletal2020} might create additional difficulties.
However, observing the eps Eri system could finally confirm the existence of the planetary companion and constrain its orbital solution, either by directly imaging it or by studying planet-disc interactions.
The fact that this planet remains accessible in all three CGI scenarios makes it a potential example of how high-contrast imaging with the Roman Telescope could help resolve conflicting orbital solutions.

\subsection{The potential of direct-imaging to confirm RV candidates: Barnard b and Proxima c} \label{subsubsec:discussion_selectedtargets_barnard&proxima}

\begin{table*}
\caption{Main planetary and stellar properties of the candidate exoplanets Barnard b and Proxima c.}
\label{table:discussion_candidates_database}
\centering
\resizebox{\textwidth}{!}{%
\begin{tabular}{ c  c  c  c  c  c  c  c  c  c  c  c  c  c } 
 \hline \hline
  Planet & {$d$} & {$P$} & {$a$} & {$M_p$} & {$R_p$} & {$i$} & $e$ & {$\omega_p$} & $T_{eq}$ & St. type & {$M_\star$} & {$V$} & Age \\
  & [pc] & [days] & [AU] & [$M_{J}$] & [$R_{J}$] & [deg] & & [deg] & K & & [$M_\odot$] & [mag] & [Gyr]  \\
\hline
Barnard b 		&  1.80$^{ +0.00 }_{ -0.00 }$ 		&  232.8$^{ +0.4 }_{ -0.4 }$ 		&  0.40$^{ +0.02 }_{ -0.02 }$ 		&  0.010$^{ +0.001 }_{ -0.001 }$ $^\dagger$ 		&  $-$ 		&  $-$ 		&  0.32$^{ +0.10 }_{ -0.15 }$ 		&  287.0$^{ +19.0 }_{ -22.0 }$ 		&  105$^{ +3 }_{ -3 }$ 		&  M3.5V 		&  0.16$^{ +0.00 }_{ -0.00 }$ 		&  9.5 		&  8.50$^{ +1.50 }_{ -1.50 }$ \\ 
Proxima c 		&  1.30$^{ +0.00 }_{ -0.00 }$ 		&  1900.0$^{ +96.0 }_{ -82.0 }$ 		&  1.48$^{ +0.08 }_{ -0.08 }$ 		&  0.022$^{ +0.003 }_{ -0.003 }$ $^\dagger$ 		&  $-$ 		&  $-$ 		&  $-$ 		&  $-$ 		&  39$^{ +16 }_{ -18 }$ 		&  M5.5V 		&  0.12$^{ +0.00 }_{ -0.00 }$ 		&  11.13 		&  $-$ \\ 

\hline
\end{tabular}
}
\tablefoot{$^\dagger$ indicates that the $M_p$ value corresponds to $M_p sin(i)$.\\ 
The quoted values for Barnard b are obtained from the discovery paper \citep{ribasetal2018} and the Extrasolar Planets Encyclopaedia.
For Proxima c, the planetary parameters are obtained from the discovery paper \citep{damassoetal2020proxc} and the Extrasolar Planets Encyclopaedia, while the stellar parameters are obtained from \citet{suarezmascarenoetal2020}.\\
Additional estimates for the inclination of Proxima c have been proposed by \citet{benedict-mcarthur2020} and \citet{kervellaetal2020}, suggesting also practically zero eccentricity.
The implications of these findings are discussed in the text.\\
}
\end{table*}

\begin{table}
\tiny
\caption{Detectability conditions for the exoplanet candidates Barnard b and Proxima c in the optimistic CGI scenario. }
\label{table:discussion_candidates_detectability}
\centering
    \begin{tabular}{l c c c c c c c c c c c c}
    \hline \hline
    Name		& $t_{obs}$	[days]	& $\alpha_{obs}$ [deg] & $P_{tr}$ \% & $P_{access}$ \% \\ 
    \hline
    Barnard b		& 167$^{+ 39 }_{- 49 }$ 		& [35$^{+ 23 }_{- 4 }$,120$^{+ 5 }_{- 7 }$]		& 2.3E-01		& 100.0		 \\  
    Proxima c		& 116$^{+ 59 }_{- 50 }$ 		& [27$^{+ 41 }_{- 15 }$,97$^{+ 33 }_{- 70 }$]		& 1.8E-01		& 64.84		 \\ 
    \hline
\end{tabular}
\tablefoot{For the case of Proxima c, the parameters $i$, $e$ and $\omega_p$ are assumed unconstrained and hence sampled as explained in Sect. \ref{subsec:dataprocessing_orbits}. If the values of $i$ and $e$ considered for Proxima c are compatible with the findings of \citet{benedict-mcarthur2020} or \citet{kervellaetal2020}, this planet would not be Roman-accessible (see Fig. \ref{fig:discussion_candidates}).} 
\end{table}

\begin{figure}
   \centering
   \includegraphics[width=7cm]{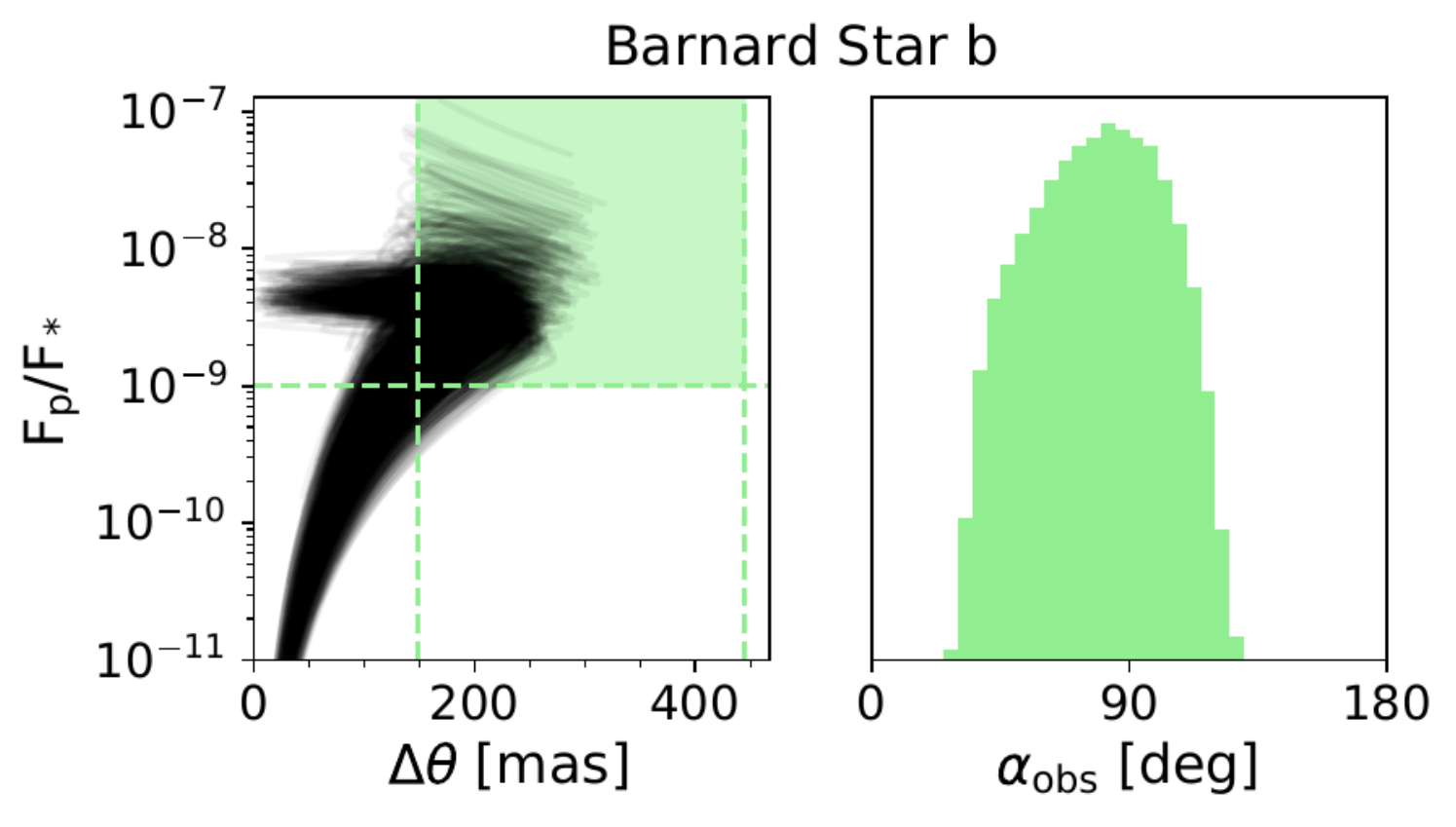} 
   \\
   \includegraphics[width=7cm]{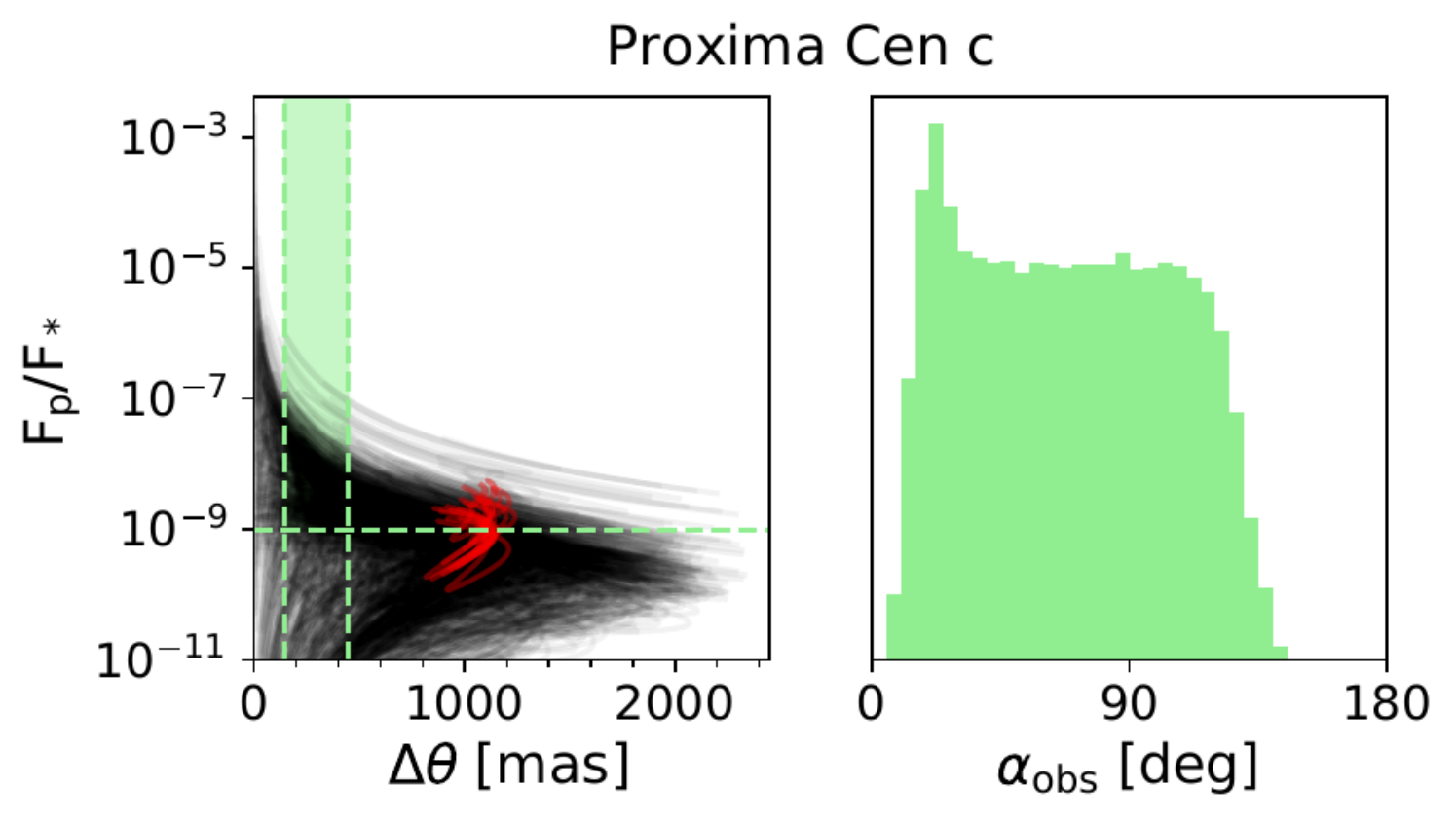} 
  \caption{\label{fig:discussion_candidates} 
   As Fig. \ref{fig:discussion_tauCete-f}, but for Barnard b and Proxima c.
   Red lines correspond to orbital configurations of Proxima c with $e<0.05$ and $i \in [14^\circ,42^\circ]$ or $[138^\circ,166^\circ]$, consistent with the estimates by \citet{benedict-mcarthur2020} and \citet{kervellaetal2020}.}
   \end{figure}

A space-based direct-imaging mission will be useful to confirm the existence of a number of targets that are often considered candidate exoplanets.
Due to the expected duration of the eventual science phase of Roman Telescope's CGI, the use of telescope time in such survey-like observations with uncertain payoff will likely not be favoured.
Nevertheless, the next generation of direct-imaging space telescopes will have among their goals the search for new exoplanets \citep{gaudietal2018, luvoirteam2018}.
In this context, we analyse the cases of Barnard b \citep{ribasetal2018} and Proxima c \citep{damassoetal2020proxc}, two super-Earth candidates orbiting the closest planet-host stars. 
The main properties of these targets, which are not included in the NASA Archive of confirmed exoplanets, and the corresponding references are listed in Table \ref{table:discussion_candidates_database}.

We find (Fig. \ref{fig:discussion_candidates}, Table \ref{table:discussion_candidates_detectability}) that both planets orbit within the optimistic Roman-accessible region of IWA, OWA and $C_{min}$ if their orbital inclinations are assumed unconstrained. 
Indeed, Barnard b is accessible in all the orbital realizations ($P_{access}$=100\%) whereas Proxima c, with larger uncertainties in the orbital parameters, has a somewhat lower probability of $P_{access}$=64.84\%.
Furthermore, Barnard b remains accessible over about 70\% of its orbital period ($t_{obs}$=167$^{+ 39 }_{- 49 }$ days) but Proxima c is only accessible over less than a tenth of its orbit ($t_{obs}$=116$^{+ 59 }_{- 50 }$ days).
The range of $\alpha_{obs}$ is particularly wide for Barnard b ($\Delta \alpha_{obs}\approx85^\circ$), which may eventually help characterize the composition and structure of its atmosphere \citep{nayaketal2017, damianoetal2020}.

The brightness of their host stars likely prohibits the observation of these planets with the Roman Telescope.
However, both stars will be within the operating range of future direct-imaging missions such as LUVOIR \citep{luvoirteam2018}, being Barnard a more suitable target ($V$=9.5 mag) than Proxima ($V$=11.13 mag).

In the sensitivity study for these candidates, we find that Barnard b has $P_{access}=100\%$ in all cases, being $i$ and $e$ the parameters with the largest impact on $\alpha_{obs}$ (Fig. \ref{fig:appendix_sensitivity_study_Barnardb}).
In the case of Proxima c, $i$ is the parameter which affects the most both $P_{access}$ and $\alpha_{obs}$. 
Indeed, only the orbits with $i\approx90^\circ$ occur to be accessible
(Fig. \ref{fig:appendix_sensitivity_study_Barnardb}).

Proxima c is indeed amenable to astrometric characterization of its orbit with existing telescopes, which strongly affects the detectability prospects for a direct-imaging mission.
\citet{benedict-mcarthur2020} obtained $i$=133$\pm$1$^\circ$ and $e$=0.04$\pm$0.01 with astrometric data from Hubble Space Telescope and SPHERE instrument at the VLT.
Correspondingly, assuming a circular orbit and using Gaia data, \citet{kervellaetal2020} proposed two solutions: a prograde orbit with $i$=152$\pm$14$^\circ$ and a retrograde orbit with $i$=28$\pm$14$^\circ$.
We find that in all these cases Proxima c would not be Roman-accessible because the angular separation is larger than the OWA during the whole orbit (red lines in Fig. \ref{fig:discussion_candidates}).

There is a growing population of exoplanet candidates, mostly detected with RV.
The examples of Barnard b and Proxima c illustrate the potential of direct-imaging missions to confirm, given the appropriate orbital conditions, the existence of these candidates.

\section{Conclusions}
\label{sec:conclusions}

The Nancy Grace Roman Space Telescope will be the first space mission capable of directly imaging exoplanets in reflected starlight.
The first measurements of this kind could therefore be available within the decade.
Designed as a technology demonstrator, it will pave the way for more ambitious direct imaging missions such as LUVOIR or HabEx.
We have shown in this work its potential for several science cases, in particular for phase-curve measurements of exoplanets.

We have analysed the complete set of confirmed exoplanets in the NASA Exoplanet Archive and computed which ones would be Roman-accessible at 575 nm in three different scenarios of CGI performance.
For that, we have compiled the planetary and stellar parameters needed to compute the evolution of the exoplanet's orbital position and brightness (Table \ref{table:NASA_database}).
To account for the uncertainties in the orbital determination and other non-orbital factors, we followed a statistical approach and computed 10000 random realizations for each exoplanet.
In each realization, the values of all parameters were independently drawn from appropriate statistical distributions within their quoted upper and lower uncertainties.
For those exoplanets lacking a value of orbital elements such as $e$, $i$ or $\omega_p$, we drew their values from uniform distributions assuming an isotropic distribution of possible orbital orientations.
In the cases without a value of the planet radius, we derived it by means of published $M_p$-$R_p$ relationships covering a range of masses from less than that of Mercury to $60\, M_J$.
From the posterior distribution of $\Delta \theta$ or $F_p/F_\star$, we derived the overall probability of the planet to be Roman-accessible, its transit probability and the values of $t_{obs}$, $\alpha_{obs}$ and $T_{eq(obs)}$.

As of September 2020, 26 exoplanets orbiting stars brighter than $V$=7 mag have $P_{access}>25\%$ in the optimistic CGI configuration. 
This number is reduced to 10 and 3 in the intermediate and pessimistic scenarios, respectively.
Only HD 219134 h, 47 UMa c and eps Eri b are Roman-accessible in all three scenarios.
We note that our assumed scenarios do not correspond to officially expected CGI specifications but rather to a range of plausible coronagraph performances according to current predictions. 
For instance, the best official estimates of the IWA currently match the value in our optimistic scenario, while the official OWA is slightly less restrictive than the one we assume.
The best official estimates of $C_{min}$ are more restrictive than the value assumed in our optimistic scenario but somewhat more favourable than the one in our intermediate scenario (see Sect. \ref{sec:directimagingtechnical}).
Additional factors not considered in this work will reduce the number of accessible targets and therefore a high value of $P_{access}$ does not guarantee a detection of the planet, which will be restricted by mission schedule and final instrument performance.
For reference, we list in our output catalogue (Table \ref{table:output_catalogue}) the up to 76 exoplanets that would be accessible in the optimistic scenario if the host-star magnitude was not a limitation.

The catalogue presented here is expected to evolve as follow-up observations are performed, and will be updated in future work as more information about the mission is available.
One of the next steps to be performed with our methodology is to simulate an optimized observing schedule for a direct-imaging telescope, including noise sources and restrictions from mission timeline.
A similar approach was discussed in \citet{brown2015} under the assumptions of no orbital uncertainties except for $i$, and $R_p=R_J$ for all considered planets.
That work concluded that successful observations of any suitable exoplanet may be restricted to windows of only a few days.
Nevertheless, the detectability criteria in that work as well as the resulting target list were shaped by the science requirement of measuring $M_p$ with a fractional uncertainty of 0.10.
Relaxing this requirement will broaden the list of observable targets and their detectability windows.
On the other hand, accounting for all the parameter uncertainties that we consider in our method will surely increase the uncertainties in the planning.
The about 3000 exoplanets discovered between the compilation of the input catalogue in \citet{brown2015} and ours also increase the options to find suitable targets as the launch of the Roman Telescope approaches.

A population study was carried out for the set of 26 Roman-accessible exoplanets in the optimistic scenario.
We compared their properties with those of the complete population of confirmed exoplanets, and with the exoplanets that have been observed in transit (Sect. \ref{subsec:results_populationstudy}).
As expected, we found that the subset of Roman-accessible planets is biased towards 
massive objects on long-period orbits with high eccentricities.
We also noted a lack of F, K and M stars in the hosts of Roman-accessible planets, caused partially by the threshold specified at $V$=7 mag.
Overall, this suggests that the Roman Telescope will probe a population of exoplanets that differs in various ways from those accessible to atmospheric characterization with current techniques.

In the optimistic CGI scenario, exoplanets will be accessible mainly near quadrature ($\alpha$=90$^\circ$) and many of them could reach minimum values of $\alpha_{obs}$ of about 30$^\circ$ or 40$^\circ$.
These phases are remarkably brighter than those generally used to estimate planet detectability and S/N, usually $\alpha=90^\circ$ or up to 60$^\circ$ in optimistic works \citep[e.g.][]{lacyetal2019}.
This may have a favourable impact on the computation of integration times.
We found several exoplanets suitable for phase curve measurements in reflected starlight with ranges of observable phases $\Delta \alpha_{obs}\gtrsim 70^\circ$.
The primary limitation to access smaller phase angles is the IWA of the coronagraph, whereas high phases will be mainly limited by the $C_{min}$ of the instrument.
This effect also narrows the intervals of $\Delta \alpha_{obs}$ in more conservative CGI scenarios.

Computing the range of $\alpha_{obs}$ is not only useful to compute more accurate levels of S/N, but also to understand the potential for atmospheric characterization.
We have shown that in the optimistic CGI scenario, $\alpha_{obs}$ could range between about $30^\circ$ and $120^\circ$ for some targets.
The atmospheric-modeling community may use these values to study whether the atmospheric retrievals of an exoplanet would benefit from multiple observations at different phases.
Analysing the impact of partial wavelength coverage on the atmospheric characterization is also ongoing theoretical work \citep[e.g.][]{batalhaetal2018, damianoetal2020}.
Such studies will benefit from our findings on the detectability at different CGI filters (Table \ref{table:results_detectability_multiwav}).
In addition, our statistical method provides both the $T_{eq}$ of each planet along its orbit and the range of observable temperatures $T_{eq(obs)}$.
Respectively, $T_{eq}$ and $T_{eq(obs)}$ are relevant parameters to model the structure of (exo)planetary atmospheres \citep[e.g.][]{hu2019} and to search for atmospheric variability.

Up to 13 of the Roman-accessible exoplanets are part of multiplanetary systems, with 
the systems 47 UMa and tau Cet hosting two Roman-accessible exoplanets each, in the optimistic scenario.
In particular, the detectability of tau Cet e and f is severely reduced in more pessimistic CGI configurations (Table \ref{table:output_catalogue}).
Nevertheless, the possibility of observing two super-Earths inside the optimistic habitable zone of their star motivates follow-up measurements of this system before the Roman Telescope is launched.

55 Cnc d, pi Men b and HD 219134 h are Roman-accessible planets that have a transiting inner companion.
These are especially valuable targets because spectroscopic observations of both planets could eventually be performed.
There are constraints on the orbital inclination of the planets in some of these systems. 
For pi Men b, such constraints are based on astrometry, while for for 55 Cnc d they come from dynamical stability analyses.
We showed that an estimate of $i$ reduces the dispersion of possible orbital solutions, thereby improving the accuracy of the computed $P_{access}$. 
The characterization of these outer planets in reflected starlight will foreseeably set valuable constraints on the possible structure of the systems and their history.

For pi Men b, we also discussed how a correct value of the argument of periastron of the exoplanet affects the prospects for phase-curve measurements.
The lack of a homogeneous criterion to report $\omega$ in the literature has resulted in multiple definitions that may yield inconsistent results.
The main exoplanet catalogues list the $\omega$ values as reported in the original references, regardless of the definitions actually used there. 
Shifts in $\omega$ by 180$^\circ$ (the usual outcome of different definitions)
do not affect the maximum angular separation. They do however affect the computed phase angles and therefore $F_p/F_\star$ (Appendix \ref{sec:appendix_effectomega}).
The future prioritization of targets for direct imaging missions will benefit from consistently reported values of $\omega_p$, as we do in this work.

Finally, we addressed the potential of direct-imaging measurements to confirm the existence of exoplanets that are controversial or remain candidates.
We showed that eps Eri b could be accessible in reflected starlight and confirm the measured RV signal.
We also found the candidate super-Earths Barnard b and Proxima c to orbit in the accessible $\Delta\theta-F_p/F_\star$ region of the Roman Telescope but will be undetectable due to the faint magnitude of their host stars.
However, these examples show the relevance of determining the orbital inclination, such as in the case of Proxima c, and its impact on the detectability prospects.
We conclude that in general direct-imaging missions will strongly rely on preliminary observations with other techniques such as RV or astrometry.

Although planned as a technology demonstrator, our work here has shown some of the possibilities of the Roman Telescope's coronagraph during an eventual phase of science operations.
It would access a population of exoplanets that has not been previously observed, widening our understanding of exoplanet diversity.
Moreover, it would be able to perform phase-curve measurements of these planets in reflected starlight, providing insight into exoplanetary atmospheres that cannot be studied with other techniques.

\begin{acknowledgements}
      The authors acknowledge the support of the DFG priority program SPP 1992 “Exploring the Diversity of Extrasolar Planets (GA 2557/1-1)”.
      OCG acknowledges the support of COST Action 18104 - Revealing the Milky Way with Gaia.
      NCS acknowledges the support by FCT - Fundação para a Ciência e a Tecnologia through national funds and by FEDER through COMPETE2020 - Programa Operacional Competitividade e Internacionalização by these grants: UID/FIS/04434/2019; UIDB/04434/2020; UIDP/04434/2020; PTDC/FIS-AST/32113/2017 \& POCI-01-0145-FEDER-032113; PTDC/FIS-AST/28953/2017 \& POCI-01-0145-FEDER-028953.
      This research has made use of the NASA Exoplanet Archive, which is operated by the California Institute of Technology, under contract with the National Aeronautics and Space Administration under the Exoplanet Exploration Program.
      We thank V. Bailey for the useful information about CGI filters.
      For their help confirming the criteria used to determine the argument of periastron in works that were quoted in the NASA Exoplanet Archive, we thank R. V. Baluev, S. Curiel, F. Feng, D. Fischer, D. Gandolfi, A. Hatzes, D. Mawet, M. Pinamonti, I. Ribas, P. Robertson, S. Vogt, J. Wright and J. W. Xuan.

\end{acknowledgements}

%\newpage

\begin{appendix} %First appendix
 
\newpage
\section{Equations of motion} \label{sec:appendix_equations_orbit}

Assuming an elliptic orbit, we can define a coordinate system with $x$ and $y$ axes co-planar to the orbit.
The $x$ axis is in the direction of the ellipse major axis, positive towards the orbital periastron; $y$ axis is perpendicular to $x$; $z$ is perpendicular to the orbital plane.
Expressed in polar coordinates with respect to an arbitrary reference direction which subtends an angle $\omega_p$ with the $x$ axis:
\begin{equation} \label{eq:kepler_xy_polar}
\begin{aligned}
x = r\, cos f \\
y = r\, sin f \\
z = 0
\end{aligned}
\end{equation}
$\omega_p$ is referred to as the argument of periastron.

The orbit can be represented in three dimensions with a new coordinate system with origin in the star, as shown in Fig. \ref{fig:sketch_orbit_XYZ}.
The $X$, $Y$ and $Z$ axes form a triad such that $X$ lays in the direction of the reference line, $Y$ is in the reference plane and $Z$ is perpendicular to both.
We will assume that the direction to the observer is $-Z$.
We note here that our assumption on the observer's direction is consistent with \citet{hatzes2016} but differs with respect to \citet{murray-correia2010} or \citet{winn2010}, who place the observer in $+Z$.
A vector ($x$,$y$,$z$) is expressed in the new axes ($X$,$Y$,$Z$) by applying three rotations \citep{murray-correia2010}.

\begin{equation} \label{eq:kepler_rotations}
\small
\begin{aligned}
\begin{split}
\begin{pmatrix}
X\\
Y\\
Z
\end{pmatrix}
=
P_z(\Omega)\, P_x(i)\, P_z(\omega_p)\, 
\begin{pmatrix}
x\\
y\\
z
\end{pmatrix}
=
\\
\begin{pmatrix}
cos\, \Omega & -sin\, \Omega & 0 \\
sin\, \Omega & cos\, \Omega & 0 \\
0 & 0 & 1
\end{pmatrix}
\begin{pmatrix}
1 & 0 & 0 \\
0 & cos\, $i$ & -sin\, $i$ & 0 \\
0 & sin\, $i$ &\, cos\, $i$ & 0
\end{pmatrix}
\begin{pmatrix}
cos\, \omega_p & -sin\, \omega_p & 0 \\
sin\, \omega_p & cos\, \omega_p & 0 \\
0 & 0 & 1
\end{pmatrix}
\,
\begin{pmatrix}
x\\
y\\
z
\end{pmatrix}
\end{split}
\end{aligned}
\end{equation}

Here the angle $i$ corresponds to the orbital inclination and $\Omega$ is the longitude of the ascending node.
The longitude of the ascending node is the angle between the reference direction and the ascending node (the point at which the orbital plane intersects the reference plane moving towards positive values of $Z$).
$\Omega$ determines the position of the orbit in the absolute reference frame of the sky.
In this work, we will assume $\Omega = 0^\circ$ without loss of generality which is equivalent to reorienting the XY axes in the plane of the sky.

\begin{figure}
   \centering
   \includegraphics[width=9.cm]{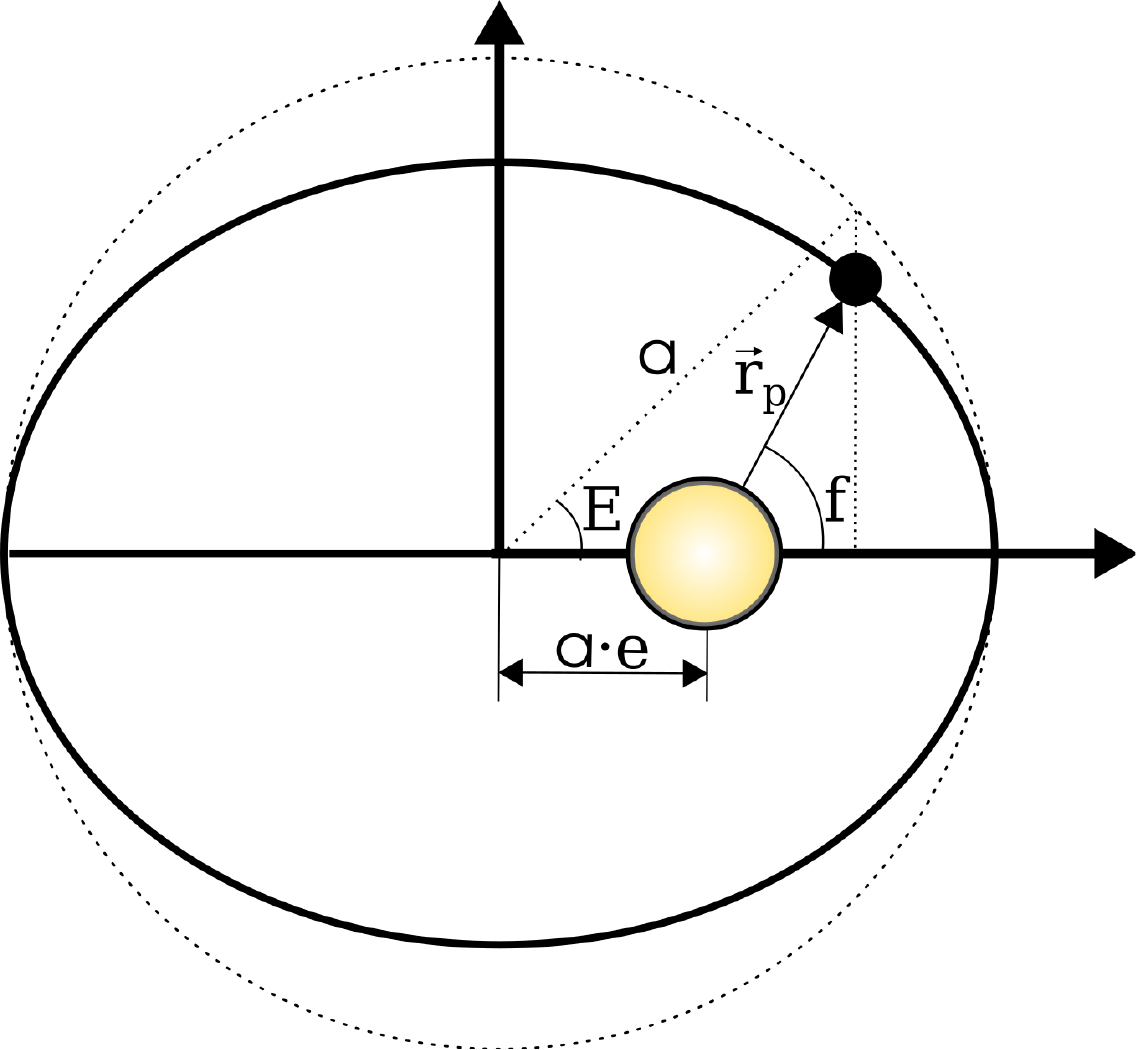}
      \caption{\label{fig:sketch_orbit_Eanomaly} 
      Sketch of the elliptic orbit and the auxiliary circle that defines the eccentric anomaly $E$.}
   \end{figure}

The orbital position of a planet at a certain time can be computed through Kepler's equation \citep{murray-dermott1999}: 
\begin{equation} \label{eq:keplers_equation}
M = E - e\, sin\,E
\end{equation}
where $e$ is the eccentricity of the orbit, $E$ is called the eccentric anomaly and $M$, the mean anomaly. $M$ is defined as:
\begin{equation} \label{eq:mean_anomaly}
M = \frac{2\pi}{P}\, (t-t_p)
\end{equation}
Here, $t$ is the time for which we are computing the position, $t_p$ is the time of periastron passage and $P$ is the orbital period of the planet.
   
$E$ is defined in terms of the true anomaly $f$, the orbital semi-major axis $a$, the eccentricity and the planet-star distance given in Eq. (\ref{eq:kepler_planet-star_distance}).
From the sketch of the orbit in Figure \ref{fig:sketch_orbit_Eanomaly}
\begin{equation} \label{eq:relating_true_eccentric_anomaly_cos}
\begin{split}
a\, cos E = a\, e + r\, cosf;\\
cos\, E = \frac{e+cos f}{1+e\, cos f}
\end{split}
\end{equation}

With that, $sin\, E$ can be computed as $\sqrt{1-cos^2\,E}$:
\begin{equation} \label{eq:relating_true_eccentric_anomaly_sin}
sin\, E = \frac{sin f\, \sqrt{1-e^2}}{1+e\, cos f}
\end{equation}

$E$ can be re-expressed in terms of the true anomaly as:
\begin{equation} \label{eq:eccentric_anomaly}
E = 2\, arctan \left( \sqrt{\frac{1-e}{1+e}}\, tan\left( \frac{f}{2} \right) \right)
\end{equation}

Substituting Eqs. (\ref{eq:mean_anomaly}), (\ref{eq:relating_true_eccentric_anomaly_sin}) and (\ref{eq:eccentric_anomaly}) into Eq. (\ref{eq:keplers_equation}) we obtain the relation between the true anomaly and time, which is given in Eq. (\ref{eq:kepler_time_dependence}).
In this work, we are not using an absolute reference frame for all of the exoplanets together with the mission timeline and therefore $t_p$ is not relevant for the detectability.

\clearpage
\section{$\omega_p$ VS. $\omega_\star$: the impact on the detectability.} \label{sec:appendix_effectomega}
\renewcommand{\thefigure}{B.\arabic{figure}}
\begin{figure*}
	\centering
	\includegraphics[width=7cm]{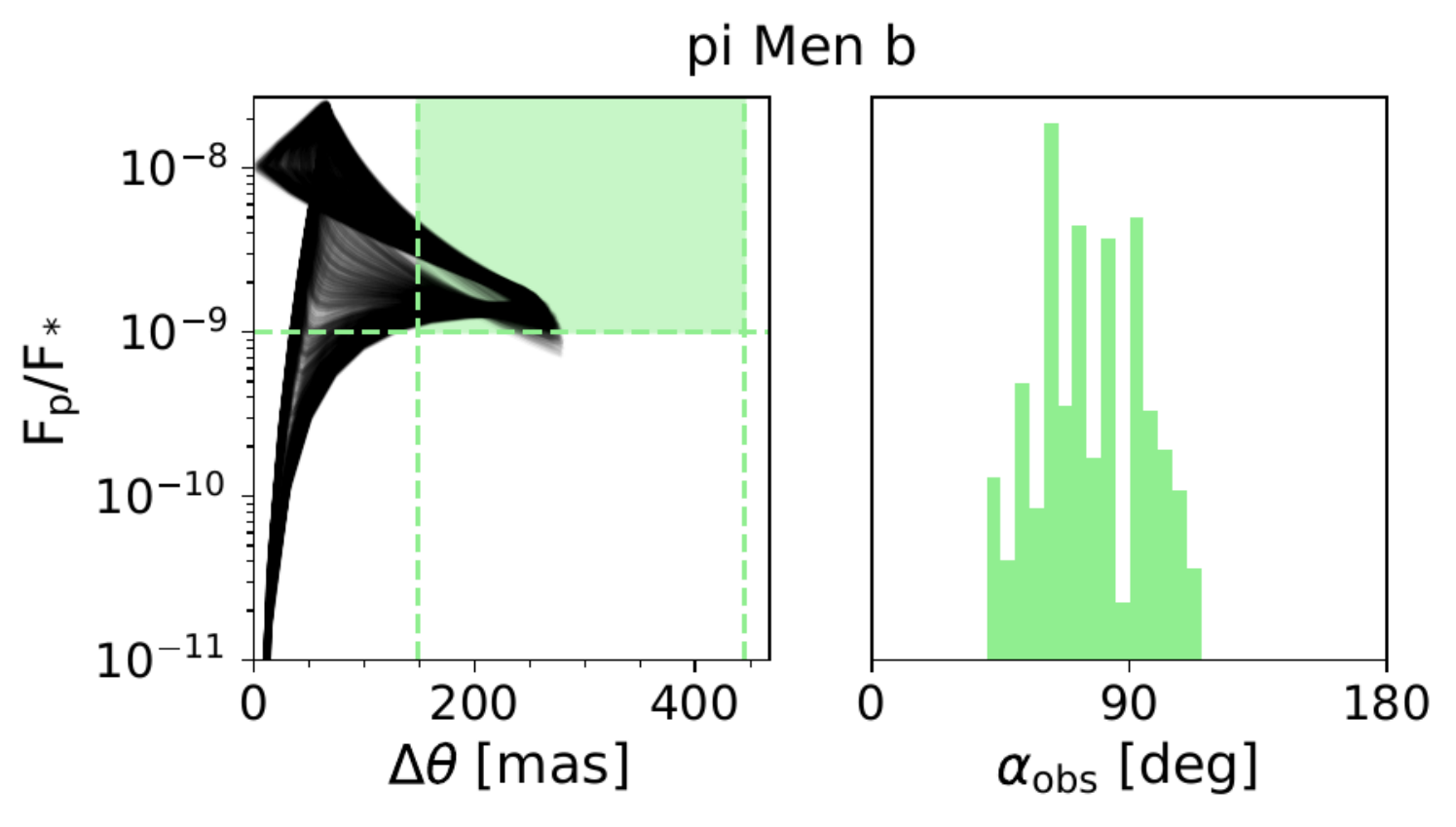} 
	\includegraphics[width=7cm]{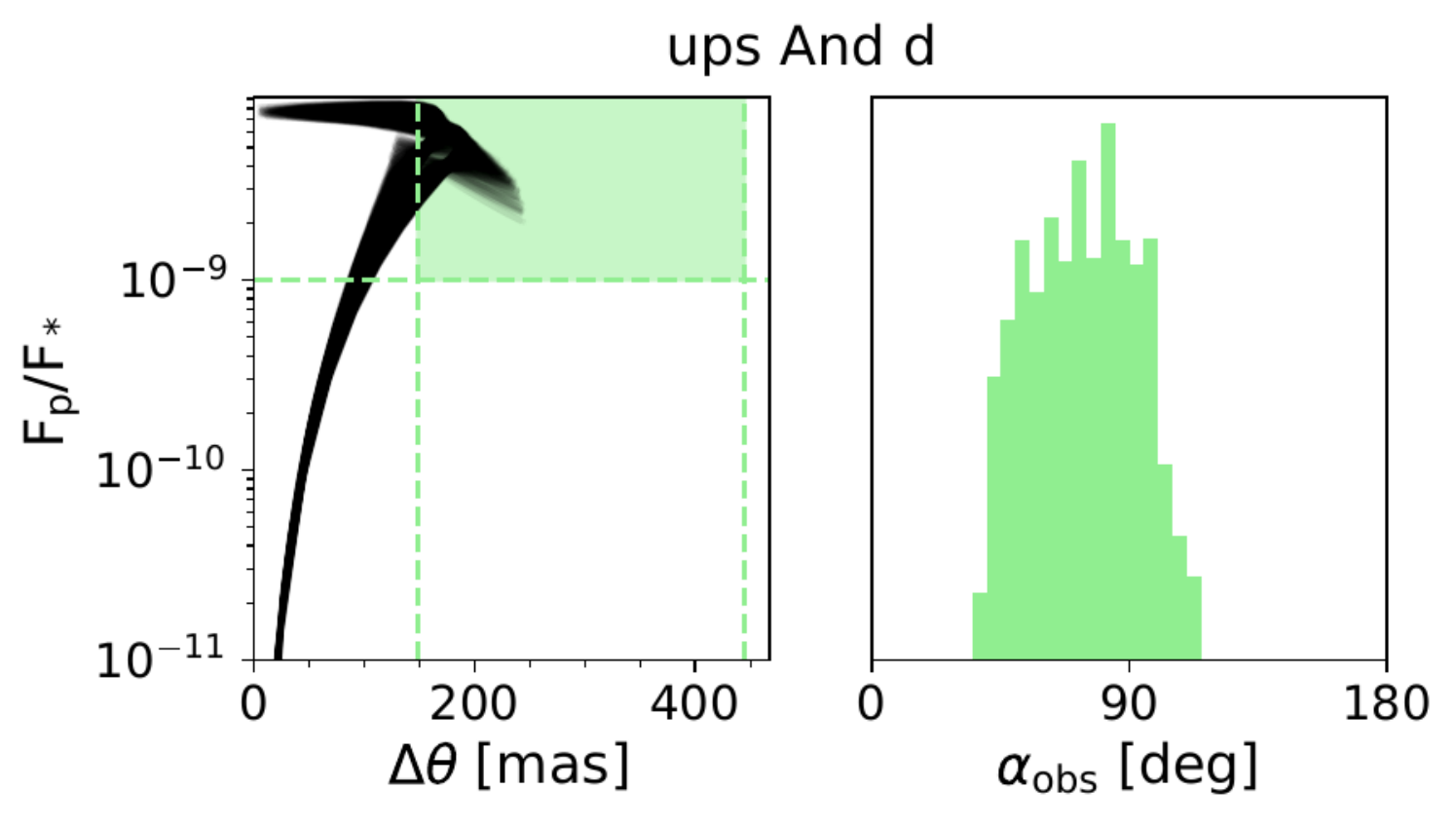} 
	   \\
	\includegraphics[width=7cm]{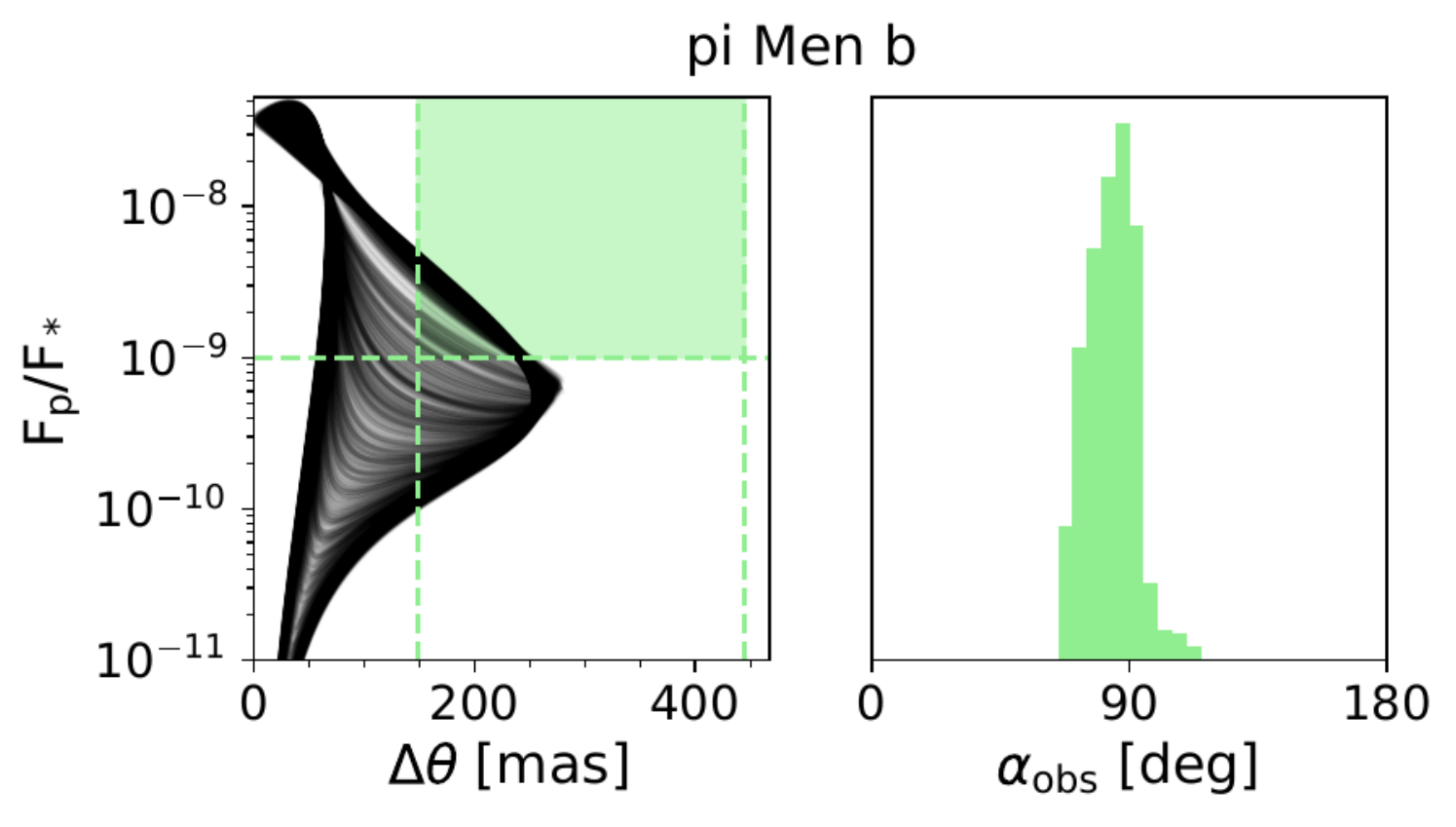} 
	\includegraphics[width=7cm]{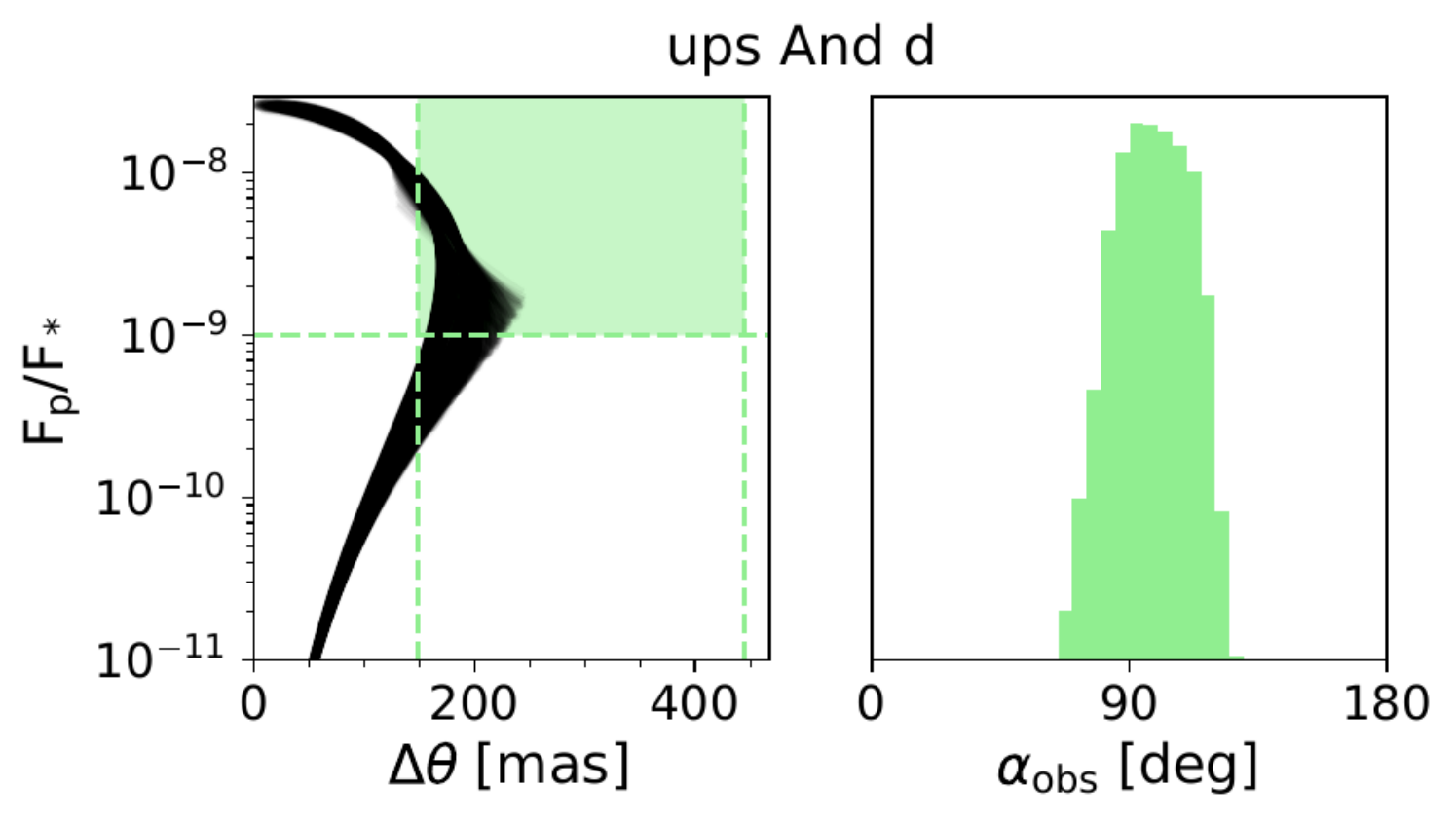} 
	   \\
	\caption{As Fig. \ref{fig:unknownorbits_orbits&histograms}, for pi Men b and ups And d in the optimistic CGI configuration. 
	Upper row: using the value of $\omega$ quoted in the NASA Archive as if it was $\omega_p$. 
	Lower row: using the value of $\omega_p$ after our standardisation process, where we add 180$^\circ$ to the $\omega$ quoted in the NASA Archive which is indeed $\omega_\star$.} \label{fig:appendix_effectomega}
\end{figure*}

As discussed in Sect. \ref{subsubsec:dataprocessing_orbits_longperiast}, the argument of periastron is not consistently reported in literature and this may change by 180$^\circ$ the quoted values of $\omega$ in the exoplanet catalogues.

In Fig. \ref{fig:appendix_effectomega} (upper row) we show, for pi Men b and ups And d, the orbital realizations and ranges of $\alpha_{obs}$ (optimistic CGI config.) for the value of $\omega$ originally reported in the Planet Columns of the NASA Exoplanet Archive.
For comparison, we show (lower row) the results for the value of $\omega_p$ after our standardisation process.
If the originally reported value of $\omega$ is used in our simulations, both pi Men b and ups And d show a wide range of $\alpha_{obs}$: $[42^{+16}_{-3}, 111^{+2}_{-7}]$ and $[38^{+19}_{-1}, 114^{+1}_{-5}]$, respectively.
With the value of $\omega_p$ used in this work (Table \ref{table:NASA_database}), the range of $\alpha_{obs}$ decreases in both cases to $[69^{+7}_{-2}, 95^{+1}_{-1}]$ for pi Men b and $[69^{+5}_{-1}, 124^{+1}_{-5}]$ for ups And d.

Hence, if $\omega_{\star}$ was mistaken for $\omega_p$, these two planets would be found among the ones with better prospects for phase-curve measurements.
Given the brightness of both host stars ($V$=5.67 mag for pi Men and 4.10 mag for ups And), these planets could be mistakenly prioritised for atmospheric characterization attempts.

\clearpage
\section{Sensitivity study: the impact of orbital uncertainties on the detectability.} \label{sec:appendix_sensitivity_study}
\renewcommand{\thefigure}{C.\arabic{figure}}
\begin{figure*}
	\centering
	\includegraphics[width=17.cm]{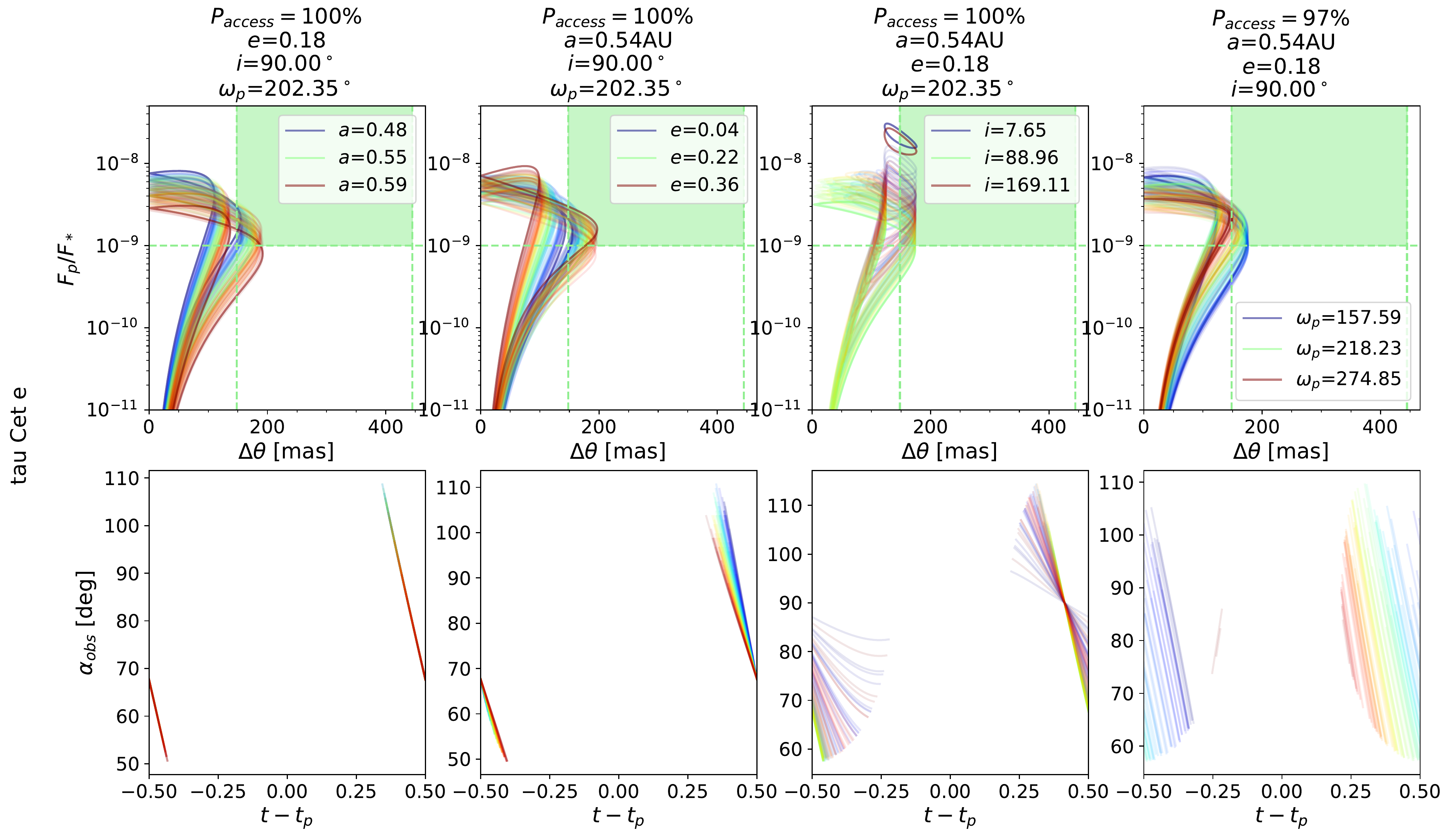} 
\caption{Sensitivity study for tau Cet e. 
Top row shows $F_p/F_\star$-$\Delta \theta$ diagrams in which all orbital parameters remain fixed (as given in the corresponding title) except for the one indicated in the legend, which varies within its upper and lower uncertainties as quoted in the input catalogue (Table \ref{table:NASA_database}).
In case any of the orbital parameters is unknown, for this sensitivity study we fix it to $i=90^\circ$, $e=0$ or $\omega_p=180^\circ$.
Lines with intermediate colours correspond to orbital realizations with intermediate values of that parameter.
A total of 100 orbital realizations are shown.
In the same colour code, bottom row shows the intervals of $\alpha_{obs}$ that would be accessible for each of the orbital realizations in the panel above. 
The optimistic CGI configuration is assumed.
} \label{fig:appendix_sensitivity_study_tauCete}
\end{figure*}

\renewcommand{\thefigure}{C.\arabic{figure}}
\begin{figure*}
	\centering
	\includegraphics[width=17.cm]{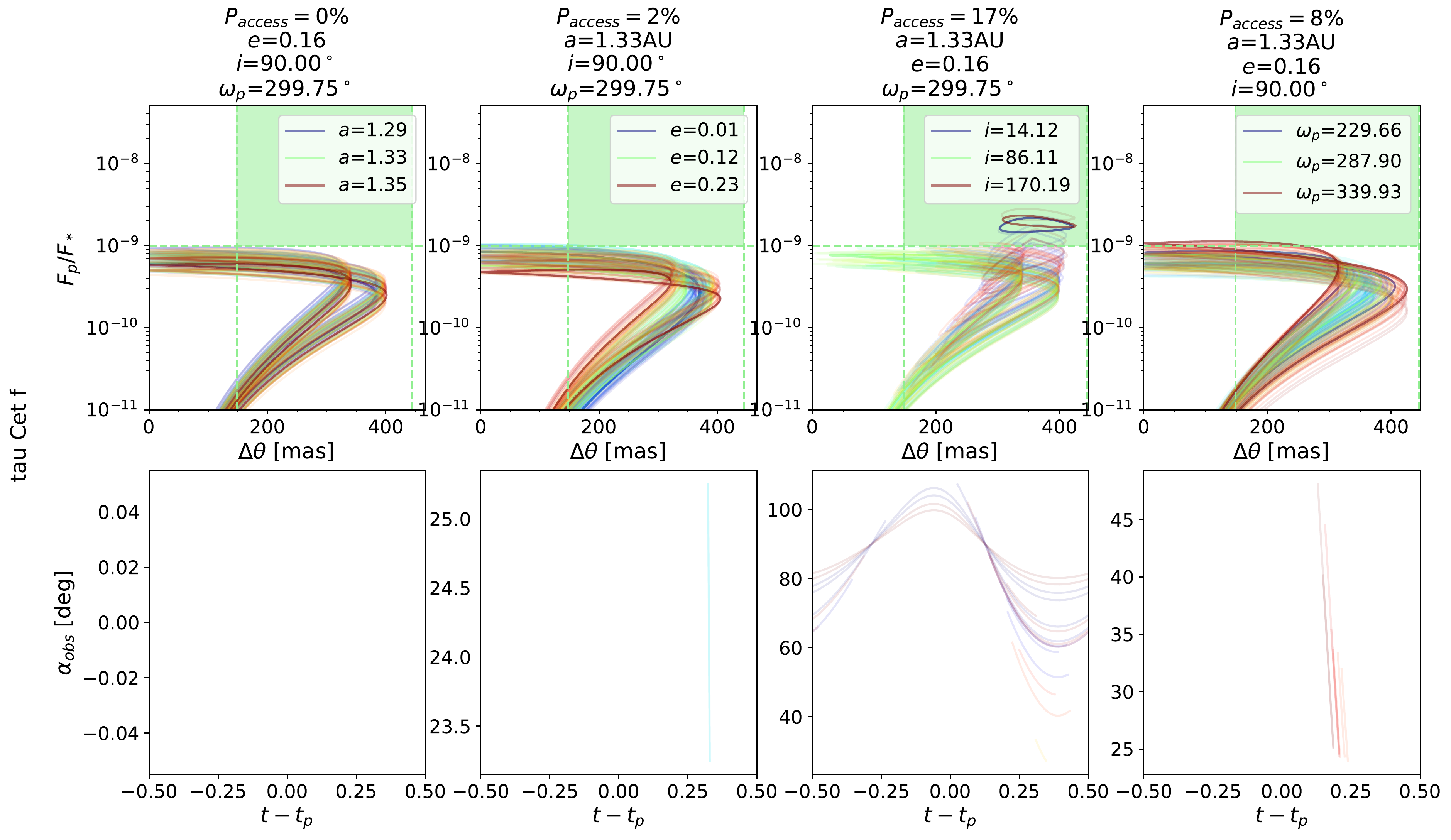} 
\caption{Same as Fig. \ref{fig:appendix_sensitivity_study_tauCete} but for tau Cet f.} \label{fig:appendix_sensitivity_study_tauCetf}
\end{figure*}

\renewcommand{\thefigure}{C.\arabic{figure}}
\begin{figure*}
	\centering
	\includegraphics[width=17.cm]{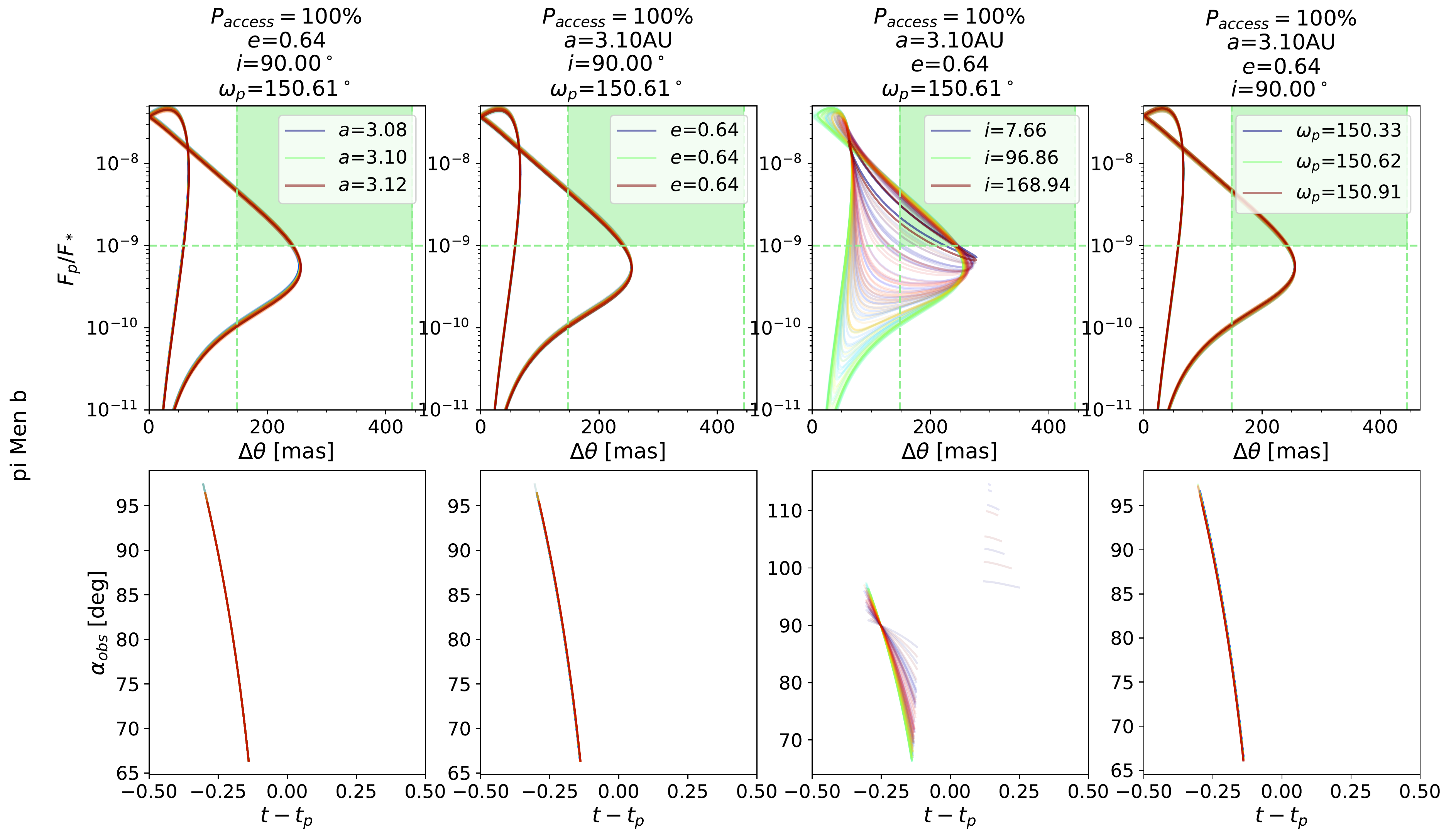} 
\caption{Same as Fig. \ref{fig:appendix_sensitivity_study_tauCete} but for pi Men b.} \label{fig:appendix_sensitivity_study_piMenb}
\end{figure*}

\renewcommand{\thefigure}{C.\arabic{figure}}
\begin{figure*}
	\centering
	\includegraphics[width=17.cm]{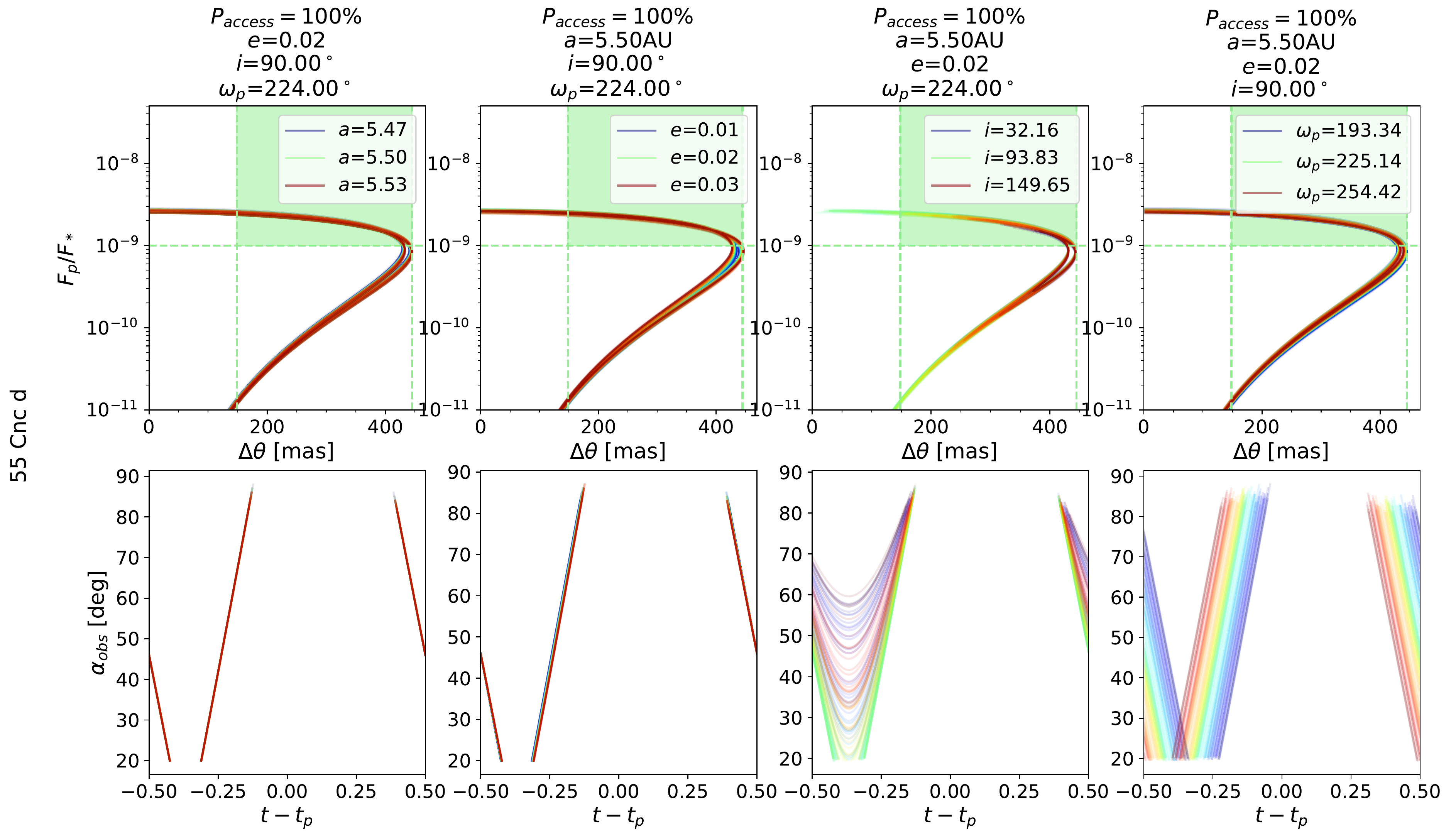} 
\caption{Same as Fig. \ref{fig:appendix_sensitivity_study_tauCete} but for 55 Cnc d.} \label{fig:appendix_sensitivity_study_55Cncd}
\end{figure*}

\renewcommand{\thefigure}{C.\arabic{figure}}
\begin{figure*}
	\centering
	\includegraphics[width=17.cm]{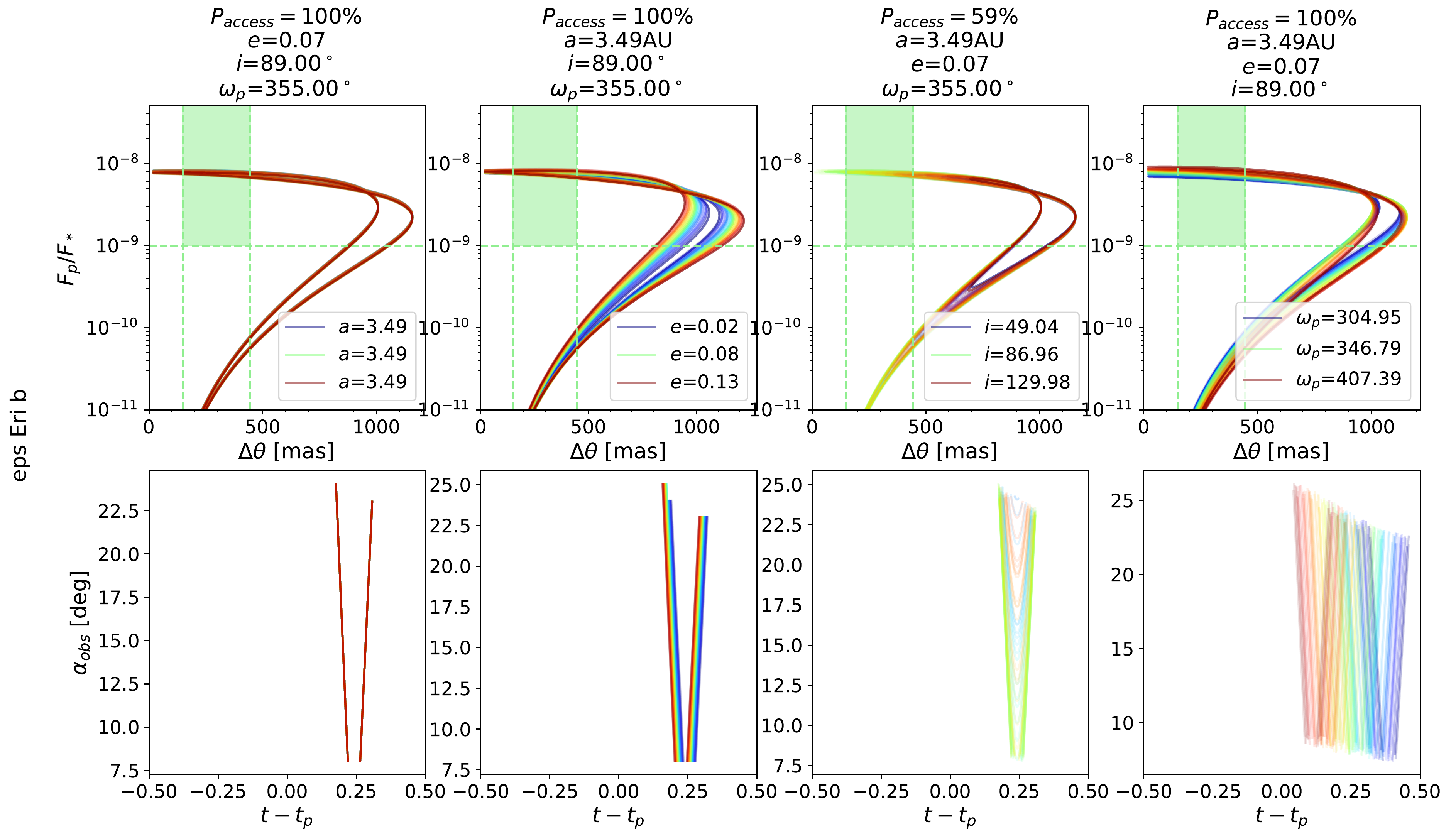} 
\caption{Same as Fig. \ref{fig:appendix_sensitivity_study_tauCete} but for eps Eri b.} \label{fig:appendix_sensitivity_study_epsErib}
\end{figure*}

\renewcommand{\thefigure}{C.\arabic{figure}}
\begin{figure*}
	\centering
	\includegraphics[width=17.cm]{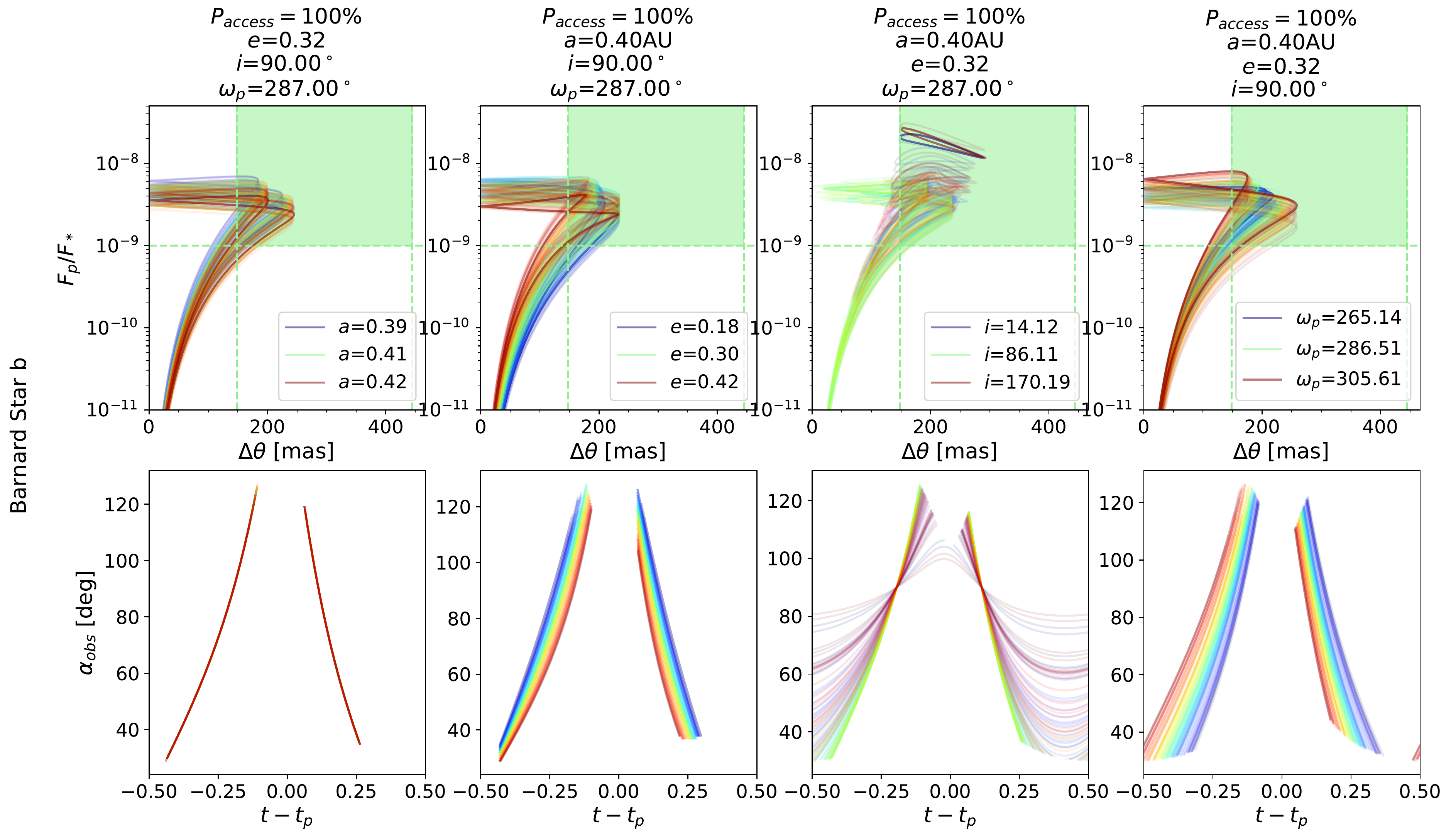} 
\caption{Same as Fig. \ref{fig:appendix_sensitivity_study_tauCete} but for Barnard b.} \label{fig:appendix_sensitivity_study_Barnardb}
\end{figure*}

\renewcommand{\thefigure}{C.\arabic{figure}}
\begin{figure*}
	\centering
	\includegraphics[width=17.cm]{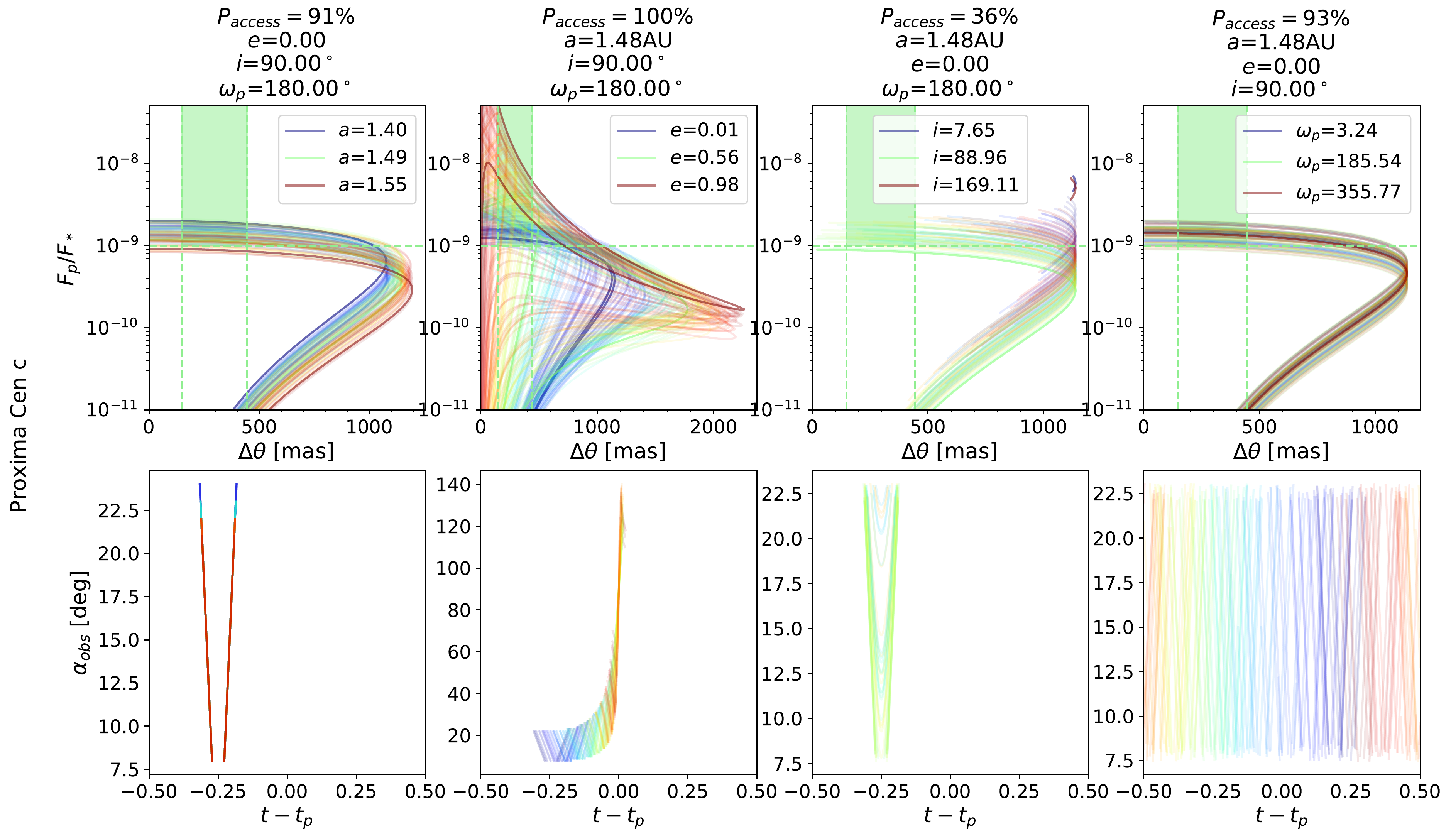} 
\caption{Same as Fig. \ref{fig:appendix_sensitivity_study_tauCete} but for Proxima c.} \label{fig:appendix_sensitivity_study_Proximac}
\end{figure*}

\end{appendix}

\end{document}